\documentclass[reprint,prd,nofootinbib,superscriptaddress]{revtex4}
\usepackage{amssymb,wrapfig,amsmath,hyperref,graphicx,epsfig,bm,color,verbatim,slashed,mathtools,xparse,adjustbox,dcolumn,float,relsize,tikz,cancel}
\usepackage{caption,booktabs}
\usepackage{subcaption}
\usepackage{float}
\usepackage[compat=1.1.0]{tikz-feynman}
\usepackage{subcaption}
\usepackage[english]{babel}
\usepackage[utf8]{inputenc}
\usepackage[mathletters]{ucs}

\newcolumntype{C}{>{$}c<{$}}
\AtBeginDocument{
\heavyrulewidth=.08em
\lightrulewidth=.05em
\cmidrulewidth=.03em
\belowrulesep=.65ex
\belowbottomsep=0pt
\aboverulesep=.4ex
\abovetopsep=0pt
\cmidrulesep=\doublerulesep
\cmidrulekern=.5em
\defaultaddspace=.5em
}

\pdfoutput=1
\newcommand{\Steve}[1]{{#1}}
\newcommand{\Antonio}[1]{{#1}}
\newcommand{\Huchan}[1]{{#1}}
\newcommand{\func}{\operatorname}
\DeclareUnicodeCharacter{2212}{-}
\input{tcilatex}
\begin{document}

\title{$Z$ mediated Flavour Changing Neutral Currents with a Fourth Vector-Like Family}
\author{A. E. C\'{a}rcamo Hern\'{a}ndez}
\email{antonio.carcamo@usm.cl}
\affiliation{Departamento de F\'{\i}sica, Universidad T\'{e}cnica Federico Santa Mar\'{\i}a,\\
 Casilla 110-V, Valpara\'{\i}so, Chile }
\affiliation{{Centro Cient\'{\i}fico-Tecnol\'ogico de Valpara\'{\i}so, Casilla 110-V,
Valpara\'{\i}so, Chile}}
\affiliation{{Millennium Institute for Subatomic Physics at High-Energy Frontier
(SAPHIR), Fern\'andez Concha 700, Santiago, Chile}}

\author{S. F. King}
\email{king@soton.ac.uk}
\affiliation{School of Physics and
Astronomy, University of Southampton,\\
SO17 1BJ Southampton, United Kingdom }

\author{H. Lee}
\email{hl2n18@soton.ac.uk}
\affiliation{School of Physics and
Astronomy, University of Southampton,\\
SO17 1BJ Southampton, United Kingdom }

\date{\today }

\begin{abstract}
We discuss $Z$ mediated flavour changing neutral currents within a model where the hierarchical quark and lepton masses 
are explained via a fourth vector-like family, together with a scalar sector consisting of two Higgs doublets 
augmented by a gauge singlet scalar field that spontaneously breaks an extra global $U(1)^{\prime}$ symmetry.  
The $Z$ mediated flavor violating interactions arise from the mixings between the SM fermions and the vector-like fermions,
where the mixing is discussed in an analytic approximation and also exactly numerically.
We first discuss charged lepton flavor violating (CLFV) $\tau \rightarrow \mu \gamma, \tau \rightarrow 3\mu$ and $Z \rightarrow \mu \tau$ decays and find that they cannot significantly constrain the masses of charged vector-like leptons. However, the $790\func{GeV}$ mass bound arising from collider searches on vector-like lepton doublets can set further constraints on the model parameter space. 
We also consider rare $t \rightarrow c Z$ decays as well as unitarity violation in the CKM mixing in order to constrain the quark sector of the model under consideration. 

\footnotesize
DOI:\href{https://doi.org/10.1103/PhysRevD.105.015021}{10.1103/PhysRevD.105.015021}
\normalsize
\end{abstract}

\maketitle

\section{INTRODUCTION} \label{sec:I}

A great success of the energy frontier is the discovery of the Higgs particle by ATLAS and CMS collaborations at the Large Hadron Collider (LHC) on 4th July 2012~\cite{ATLAS:2012yve,CMS:2012qbp}. After that discovery, no new particle has been found so far by the experiments at 
LHC with $13\func{TeV}$ proton-proton centre of mass energy. This highlights the fact that  
not just the energy frontier but also the luminosity (intensity) frontier should be considered as of equal importance 
in the search for physics beyond the Standard Model (SM). 
For example, one may consider observables mediated by flavour-changing-neutral-currents (FCNCs), which are quite sensitive to new physics, since such FCNC observables are extremely suppressed in the Standard Model
(SM) due to the well-known Glashow-Illiopoulos-Maiani (GIM) mechanism. Another example of a highly suppressed process is provided by 
the branching ratio of $\mu \rightarrow e \gamma$ decay mediated by  
massive neutrinos at the one-loop level~\cite{Calibbi:2017uvl}:
\begin{equation}
\func{BR}\left( \mu \rightarrow e \gamma \right) \approx 10^{-55}.
\end{equation}
The experimentally known sensitivity for the branching ratio of $\mu \rightarrow e \gamma$ is
\begin{equation}
\func{BR}\left( \mu \rightarrow e \gamma \right)_{\func{EXP}} = 4.2 \times 10^{-13}.
\end{equation}
The large gap between the tiny rates of the flavour violating decays predicted by the SM
 and their experimental upper limits has motivated the construction of many flavour models with extended scalar, quark and leptonic spectrum aimed at enhancing those rates by several orders of magnitude up to an observable level within the reach of the sensitivity of the future experiments. A similar situation occurs for other rare FCNC decays such as, for instance $Z\to\mu\tau$ and $t\rightarrow cZ$, which are very suppressed in the SM, \Antonio{but in extensions of the SM,} can acquire sizeable values, within the reach of the future experimental sensitivity.
Although various models with a heavy $Z^\prime$ boson have also 
received a lot of attention by the particle physics community as a new source of FCNCs, its properties, being not fully constrained, do not lead to definite predictions. For this reason we shall restrict ourselves to the SM $Z$ couplings in this paper.

In this paper we focus on the SM $Z$ FCNC interactions induced by tree-level gauge boson exchange in a model in which the fermion sector of the SM is enlarged with a fourth vector-like family. An interesting feature of this approach is all coupling constants of $Z$ interactions in this work are fixed by the known values of the SM $Z$ gauge boson interactions, together with mixing parameters.
Our main motivation for adding a fourth vector-like family is to explain quark and lepton mass hierarchies. We first forbid the SM Yukawa couplings with a global 
$U(1)'$ symmetry, then allow them to be generated effectively via mixing with the fourth vector-like family, a mechanism somewhat analogous to the seesaw mechanism for neutrino masses.
Consequently, the SM charged fermion masses are inversely proportional to the masses of the heavy vector-like leptons and directly proportional to the product of the couplings of Yukawa interactions that mix SM charged fermions with vector-like fermions. This implies that a small hierarchy in those couplings can yield a quadratically larger hierarchy in the
effective couplings. Combined with a moderate hierarchy in the vector-like masses, this allows us to naturally explain the SM charged fermion mass hierarchy and to predict the mass scale of  vector-like fermions.\\~\\
A similar model was discussed in
our previous works~\cite{Hernandez:2021tii}, although with two vector-like families, but the effect on the $Z$ and $W$ boson couplings was not studied.
\Steve{In our previous work~\cite{Hernandez:2021tii}, whose purpose was to explain the muon and electron anomalous magnetic moments simultaneously, the main focus was on the 2HDM scalar sector, and the FCNCs arising from the $Z$ and $W$ boson couplings were not considered, 
since the full mass matrices were not accurately diagonalised, and hence such effects were beyond the approximations used there.
By contrast, the main goal of this work is to study the SM $Z$ and $W$ contributions to the FCNC observables at leading order to constrain the masses of vector-like fermions, and to explore other possible phenomenological signatures. The SM $W$ contributions to the CKM mixing matrix with the extended quark sector are also studied for the first time in this work. In order for these effects to be considered reliably, the mass matrices of each fermion sector are accurately diagonalized, both numerically and analytically, unlike the previous work where simple approximations were used which masked the 
effects we consider here. The results in this work are sufficiently accurate to enable the contribution of the $Z$ and $W$ boson couplings
to physics beyond the SM to be reliably considered for the first time.}

In order to make the results completely transparent, we shall study the $Z$ and $W$ boson couplings in the presence of only one vector-like family 
so that mass matrices of this work can be straightforwardly diagonalized using both analytical and numerical methods. Since only one vector-like fermion family is used, the first generation of SM charged fermions do not acquire masses, which nonetheless is 
a very good approximation considering 
the SM fermions belonging to the first family 
are very light. Consequently, we restrict our attention to 
the second and third generations of SM fermions, 
as well as to several observables related to FCNC processes involving the second and third SM families.
In our approach, then, the SM is a low effective energy theory arising after 
integrating out a single heavy fourth vector-like family. 
In order to dynamically generate the hierarchical structure of SM fermion masses, the fermionic mass matrices given in \cite{Hernandez:2021tii} as well as the ones obtained in this work must be accurately and completely diagonalized, which, as mentioned above, has not been done previously.
The mass matrices for the charged lepton and up-quark sectors share the same structure, whereas the one for 
the down-type quark sector involves 
an additional non-zero element in a particular basis, although we later show that the results are basis independent. 
This reasoning does not apply to the neutrino sector, since this sector is treated independently. 
This different
feature of the down-type mass matrix, in the preferred basis, allows us to achieve
all mixings among the three generations of SM fermions even though the first one remains massless, and 
this leads to a prediction for the Cabbibo-Kobayashi-Maskawa (CKM) mixing matrix. In addition, due to the mixings between the SM quarks and the vector-like quarks, the CKM quark mixing matrix originating from the $W$ couplings
is not unitary, thus implying the need of relaxing the unitarity condition of the CKM mixing matrix, and we also study this feature.
\\~\\
This paper is organized as follows. In Section~\ref{sec:II} we introduce our model to explain the origin of the SM fermion's mass with a fourth vector-like family. In Section~\ref{sec:III} the mass matrices in both quark and lepton sectors are constructed and diagonalized using the mixing formalism. \Antonio{In Section~\ref{sec:IV} the  
$Z$ gauge boson interactions with fermions  
are determined from the mixing matrices used in the mass matrix diagonalization.  
Several FCNC observables for both lepton and quark sectors are analyzed in detail in Sections~\ref{sec:V} and \ref{sec:VI}, respectively. 
We state our conclusions in Section~\ref{sec:VII}. Several technical details are relegated to the Appendices. The perturbative analytical diagonalization of the mass matrices for the charged lepton, up type quark and down type quark sectors are discussed in detail in Appendices \ref{app:A}, \ref{app:B} and \ref{app:C}, respectively. The comparison between the numerical and approximate analytic diagonalization of the mass matrices for charged leptons and quarks is made in Appendices \ref{app:D} and \ref{app:E}, respectively.} 

\section{AN EXTENDED MODEL
WITH A FOURTH VECTOR-LIKE FAMILY} \label{sec:II}

The origin of the pattern of SM fermion masses is interesting open question, not addressed by the SM. The mass parameters of the SM have been experimentally determined with good precision, 
and these experimentally observed mass parameters show a  
strong hierarchical structure of the SM fermion masses. The most extreme hierarchy is exhibited between the SM neutrino Yukawa coupling of about $10^{-12}$ and the top quark Yukawa coupling of about $1$. Regarding the tiny neutrino masses, many particle physicists regard their masses as most likely explained by the see-saw mechanism rather than by the Yukawa interactions, thus predicting 
the presence of the heavy right-handed neutrinos. The reason why the see-saw mechanism has received a large amount of attention by the particle physics community is that 
it provides a dynamical explanation of the tiny active neutrino masses. For a similar reason, it is interesting to speculate about the existence of a dynamical mechanism that produces the masses of all SM fermions via the exchange of heavy fermionic degrees of freedom thus implying that
 the SM is an effective low energy theory arising from some spontaneous breaking at higher energy scales  
 of a more complete underlying theory. 
  In order to specify a possible candidate of an underlying theory responsible for the generation of the SM fermion mass hierarchy,
  we shall consider a minimal extension 
  of the SM 
  consistent with the SM current experimental bounds. With this motivation in mind, we enlarge the SM fermion and scalar sectors by including a fourth vector-like family and an extra $SU(2)$ scalar doublet as well as a scalar singlet, respectively. Furthermore, we extend the SM gauge symmetry by adding a $U(1)^\prime$ global symmetry. The particle content of the proposed model is shown in Table \ref{tab:BSM_model}.
\begin{table}[H]
\resizebox{\textwidth}{!}{
\centering\renewcommand{\arraystretch}{1.3} 
\begin{tabular}{*{21}{c}}
\toprule
\toprule
Field & $Q_{iL}$ & $u_{iR}$ & $d_{iR}$ & $L_{iL}$ & $e_{iR}$ & $Q_{kL}$ & $%
u_{kR}$ & $d_{kR}$ & $L_{kL}$ & $e_{kR} $ & $\nu_{kR}$ & $\widetilde{Q}_{kR}$
& $\widetilde{u}_{kL}$ & $\widetilde{d}_{kL}$ & $\widetilde{L}_{kR}$ & $%
\widetilde{e}_{kL}$ & $\widetilde{\nu}_{kR}$ & $\phi$ & $H_u$ & $H_d$ \\ 
\midrule
$SU(3)_C$ & $\mathbf{3}$ & $\mathbf{3}$ & $\mathbf{3}$ & $\mathbf{1}$
& $\mathbf{1}$ & $\mathbf{3}$ & $\mathbf{3}$ & $\mathbf{3}$ & $\mathbf{1}$ & 
$\mathbf{1}$ & $\mathbf{1}$ & $\mathbf{3}$ & $\mathbf{3}$ & $\mathbf{3}$ & $%
\mathbf{1}$ & $\mathbf{1}$ & $\mathbf{1}$ & $\mathbf{1}$ & $\mathbf{1}$ & $%
\mathbf{1}$ \\ 
$SU(2)_L$ & $\mathbf{2}$ & $\mathbf{1}$ & $\mathbf{1}$ & $\mathbf{2}$
& $\mathbf{1}$ & $\mathbf{2}$ & $\mathbf{1}$ & $\mathbf{1}$ & $\mathbf{2}$ & 
$\mathbf{1}$ & $\mathbf{1}$ & $\mathbf{2}$ & $\mathbf{1}$ & $\mathbf{1}$ & $%
\mathbf{2}$ & $\mathbf{1}$ & $\mathbf{1}$ & $\mathbf{1}$ & $\mathbf{2}$ & $%
\mathbf{2}$ \\ 
$U(1)_Y$ & $\frac{1}{6}$ & $\frac{2}{3}$ & $-\frac{1}{3}$ & $-\frac{1%
}{2}$ & $1$ & $\frac{1}{6}$ & $\frac{2}{3}$ & $-\frac{1}{3}$ & $-\frac{1}{2}$
& $-1$ & $0$ & $\frac{1}{6}$ & $\frac{2}{3}$ & $-\frac{1}{3}$ & $-\frac{1}{2}
$ & $-1$ & $0$ & $0$ & $\frac{1}{2}$ & $-\frac{1}{2}$ \\ 
$U(1)^\prime$ & $0$ & $0$ & $0$ & $0$ & $0$ & $1$ & $-1$ & $-1$ & $1$
& $-1$ & $-1$ & $1$ & $-1$ & $-1$ & $1$ & $-1$ & $-1$ & $1$ & $-1$ & $-1$ \\ 
\bottomrule
\bottomrule
\end{tabular}}%
\caption{Particle assigments under the $SU(3)_C\times SU(2)_L\times U(1)_Y\times U(1)^\prime$ symmetry of the  
extended 2HDM theory 
with fourth vector-like family. 
The index $i=1,2,3$ denotes the 
 the $i$th SM fermion generation and $k=4$ stands for the 
 fourth vector-like family. }
\label{tab:BSM_model}
\end{table}
Our proposed theory is a minimal extended 2 Higgs Doublet Model (2HDM) where the SM fermion sector is enlarged by the inclusion of a fourth vector-like family and the scalar sector is augmented by an extra $SU(2)$ scalar doublet
and a singlet flavon and lastly the SM gauge symmetry is extended by the $U(1)^\prime$ global symmetry. As this model features the global $U(1)^\prime$ symmetry, there is no a  
neutral $Z^\prime$ gauge boson in the particle spectrum. 
Furthermore,  
the up-type quarks feature Yukawa interactions with the 
up-type SM Higgs $H_u$, whereas the extra scalar doublet $H_d$ couples with the SM down-type quarks and charged leptons. 
Our proposed model is especially motivated by the hierarchical structure of the SM and, in order to implement 
this hierarchy, 
we forbid the SM-type Yukawa interactions 
by appropiate $U(1)^\prime$ charge assignments of the scalar and fermionic fields. Then, for the above specified particle content, the following effective Yukawa interactions arise: 
\begin{equation}
\mathcal{L}_{\func{eff}}^{\func{Yukawa}} =y_{ik}^\psi (M_{\psi^\prime}^{-1})_{kl} {x_{lj}^{\psi^\prime}
\left\langle \phi \right\rangle} \overline{%
\psi}_{iL} \widetilde{H} \psi_{jR} + {x_{ik}^{\psi^\prime} \left\langle \phi
\right\rangle}(M_{\psi}^{-1} )_{kl}  y_{lj}^\psi  \overline{\psi}_{iL} \widetilde{H}
\psi_{jR} + \func{h.c.}  \label{eqn:the_effective_Yukawa_Lagrangian}
\end{equation}
where the indices $i,j=1,2,3$ and $k,l=4$ whereas $\psi,\psi^\prime = Q, u, d, L, e$ and $M$ means heavy vector-like mass. The masses of all SM fermions can be explained by this effective Lagrangian of Equation~\ref{eqn:the_effective_Yukawa_Lagrangian}, emphasizing their relative different masses are explained by the factor $\langle \phi \rangle/M \ll 1 \text{ ( apart from top quark )}$, except for the neutrinos which requires an independent approach to their mass. Feynman diagrams corresponding for the effective Lagrangian are shown in Figure \ref{fig:mass_insertion_diagrams}:
\begin{figure}[H]
\centering
\begin{subfigure}{0.48\textwidth}
	\includegraphics[width=1.0\textwidth]{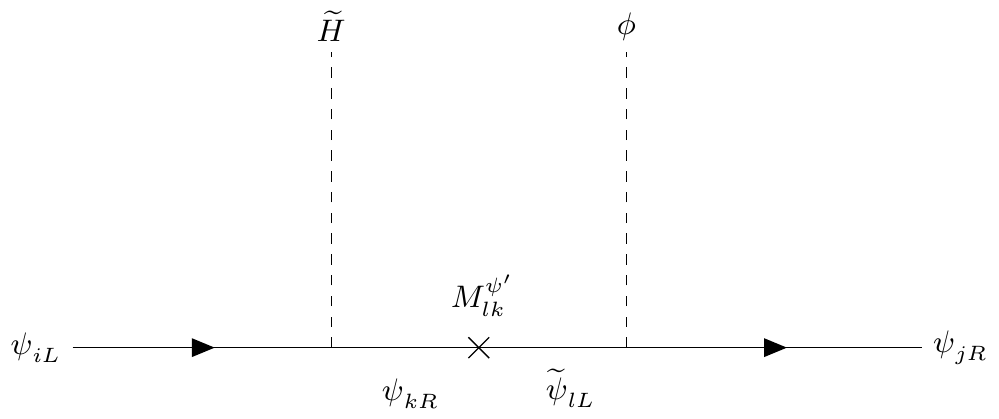}
\end{subfigure} \hspace{0.1cm} 
\begin{subfigure}{0.48\textwidth}
	\includegraphics[width=1.0\textwidth]{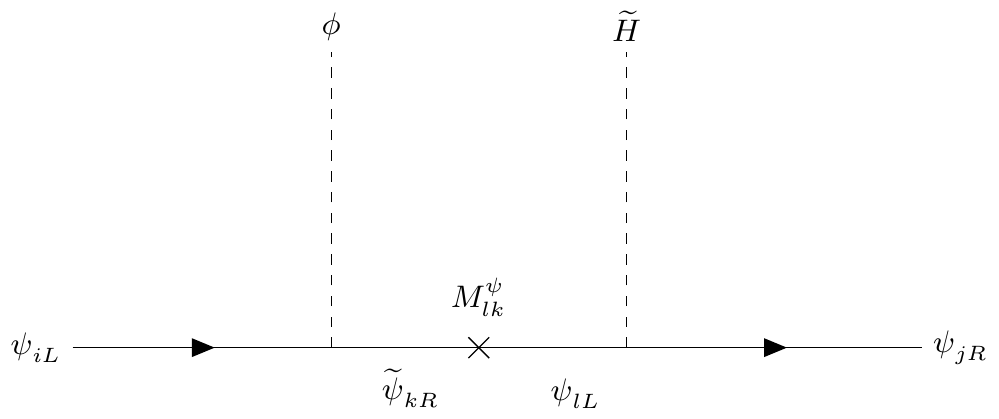}
\end{subfigure}
\caption{Feynman diagrams leading to the effective Yukawa
interactions, where $\protect\psi,\protect\psi^\prime = Q,u,d,L,e$ (neutrinos
will be treated separately), $i,j=1,2,3$, $k,l=4$, $M_{lk}$ is vector-like
mass and $\widetilde{H} = i\protect\sigma_2 H^*, H = H_{u,d}$}
\label{fig:mass_insertion_diagrams}
\end{figure}
The theory considered in this paper corresponds to the one given in  
one of our previous works~\cite{Hernandez:2021tii}, however one vector-like family is used instead of two so that the mass matrices for both quark and lepton sectors can be diagonalized much more economically than in our previous model of \cite{Hernandez:2021tii} at cost of having massless 
the first generation SM charged fermions (One of our main purposes is to diagonalize mass matrices for the quark and lepton sectors without any assumptions) and this is actually a good approximation taking into account that the first generation of SM charged fermions are very light.
\subsection{Effective Yukawa interactions for the SM fermions} \label{sec:II_1}
\begin{figure}[H]
\centering
\begin{subfigure}{0.48\textwidth}
	\includegraphics[width=1.0\textwidth]{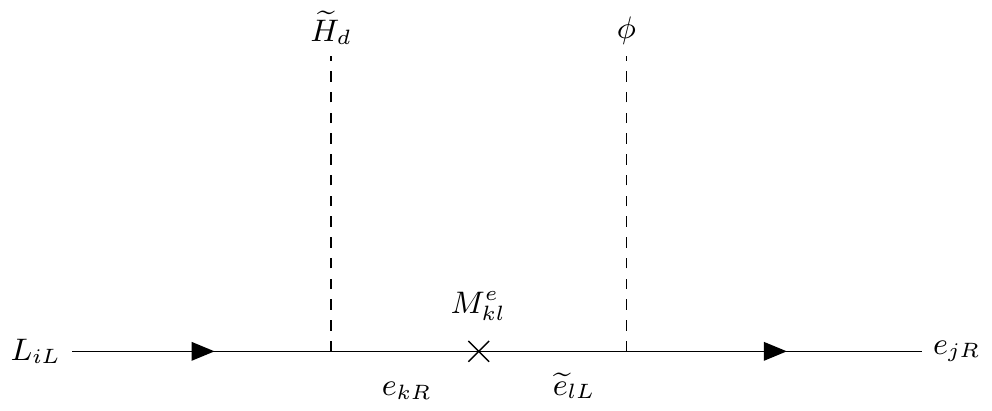}
\end{subfigure} \hspace{0.1cm} 
\begin{subfigure}{0.48\textwidth}
	\includegraphics[width=1.0\textwidth]{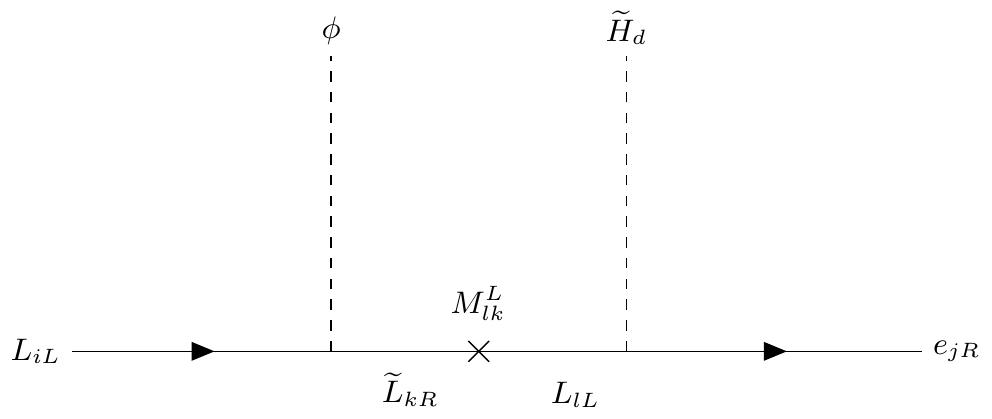}
\end{subfigure}
\caption{Feynman diagrams contributing to the charged lepton's effective Yukawa interactions in the mass insertion formalism. Here $i,j=1,2,3$ and $k,l=4$ and $M_{lk}$ is vector-like mass.}  
\label{fig:diagrams_charged_leptons_mass_insertion}
\end{figure}

\begin{figure}[H]
\centering
\begin{subfigure}{0.48\textwidth}
	\includegraphics[width=1.0\textwidth]{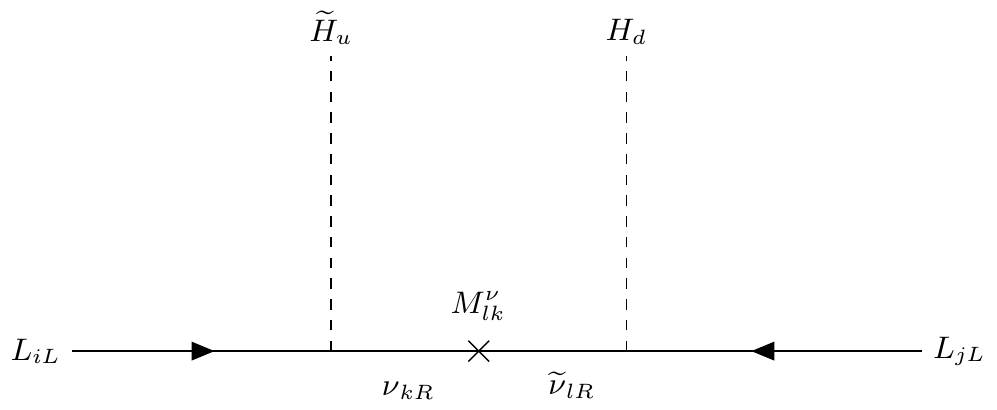}
\end{subfigure}
\caption{Type Ib seesaw diagram~\cite{Hernandez:2021tii,Hernandez-Garcia:2019uof} which leads to the effective Yukawa
interactions for the Majorana neutrinos in mass insertion formalism,
where $i,j=1,2,3$ and $k,l=4$ and $M_{lk}$ is vector-like mass.}
\label{fig:diagrams_neutrinos_mass_insertion}
\end{figure}

The renormalizable interactions of the quark sector in this model are given by:
\begin{equation}
\begin{split}
\mathcal{L}_{q}^{\func{Yukawa+Mass}} &= y_{ik}^{u} \overline{Q}_{iL} 
\widetilde{H}_u u_{kR} + x_{ki}^{u} \phi \overline{\widetilde{u}}_{kL}
u_{iR} + x_{ik}^Q \phi \overline{Q}_{iL} \widetilde{Q}_{kR} + y_{ki}^u 
\overline{Q}_{kL} \widetilde{H}_u u_{iR} \\
&+ y_{ik}^{d} \overline{Q}_{iL} \widetilde{H}_d d_{kR} + x_{ki}^{d} \phi 
\overline{\widetilde{d}}_{kL} d_{iR} + y_{ki}^d \overline{Q}_{kL} \widetilde{%
H}_d d_{iR} \\
&+ M_{kl}^{u} \overline{\widetilde{u}}_{lL} u_{kR} + M_{kl}^{d} \overline{%
\widetilde{d}}_{lL} d_{kR} + M_{kl}^Q \overline{Q}_{kL} \widetilde{Q}_{lR} + 
\func{h.c.}  \label{eqn:general_Quark_Yukawa__Mass_Lagrangian}
\end{split}%
\end{equation}
where $i,j=1,2,3$, $k,l=4$ and $\widetilde{H}=i\sigma _{2}H^{\ast }$. After the $U(1)^\prime$ symmetry is spontaneously broken by the vacuum expectation value (vev) of the 
 the singlet flavon $\phi$, and the heavy vector-like fermions are integrated out, the renormalizable Yukawa terms at higher energy scale give rise to 
 the effective Yukawa interactions which explain the current SM fermion mass hierarchy. The Feynman diagrams corresponding to the effective Yukawa interactions of the quark sector are shown in Figure~\ref{fig:diagrams_quark_mass_insertion}:
\begin{figure}[H]
\centering
\begin{subfigure}{0.48\textwidth}
	\includegraphics[width=1.0\textwidth]{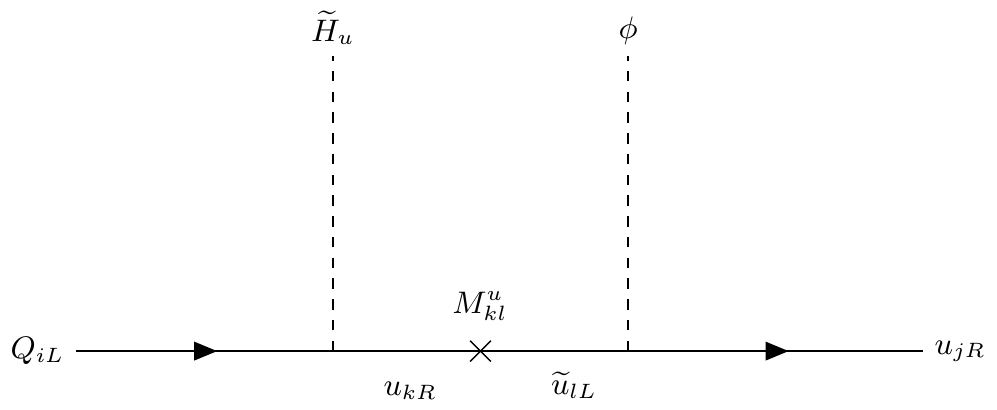}
\end{subfigure} \hspace{0.1cm} 
\begin{subfigure}{0.48\textwidth}
	\includegraphics[width=1.0\textwidth]{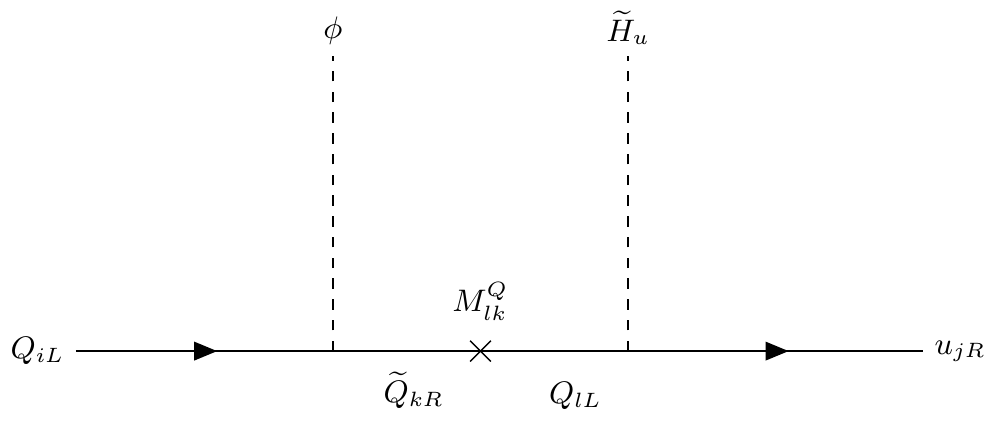}
\end{subfigure}
\begin{subfigure}{0.48\textwidth}
	\includegraphics[width=1.0\textwidth]{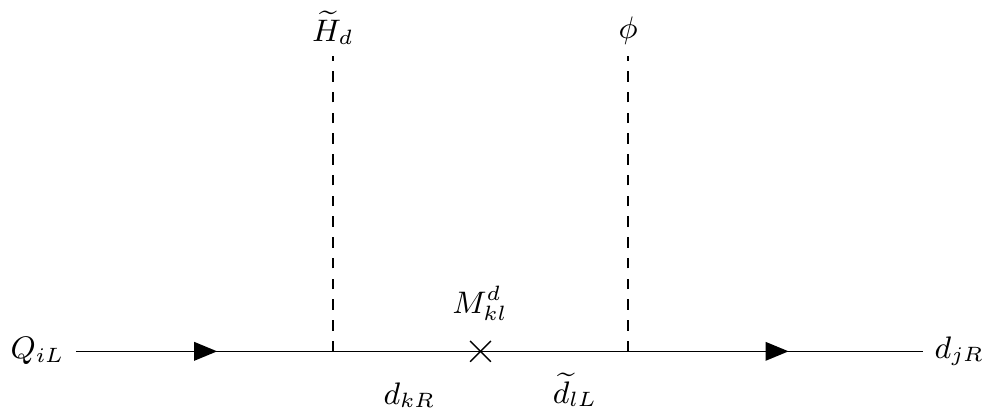}
\end{subfigure} \hspace{0.1cm} 
\begin{subfigure}{0.48\textwidth}
	\includegraphics[width=1.0\textwidth]{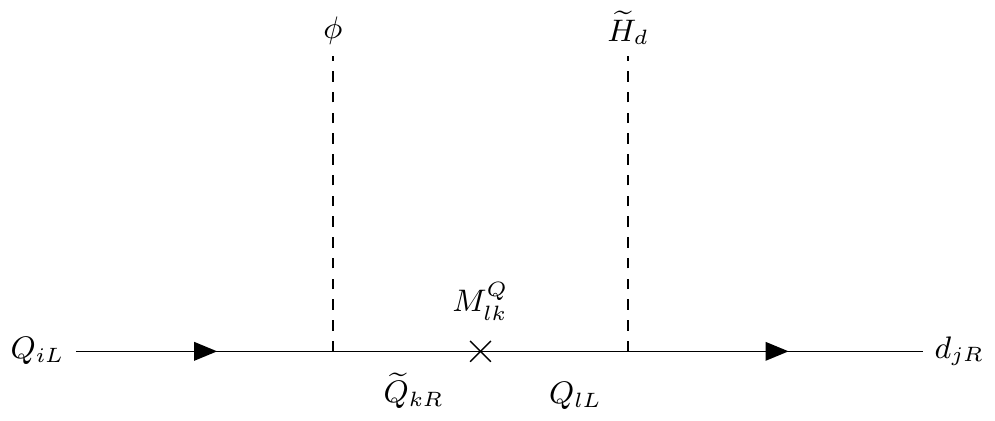}
\end{subfigure}
\caption{Feynman diagrams contributing to the up and down type quark's effective Yukawa
interactions in the mass insertion formalism. Here $i,j=1,2,3$ and $k,l=4$ and $M_{lk}$ is vector-like mass.}  
\label{fig:diagrams_quark_mass_insertion}
\end{figure}
The same approach can be applied to the SM charged lepton sector and the renormalizable charged lepton Yukawa interactions are given by:
\begin{equation}
\begin{split}
\mathcal{L}_{e}^{\func{Yukawa+Mass}} &= y_{ik}^{e} \overline{L}_{iL} 
\widetilde{H}_{d} e_{kR} + x_{ki}^{e} \phi \overline{\widetilde{e}}_{kL}
e_{iR} + x_{ik}^L \phi \overline{L}_{iL} \widetilde{L}_{kR} + y_{ki}^e 
\overline{L}_{kL} \widetilde{H}_{d} e_{iR} 
\\
&+ M_{kl}^{e} \overline{\widetilde{%
e}}_{lL} e_{kR} + M_{kl}^L \overline{L}_{kL} \widetilde{L}_{lR} + \func{h.c.},
\end{split}
\label{eqn:general_charged_lepton_Yukawa_Mass_Lagrangian}
\end{equation}
and its following effective Yukawa interactions read off in Figure~\ref{fig:diagrams_charged_leptons_mass_insertion}.
It is possible to generate the masses of all SM charged fermions by the same method relying on 
effective Yukawa interactions. However, this is not the case for the SM light active neutrinos as they need to be independently treated since the simplest mechanism responsible for generating their tiny masses requires the inclusion of 
Majorana particles in the leptonic spectrum. In order to make the SM neutrinos massive, we made use of 
two important assumptions, one of which is that the SM neutrinos are Majorana particles and the other is they get masses via the type 1b seesaw mechanism~\cite{Hernandez:2021tii,Hernandez-Garcia:2019uof} mediated by the heavy vector-like neutrinos without considering the right-handed neutrinos $\nu_{iR}$. The renormalizable Yukawa interactions for the neutrino sector are given by:
\begin{equation}
\begin{split}
\mathcal{L}_{\nu}^{\func{Yukawa+Mass}} = y_{ik}^{\nu} \overline{L}_{iL} 
\widetilde{H}_u \nu_{kR} + x_{ik}^L \overline{L}_{iL} H_d \overline{%
\widetilde{\nu}}_{kR} + M_{kl}^{M} \overline{\widetilde{\nu}}_{lR} \nu_{kR}
+ \func{h.c.}
\end{split}
\label{eqn:general_neutrinos_Yukawa_Mass_Lagrangian}
\end{equation}
It is worth mentioning that 
the nature of the vector-like mass appearing in Equation~\ref{eqn:general_neutrinos_Yukawa_Mass_Lagrangian} is that the vector-like mass is different than the Majorana mass since the particles involved in vector-like mass terms 
are different, whereas the ones appearing in a Majorana mass terms does not. However, they share 
the common feature that both break the lepton number, which is confirmed 
 by checking each lepton number of $\nu_{kR}$ and $\widetilde{\nu}_{kR}$ in the two Yukawa interactions of Equation~\ref{eqn:general_neutrinos_Yukawa_Mass_Lagrangian}. We call this mechanism ``type 1b seesaw mechanism" and it can allow a different Yukawa interaction at each vertex as seen in Figure~\ref{fig:diagrams_neutrinos_mass_insertion}.
Allowing a different Yukawa interaction at each vertex of Figure~\ref{fig:diagrams_neutrinos_mass_insertion} means that one of the Yukawa interactions can have a very suppressed coupling constant, which can lower the expected order of magnitude of the right-handed Majorana neutrinos masses of the usual type I seesaw mechanism from $10^{14}\func{GeV}$ up to the $\func{TeV}$ scale. The most relevant 
features of the 
model considered in this paper are: 
\begin{enumerate}
\item 
It allows a dynamical and natural explanation of the origin of the observed SM fermion mass hierarchy
\item The model under consideration  
is economical in the sense that it includes a common mechanism for generating the masses of the SM charged fermions via 
 effective Yukawa interactions resulting after integrating out the heavy vector-like fermions.
 \item The expected right-handed neutrinos can have a much smaller mass compared to the ones mediating the usual type I seesaw mechanism, thus allowing to test our model at colliders as well as via charged lepton flavor violating processes.
\end{enumerate}
Now that we have discussed how the SM fermions get massive via the effective Yukawa interactions, the next task is to construct their mass matrices in the flavor basis and then to diagonalize those and this will be discussed in the next section.
\section{EFFECTIVE YUKAWA MATRICES USING A MIXING FORMALISM}
\label{sec:III}
The effective Yukawa interactions discussed in section~\ref{sec:II} give rise to the following 
 mass matrix for fermions written in the flavor basis: 
\begin{equation}
M^{\psi }=\left( 
\begin{array}{c|ccccc}
& \psi _{1R} & \psi _{2R} & \psi _{3R} & \psi _{4R} & \widetilde{\psi }_{4R} \\[0.5ex] \hline
\overline{\psi }_{1L} & 0 & 0 & 0 & y_{14}^{\psi }\langle \widetilde{H}%
^{0}\rangle & x_{14}^{\psi }\langle \phi \rangle \\[1ex]
\overline{\psi }_{2L} & 0 & 0 & 0 & y_{24}^{\psi }\langle \widetilde{H}%
^{0}\rangle & x_{24}^{\psi }\langle \phi \rangle \\[1ex] 
\overline{\psi }_{3L} & 0 & 0 & 0 & y_{34}^{\psi }\langle \widetilde{H}%
^{0}\rangle & x_{34}^{\psi }\langle \phi \rangle \\[1ex]
\overline{\psi }_{4L} & y_{41}^{\psi }\langle \widetilde{H}%
^{0}\rangle & y_{42}^{\psi }\langle \widetilde{H}^{0}\rangle
& y_{43}^{\psi }\langle \widetilde{H}^{0}\rangle & 0 & 
M_{44}^{\psi } \\[1ex]
\overline{\widetilde{\psi }}_{4L} & x_{41}^{\psi ^{\prime }}\langle
\phi \rangle & x_{42}^{\psi ^{\prime }}\langle \phi \rangle
& x_{43}^{\psi ^{\prime }}\langle \phi \rangle & M_{44}^{\psi
^{\prime }} & 0 \\ 
\end{array}%
\right) ,  
\label{eqn:general_55_mass_matrix}
\end{equation}
where $\psi,\psi^{\prime} = Q,u,d,L,e$ and the zeros in the upper-left $3 \times 3$ block mean that the SM fermions 
acquire masses only via their mixing 
 with the fourth vector-like family. The other zeros appearing in the diagonal positions are forbidden by the $U(1)^{\prime}$ charge conservation. This mass matrix was obtained for the first time in 
\cite{King:2018fcg} and it can reveal the hierarchical structure of the SM since this mass matrix involves three different mass scales $\langle H^0 \rangle, \langle \phi \rangle$ and $M$. In order to dynamically reproduce the hierarchical structure of the SM fermion masses,  
  we need to maximally rotate this mass matrix and the resulting maximally rotated mass matrix should be a starting point for our analysis in this work. For the fully rotated mass matrix, the up-quark and charged lepton sectors share the same structure, whereas the down-type quark mass matrix has an  
  additional element since one of the quark doublet rotations was already used in the up quark sector, as it will be shown below.  
\Huchan{\Antonio{Regarding} the diagonalization for each fermion sector, we will employ two methods for comparison; one of which is the numerical SVD diagonalization and the \Antonio{other is the analytical perturbative 
step-by-step diagonalization}. We will make use of the numerical SVD diagonalization for the \Antonio{exact} diagonalization as well as for our numerical scans in main body of this work, however it is important to look at the analytical approximated step-by-step diagonalization since   
\Antonio{it provides an analytical understanding  
on how the SM $Z$ gauge boson} can induce the flavor violating interactions at tree-level and this analytic diagonalization will be covered in Appendix~\ref{app:A} to \ref{app:C}. Lastly, we have found that the analytic diagoanlization for each fermion sector is quite close to its numerical result with very small differences and this feature will be discussed in detail \Antonio{in} Appendix~\ref{app:D} to \ref{app:E}.}
\subsection{Diagonalizing the charged lepton mass matrix} \label{sec:III_1}
After all scalars of our proposed model acquire  
their vevs ($v_{d} = \langle H_d^0 \rangle$ and $v_{\phi} = \langle \phi \rangle$) from Equation~\ref{eqn:general_55_mass_matrix}, we otain the following 
fully rotated mass matrix for the charged lepton sector 
\begin{equation}
M^{e }=\left( 
\begin{array}{c|ccccc}
& e _{1R} & e _{2R} & e _{3R} & e _{4R} & \widetilde{L }_{4R} \\[0.5ex] \hline
\overline{L }_{1L} & 0 & 0 & 0 & 0 & 0 \\[1ex]
\overline{L }_{2L} & 0 & 0 & 0 & y_{24}^{e } v_{d} & 0 \\[1ex] 
\overline{L }_{3L} & 0 & 0 & 0 & y_{34}^{e } v_{d} & x_{34}^{L } v_{\phi} \\[1ex]
\overline{L }_{4L} & 0 & 0 & y_{43}^{e } v_{d} & 0 & M_{44}^{L } \\[1ex]
\overline{\widetilde{e }}_{4L} & 0 & x_{42}^{e} v_{\phi}
& x_{43}^{e} v_{\phi} & M_{44}^{e} & 0 \\ 
\end{array}%
\right), 
\label{eqn:cl_1}
\end{equation}
where we use this fully rotated basis as a starting point in order to easily explain the observed SM fermion mass hierarchy. 
This rotated basis is exactly consistent with the one given in~\cite{King:2018fcg} and we need to explain how the mass matrix of Equation~\ref{eqn:cl_1} is fully rotated. First of all, we rotate the left-handed leptonic fields $L_{1L}$ and $L_{3L}$ to turn off the entry $x_{14}^{L} v_{\phi}$ and then rotate $L_{2L}$ and $L_{3L}$ to trun off the next $x_{24}^{L} v_{\phi}$ entry. Next, we can rotate again the leptonic fields $L_{1L}$ and $L_{2L}$ to turn off the $y_{14}^{e} v_d$ entry. These rotations can be applied to the right-handed leptonic fields $e_{1,2,3R}$ in order to make the zeros appearing in the lower-left $2 \times 3$ block. This fully rotated mass matrix of Equation~\ref{eqn:cl_1} is our starting point to implement both the hierarchical structure of the SM fermion masses and to analyze the flavor violating interactions mediated by the SM $Z$ gauge boson. \Huchan{Before diagonalizing the mass matrix of Equation~\ref{eqn:cl_1}, it is convenient to rearrange the mass matrix by switching the Yukawa terms by mass parameters and then by swapping the fourth and fifth column in order to make the heavy vector-like masses locate in the diagonal positions as given in Equation~\ref{eqn:cl_2}}. 
\begin{equation}
M^{e }
=
\left( 
\begin{array}{c|ccccc}
& e _{1R} & e _{2R} & e _{3R} & e _{4R} & \widetilde{L }_{4R} \\[0.5ex] \hline
\overline{L }_{1L} & 0 & 0 & 0 & 0 & 0 \\[1ex]
\overline{L }_{2L} & 0 & 0 & 0 & m_{24} & 0 \\[1ex] 
\overline{L }_{3L} & 0 & 0 & 0 & m_{34} & m_{35} \\[1ex]
\overline{L }_{4L} & 0 & 0 & m_{43} & 0 & M_{45}^{L } \\[1ex]
\overline{\widetilde{e }}_{4L} & 0 & m_{52} & m_{53} & M_{54}^{e} & 0 \\ 
\end{array}%
\right)
=
\left( 
\begin{array}{c|ccccc}
& e _{1R} & e _{2R} & e _{3R} & \widetilde{L }_{4R} & e _{4R} \\[0.5ex] \hline
\overline{L }_{1L} & 0 & 0 & 0 & 0 & 0 \\[1ex]
\overline{L }_{2L} & 0 & 0 & 0 & 0 & m_{24} \\[1ex] 
\overline{L }_{3L} & 0 & 0 & 0 & m_{35} & m_{34} \\[1ex]
\overline{L }_{4L} & 0 & 0 & m_{43} & M_{45}^{L } & 0 \\[1ex]
\overline{\widetilde{e }}_{4L} & 0 & m_{52} & m_{53} & 0 & M_{54}^{e} \\ 
\end{array}%
\right), 
\label{eqn:cl_2}
\end{equation}
\Huchan{\Antonio{We use two methods for diagonalizing 
the rotated mass matrix of Equation~\ref{eqn:cl_2}, 
one of which corresponds to the 
numerical diagonalization carried out by the singular value decomposition (SVD) and the other} is an approximated analytical step-by-step diagonalization. We make use of the numerical SVD diagonalization for the \Antonio{exact diagonalization and perform numerical scans in main body of this work}, however it is worth discussing the analytic step-by-step diagonalization as it gives an \Antonio{analytical understanding} 
on how the SM $Z$ gauge boson can induce the flavor violating interactions at tree-level with the $SU(2)$ violating mixings, which will be defined in the analytic diagonalization covered in Appendix~\ref{app:A}. \Antonio{From the comparison between the analytic and numerical computations, we found that the former 
 works quite well and yields resuls close the ones obtained from the latter. The comparisons between the analytic and numerical computations for both lepton and quark sectors will be discussed in detail in   
 Appendices \ref{app:D} and \ref{app:E}, respectively}. The charged lepton sector can be diagonalized by \Antonio{peforming} the SVD diagonalization as follows:}
\begin{equation}
M^{e \prime} = \func{diag}\left( 0, m_{\mu}, m_{\tau}, M_{E_4}, M_{\widetilde{E}_4} \right) = V^{L} M^{e} (V^{e})^{\dagger},
\end{equation}
\Huchan{where $V^{L} (V^{e})$ is the mixing matrix for the left-handed (right-handed) leptonic fields, \Antonio{defined} as follows:}
\begin{equation}
\begin{split}
\begin{pmatrix}
e_{L} \\[0.5ex]
\mu_{L} \\[0.5ex]
\tau_{L} \\[0.5ex]
E_{4L} \\[0.5ex]
\widetilde{E}_{4L}
\end{pmatrix}
=
V^L  
\begin{pmatrix}
e_{1L} \\[0.5ex]
e_{2L} \\[0.5ex]
e_{3L} \\[0.5ex]
e_{4L} \\[0.5ex]
\widetilde{e}_{4L}
\end{pmatrix},
\qquad
\begin{pmatrix}
e_{R} \\[0.5ex]
\mu_{R} \\[0.5ex]
\tau_{R} \\[0.5ex]
\widetilde{E}_{4R} \\[0.5ex]
E_{4R}
\end{pmatrix}
=
V^e
\begin{pmatrix}
e_{1R} \\[0.5ex]
e_{2R} \\[0.5ex]
e_{3R} \\[0.5ex]
\widetilde{e}_{4R} \\[0.5ex]
e_{4R}
\end{pmatrix},
\label{eqn:cl_mixing}
\end{split}
\end{equation}
\Huchan{and the numerical mixing matrices $V^{L,e}$ can be expressed by an analytic expression consisting of a series of $V_{ij}^{L,e}$ which describes mixing between $i$th and $j$th fermion where $i,j=1,2,3,4,5$ and this will be discussed in Appendix~\ref{app:A}.}
\subsection{Diagonalizing the up-type quark mass matrix} \label{sec:III_2}
The initial mass matrix for the up-type quark sector in the flavor basis is given by:
\begin{equation}
M^{u }
=
\left( 
\begin{array}{c|ccccc}
& u _{1R} & u _{2R} & u _{3R} & u _{4R} & \widetilde{Q }_{4R} \\[0.5ex] \hline
\overline{Q }_{1L} & 0 & 0 & 0 & 0 & 0 \\[1ex]
\overline{Q }_{2L} & 0 & 0 & 0 & y_{24}^{u } v_{u} & 0 \\[1ex] 
\overline{Q }_{3L} & 0 & 0 & 0 & y_{34}^{u } v_{u} & x_{34}^{Q } v_{\phi} \\[1ex]
\overline{Q }_{4L} & 0 & 0 & y_{43}^{u } v_{u} & 0 & M_{44}^{Q } \\[1ex]
\overline{\widetilde{u }}_{4L} & 0 & x_{42}^{u} v_{\phi}
& x_{43}^{u} v_{\phi} & M_{44}^{u} & 0 \\ 
\end{array}%
\right)
=
\left( 
\begin{array}{c|ccccc}
& u _{1R} & u _{2R} & u _{3R} & \widetilde{Q }_{4R} & u _{4R} \\[0.5ex] \hline
\overline{Q }_{1L} & 0 & 0 & 0 & 0 & 0 \\[1ex]
\overline{Q }_{2L} & 0 & 0 & 0 & 0 & m_{24}^{u} \\[1ex] 
\overline{Q }_{3L} & 0 & 0 & 0 & m_{35}^{u} & m_{34}^{u} \\[1ex]
\overline{Q }_{4L} & 0 & 0 & m_{43}^{u} & M_{44}^{Q } & 0 \\[1ex]
\overline{\widetilde{u }}_{4L} & 0 & m_{52}^{u}
& m_{53}^{u} & 0 & M_{44}^{u} \\ 
\end{array}%
\right)
, \label{eqn:uq_1}
\end{equation}
The mass matrix of Equation~\ref{eqn:uq_1} in the flavor basis is exactly consistent with the one corresponding to the charged lepton sector excepting for a few substitutions $y^{e} \rightarrow y^{u}$, $v_{d} \rightarrow v_{u}$, $x^{L} \rightarrow x^{Q}$ and $x^{e} \rightarrow x^{u}$. 
\Huchan{The analytic mixing matrix for the up-quark sector is exactly same as the one for the charged lepton sector, however \Antonio{unlike the charged lepton sector, 
diagonalizing the mass matrix for the up-quark sector requires} more caution as some numerical off-diagonal elements \Antonio{of order unity.} 
This feature is resulted from the 
some off-diagonal elements arising as a result of mixing between the heavy top quark mass and the other heavy exotic up-type quark masses are of order unitiry, not small enough to obtain precise results in the perturbative diagonalization when compared to the charged lepton sector. We discuss this feature in Appendix~\ref{app:E} by comparing a numerical mixing matrix \Antonio{obtained from the SVD with the one resulting from the analytic diagonalization.} Then we can numerically diagonalize the up-type mass matrix by using the SVD diagonalization as follows:}
\begin{equation}
\begin{split}
\begin{pmatrix}
u_{L} \\[0.5ex]
c_{L} \\[0.5ex]
t_{L} \\[0.5ex]
U_{4L} \\[0.5ex]
\widetilde{U}_{4L}
\end{pmatrix}
=
V_{L}^{u}  
\begin{pmatrix}
u_{1L} \\[0.5ex]
u_{2L} \\[0.5ex]
u_{3L} \\[0.5ex]
u_{4L} \\[0.5ex]
\widetilde{u}_{4L}
\end{pmatrix}
, \qquad
\begin{pmatrix}
u_{R} \\[0.5ex]
c_{R} \\[0.5ex]
t_{R} \\[0.5ex]
\widetilde{U}_{4R} \\[0.5ex]
U_{4R}
\end{pmatrix}
=
V_{R}^{u}
\begin{pmatrix}
u_{1R} \\[0.5ex]
u_{2R} \\[0.5ex]
u_{3R} \\[0.5ex]
\widetilde{u}_{4R} \\[0.5ex]
u_{4R}
\end{pmatrix}
\label{eqn:up_mixing}
\end{split}
\end{equation}
\Antonio{where 
the symbol $L$ means left-handed doublet and $e$ denotes right-handed singlet in the charged lepton sector, however  
it is worth mentioning that the above described notation used in the lepton sector} becomes complicated in the quark sector since the mass matrices for the up- and down-type quark sectors have a different form, so we change the mixing notation by $V_{L(R)}^{u,d}$ instead of $V^{Q}$.
\Huchan{The analytic diagonalizations for the up-quark sector \Antonio{will be} discussed in Appendix~\ref{app:B}.}
\subsection{Diagonalizing the down-type quark mass matrix} \label{sec:III_3}
A nice  
feature of the model under consideration is that it can predict 
the CKM mixing matrix and this feature is mainly based on the mixings derived from the down-type quark mass matrix as we will see soon. A quite encouraging feature is that all the mixings among the three SM generations in the down-type quark sector can be accessible even though the first generation of the down-type quark sector remains massless and this feature is quite naturally attributed to this model with the vector-like family. We start from two mass matrices, one of which is for the up-quark sector whereas the another one is for the down-quark sector. 
\begin{equation}
M^{u }=\left( 
\begin{array}{c|ccccc}
& u _{1R} & u _{2R} & u _{3R} & u _{4R} & \widetilde{Q }_{4R} \\[0.5ex] \hline
\overline{Q }_{1L} & 0 & 0 & 0 & 0 & 0 \\[1ex]
\overline{Q }_{2L} & 0 & 0 & 0 & y_{24}^{u } v_{u} & 0 \\[1ex] 
\overline{Q }_{3L} & 0 & 0 & 0 & y_{34}^{u } v_{u} & x_{34}^{Q } v_{\phi} \\[1ex]
\overline{Q }_{4L} & 0 & 0 & y_{43}^{u } v_{u} & 0 & M_{44}^{Q } \\[1ex]
\overline{\widetilde{u }}_{4L} & 0 & x_{42}^{u} v_{\phi}
& x_{43}^{u} v_{\phi} & M_{44}^{u} & 0 \\ 
\end{array}%
\right), 
\quad
M^{d }=\left( 
\begin{array}{c|ccccc}
& d _{1R} & d _{2R} & d _{3R} & d _{4R} & \widetilde{Q }_{4R} \\[0.5ex] \hline
\overline{Q }_{1L} & 0 & 0 & 0 & y_{14}^{d } v_{d} & 0 \\[1ex]
\overline{Q }_{2L} & 0 & 0 & 0 & y_{24}^{d } v_{d} & 0 \\[1ex] 
\overline{Q }_{3L} & 0 & 0 & 0 & y_{34}^{d } v_{d} & x_{34}^{Q } v_{\phi} \\[1ex]
\overline{Q }_{4L} & 0 & 0 & y_{43}^{d } v_{d} & 0 & M_{44}^{Q } \\[1ex]
\overline{\widetilde{d }}_{4L} & 0 & x_{42}^{d} v_{\phi}
& x_{43}^{d} v_{\phi} & M_{44}^{d} & 0 \\ 
\end{array}%
\right), 
\label{eqn:diff_up_down}
\end{equation}
This difference between the mass matrices for the up-type and down-type quark sectors was noticed for the first time in~\cite{King:2018fcg}.  
The first property we need to focus on is the fifth column of both mass matrices is exactly same. The zeros appearing in the fifth column of both are the common elements shared by both up- and down-type quark mass matrices since the quark doublets as well as the fourth vector-like quark doublets contribute equally to both sectors. For the up-type quark sector, we were able to rotate further between $Q_{1L}$ and $Q_{2L}$ to vanish $y_{14}^{u} v_{u}$, however this rotation simply remixes $y_{14}^{d} v_{d}$ and $y_{24}^{d} v_{d}$, so both the down-type Yukawa terms survive. For the lower-left $2 \times 3$ block of the down-type quark mass matrix, the same zeros can appear since the down-type quarks $d_{1,2,3R}$ have a different mixing angle against that for the up-type quarks $u_{1,2,3R}$. The down-type mass matrix $M^{d}$ of Equation~\ref{eqn:diff_up_down} can be diagonalized by the numerical SVD diagonalization as follows:
\begin{equation}
\begin{split}
\begin{pmatrix}
d_{L} \\
s_{L} \\
b_{L} \\
D_{4L} \\
\widetilde{D}_{4L}
\end{pmatrix}
=
V_{L}^{d}  
\begin{pmatrix}
d_{1L} \\
d_{2L} \\
d_{3L} \\
d_{4L} \\
\widetilde{d}_{4L}
\end{pmatrix}
, \qquad
\begin{pmatrix}
d_{R} \\
s_{R} \\
b_{R} \\
D_{4R} \\
\widetilde{D}_{4R}
\end{pmatrix}
=
V_{R}^{d}
\begin{pmatrix}
d_{1R} \\
d_{2R} \\
d_{3R} \\
\widetilde{d}_{4R} \\
d_{4R}
\end{pmatrix}.
\label{eqn:down_mixing}
\end{split}
\end{equation}
\Huchan{The analytic diagonalization for the down-quark sector is discussed in Appendix~\ref{app:C}.}
We will see that a numerical mixing matrix derived by the SVD is quite close to one by the analytic diagonalization in Appendix~\ref{app:E}. In Appendix~\ref{app:E} we confirm that even though the numerical mixing matrix $V_{L}^{d}$ can have all mixings among the three SM generations, the $Z$ coupling constants $D_{L}^{d \prime}$ in the mass basis will have zeros in the first column and row due to some internal cancellations. Therefore the whole structure of $D_{L}^{d \prime}$ is exactly same as the other $Z$ coupling constants $D_{L,R}^{u \prime}$ and $D_{R}^{d \prime}$ in the mass basis, so we verify that the SM $Z$ physics does not get affected by any specific basis we choose. This feature will be discussed again in the next section as well as in Appendix~\ref{app:E}.
\section{THE SM $Z$ GAUGE BOSON INTERACTIONS WITH THE VECTOR-LIKE FAMILY}
\label{sec:IV}
One of our main motivations of this work is to study flavor violating 
processes mediated by the $Z$ gauge boson in order to constrain the mass range of the vector-like fermions. It is worth reminding that the neutral $Z^{\prime}$ gauge boson does not appear in the particle spectrum of this model due to the global $U(1)^{\prime}$ symmetry of the theory under consideration. It is worth mentioning that the tree-level flavor violating $Z$ decays are absent in the SM, indepently of the fermion mixings, as can be seen from Equation~\ref{eqn:SM_Z}
shown below:
\begin{equation}
\mathcal{L}_{\func{SM}}^{Z} = g Z_{\mu} J_{Z}^{\mu} = \frac{g}{c_w} Z_{\mu} \sum_{f=e,\mu,\tau} \overline{f} \gamma^{\mu} \left( T^3 - \sin^2\theta_w Q \right) f
\label{eqn:SM_Z}
\end{equation}
Factoring out the prefactor $g/c_{w}$, we can find matrices $D_{L,R}^{e}$, which determine the magnitude of the coupling constant for the $Z$ interactions to either the left-handed or right-handed SM fermions.
\begin{equation}
\begin{split}
D_{L}^{e}&= 
\left(
\begin{array}{c|ccc}
  & e_{1L} & e_{2L} & e_{3L} \\[0.5ex]
  \hline
\overline{e}_{1L} & \left( -\frac{1}{2}+\sin^2\theta_w \right) & 0 & 0 \\[1ex]
\overline{e}_{2L} & 0 & \left( -\frac{1}{2}+\sin^2\theta_w \right) & 0 \\[1ex]
\overline{e}_{3L} & 0 & 0 & \left( -\frac{1}{2}+\sin^2\theta_w \right) \\[1ex]
\end{array}
\right)
\\
D_{R}^{e}&= 
\left(
\begin{array}{c|ccc}
  & e_{1R} & e_{2R} & e_{3R} \\[0.5ex]
  \hline
\overline{e}_{1R} & \left(\sin^2\theta_w \right) & 0 & 0 \\[1ex]
\overline{e}_{2R} & 0 & \left(\sin^2\theta_w \right) & 0 \\[1ex]
\overline{e}_{3R} & 0 & 0 & \left(\sin^2\theta_w \right) \\[1ex]
\end{array}
\right)
\end{split}
\end{equation}
However, this SM $Z$ gauge boson can cause the renormalizable flavor violating interactions with the SM fermions by extending the SM fermion sector by the vector-like fermions as well as by considering the $SU(2)$ violating mixings defined in Appendix~\ref{app:A} together. This features will be discussed in detail in the following subsections.
\subsection{FCNC mediated by the SM $Z$ gauge boson in the charged lepton sector with the fourth vector-like charged leptons} \label{sec:IV_1}
We can construct an extended the SM $Z$ coupling constants in the charged lepton sector with the vector-like charged leptons in the flavor basis.
\begin{equation}
\begin{split}
D_{L}^{e} 
&= 
\left(
\begin{array}{c|ccccc}
  & e_{1L} & e_{2L} & e_{3L} & e_{4L} & \widetilde{e}_{4L} \\[1ex]
  \hline
\overline{e}_{1L} & \left( -\frac{1}{2}+\sin^2\theta_w \right) & 0 & 0 & 0 & 0 \\[1ex]
\overline{e}_{2L} & 0 & \left( -\frac{1}{2}+\sin^2\theta_w \right) & 0 & 0 & 0 \\[1ex]
\overline{e}_{3L} & 0 & 0 & \left( -\frac{1}{2}+\sin^2\theta_w \right) & 0 & 0 \\[1ex]
\overline{e}_{4L} & 0 & 0 & 0 & \left( -\frac{1}{2}+\sin^2\theta_w \right) & 0 \\[1ex]
\overline{\widetilde{e}}_{4L} & 0 & 0 & 0 & 0 & \left( \sin^2\theta_w \right) 
\end{array}
\right)
\\
D_{R}^{e}
&= 
\left(
\begin{array}{c|ccccc}
  & e_{1R} & e_{2R} & e_{3R} & \widetilde{e}_{4R} & e_{4R} \\[1ex]
  \hline
\overline{e}_{1R} & \left( \sin^2\theta_w \right) & 0 & 0 & 0 & 0 \\[1ex]
\overline{e}_{2R} & 0 & \left( \sin^2\theta_w \right) & 0 & 0 & 0 \\[1ex]
\overline{e}_{3R} & 0 & 0 & \left( \sin^2\theta_w \right) & 0 & 0 \\[1ex]
\overline{\widetilde{e}}_{4R} & 0 & 0 & 0 & \left( -\frac{1}{2}+\sin^2\theta_w \right) & 0 \\[1ex]
\overline{e}_{4R} & 0 & 0 & 0 & 0 & \left( \sin^2\theta_w \right) 
\end{array}
\right)
\label{eqn:SMZ_DLe_DRe}
\end{split}
\end{equation}
where it is worth reminding that the order of the left-handed fermions is 12345, whereas that of the right-handed fermions is 12354 (This ordering is stressed in Appendix~\ref{app:A}). An important feature of this SM $Z$ coupling constants of Equation~\ref{eqn:SMZ_DLe_DRe} is the couplings constants are naturally determined, based on the nature of the vector-like charged leptons, without imposing neither any other symmetry nor other constraints. Therefore, the SM $Z$ coupling constants of Equation~\ref{eqn:SMZ_DLe_DRe} are not the identity matrix anymore unlike the case for 
the SM charged leptons. From these considerations, it follows that there can exist non-zero  
off-diagonal elements in the mass basis by operating the $SU(2)$ violating mixings. Reminding the mixing matrices for the left- or right-handed charged leptons of Equation~\ref{eqn:cl_umm_LH_RH}, the coupling constant in the mass basis ($D_{L,R}^{e \prime}$) can be written as follows:
\begin{equation}
\begin{split}
D_{L}^{e \prime} &= V^{L} D_{L}^{e} (V^{L})^{\dagger}
\\[1ex]
 &= V_{45}^{L} V_{23}^{L} V_{25}^{L} V_{35}^{L} V_{34}^{L} D_{L}^{e} (V_{34}^{L})^{\dagger} (V_{35}^{L})^{\dagger} (V_{25}^{L})^{\dagger} (V_{23}^{L})^{\dagger} (V_{45}^{L})^{\dagger}
\\[1ex]
D_{R}^{e \prime} &= V^{e} D_{R}^{e} (V^{e})^{\dagger}
\\[1ex]
 &= V_{54}^{e} V_{23}^{e} V_{25}^{e} V_{35}^{e} V_{24}^{e} V_{34}^{e} D_{R}^{e} (V_{34}^{e})^{\dagger} (V_{24}^{e})^{\dagger} (V_{35}^{e})^{\dagger} (V_{25}^{e})^{\dagger} (V_{23}^{e})^{\dagger} (V_{54}^{e})^{\dagger}
\label{eqn:SMZ_cl_mb_LR_1}
\end{split}
\end{equation}
It is possible to make the SM $Z$ coupling constants of Equation~\ref{eqn:SMZ_cl_mb_LR_1} simpler by using the $SU(2)$ conserving mixing as it just remixes an identity matrix.
\begin{equation}
\begin{split}
D_{L}^{e \prime} &= V_{45}^{L} V_{23}^{L} V_{25}^{L} V_{35}^{L} D_{L}^{e} (V_{35}^{L})^{\dagger} (V_{25}^{L})^{\dagger} (V_{23}^{L})^{\dagger} (V_{45}^{L})^{\dagger}
\\[1ex]
D_{R}^{e \prime} &= V_{54}^{e} V_{23}^{e} V_{25}^{e} V_{35}^{e} D_{R}^{e} (V_{35}^{e})^{\dagger} (V_{25}^{e})^{\dagger} (V_{23}^{e})^{\dagger} (V_{54}^{e})^{\dagger}
\label{eqn:SMZ_cl_mb_LR_2}
\end{split}
\end{equation}
However, the following mixing matrices after the $SU(2)$ violating mixings $V_{35}^{L,e}$ must be conserved since all of them contribute to the off-diagonal elements of the coupling constants in the mass basis. It is insightful to look at the coupling constants $D_{L,R}^{e \prime}$ in the mass basis (We substitute $\left( -1/2+\sin^2\theta_w \right)$ by $a$ and $\left( \sin^2\theta_w \right)$ by $b$ and assume the mixing angles $\theta_{35,25,23,45(54)}^{L,e}$ are small enough to approximate for simplicity).
\begin{equation}
\begin{split}
D_{L}^{e \prime} &\approx 
\begin{pmatrix}
a & 0 & 0 & 0 & 0 \\[1ex]
0 & a(1 + \theta_{23}^{L2}) + b \theta_{25}^{L2} & b \theta_{25}^{L} \theta_{35}^{L} & b \theta_{25}^{L} \theta_{45}^{L} & (a-b) \theta_{25}^{L} \\[1ex]
0 & b \theta_{25}^{L} \theta_{35}^{L} & a(1+\theta_{23}^{L2}) + b \theta_{35}^{L2} & b \theta_{35}^{L} \theta_{45}^{L} & (a-b) \theta_{35}^{L} \\[1ex]
0 & b \theta_{25}^{L} \theta_{45}^{L} & b \theta_{35}^{L} \theta_{45}^{L} & a + b \theta_{45}^{L2} & (a-b) \theta_{45}^{L} \\[1ex]
0 & (a-b) \theta_{25}^{L} & (a-b) \theta_{35}^{L} & (a-b) \theta_{45}^{L} & b + a(\theta_{25}^{L2} + \theta_{35}^{L2} + \theta_{45}^{L2})
\end{pmatrix}
\\[1ex]
D_{R}^{e \prime} &\approx 
\begin{pmatrix}
b & 0 & 0 & 0 & 0 \\[1ex]
0 & b(1 + \theta_{23}^{e2}) + a \theta_{25}^{e2} & a \theta_{25}^{e} \theta_{35}^{e} & (b-a) \theta_{25}^{e} & -a \theta_{25}^{e} \theta_{54}^{e} \\[1ex]
0 & a \theta_{25}^{e} \theta_{35}^{e} & b(1+\theta_{23}^{e2}) + a \theta_{35}^{e2} & (b-a) \theta_{35}^{e}  & -a \theta_{35}^{e} \theta_{54}^{e} \\[1ex]
0 & (b-a) \theta_{25}^{e} & (b-a) \theta_{35}^{e} & a + b (\theta_{25}^{e2}+\theta_{35}^{e2}+\theta_{54}^{e2}) & (a-b) \theta_{54}^{e} \\[1ex]
0 & -a \theta_{25}^{e} \theta_{54}^{e} & -a \theta_{35}^{e} \theta_{54}^{e} & (a-b) \theta_{54}^{e} & b + a \theta_{54}^{e2}
\end{pmatrix}
\label{eqn:simple_DLep_DRep}
\end{split}
\end{equation}
There are two important features we can read off from the SM $Z$ coupling constants of Equation~\ref{eqn:simple_DLep_DRep}; the first is the diagonal elements $(a,a,a,a,b)$ and $(b,b,b,a,b)$ get hardly affected by the small mixing angles and the second is magnitude of the off-diagonal elements are very weak as the mixing angles are defined by the ratio between the Yukawa and the vector-like masses. Therefore, we can imply the flavor violating mixing mediated by the SM $Z$ gauge boson in the mass basis. Using the SM $Z$ coupling constants in the mass basis, we can draw the Feynman diagram for $\tau \rightarrow \mu \mu \mu$ and $Z \rightarrow \mu \tau$ decay at tree-level and this will be discussed in the next section.
\subsection{FCNC mediated by the SM $Z$ gauge boson in the quark sector with the fourth vector-like quarks} \label{sec:IV_2}
The quark sector have two different mass matrices for the up- and down-type quark sector. We start from the up-type quark sector first. The SM $Z$ coupling constants for the up-type quark sector are given by:
\begin{equation}
\begin{split}
D_{L}^{u} &= 
\left(
\begin{array}{c|ccccc}
  & u_{1L} & u_{2L} & u_{3L} & u_{4L} & \widetilde{u}_{4L} \\[0.5ex]
  \hline
\overline{u}_{1L} & \left( \frac{1}{2}-\frac{2}{3}\sin^2\theta_w \right) & 0 & 0 & 0 & 0 \\[1ex]
\overline{u}_{2L} & 0 & \left( \frac{1}{2}-\frac{2}{3}\sin^2\theta_w \right) & 0 & 0 & 0 \\[1ex]
\overline{u}_{3L} & 0 & 0 & \left( \frac{1}{2}-\frac{2}{3}\sin^2\theta_w \right) & 0 & 0 \\[1ex]
\overline{u}_{4L} & 0 & 0 & 0 & \left( \frac{1}{2}-\frac{2}{3}\sin^2\theta_w \right) & 0 \\[1ex]
\overline{\widetilde{u}}_{4L} & 0 & 0 & 0 & 0 & \left( -\frac{2}{3}\sin^2\theta_w \right) \\[1ex]
\end{array}
\right) \\[1ex]
D_{R}^{u} &= 
\left(
\begin{array}{c|ccccc}
  & u_{1R} & u_{2R} & u_{3R} & \widetilde{u}_{4R} & u_{4R} \\[0.5ex]
  \hline
\overline{u}_{1R} & \left( -\frac{2}{3}\sin^2\theta_w \right) & 0 & 0 & 0 & 0 \\[1ex]
\overline{u}_{2R} & 0 & \left( -\frac{2}{3}\sin^2\theta_w \right) & 0 & 0 & 0 \\[1ex]
\overline{u}_{3R} & 0 & 0 & \left( -\frac{2}{3}\sin^2\theta_w \right) & 0 & 0 \\[1ex]
\overline{\widetilde{u}}_{4R} & 0 & 0 & 0 & \left( \frac{1}{2}-\frac{2}{3}\sin^2\theta_w \right) & 0 \\[1ex]
\overline{u}_{4R} & 0 & 0 & 0 & 0 & \left( -\frac{2}{3}\sin^2\theta_w \right) \\[1ex]
\end{array}
\right)
\label{eqn:SMZ_DLu_DRu}
\end{split}
\end{equation}
The up-type quark mass matrix is exactly same as the one for the charged lepton, so we can simply follow the mixing matrices given in Equation~\ref{eqn:ana_up_mixing}. Then the SM $Z$ gauge coupling constants in the mass basis are defined by:
\begin{equation}
\begin{split}
D_{L}^{u \prime} &= V_{L}^{u} D_{L}^{u} V_{L}^{u \dagger}
\\[1ex]
&= (V_{L}^{u})_{45} (V_{L}^{u})_{23} (V_{L}^{u})_{25} (V_{L}^{u})_{35} (V_{L}^{u})_{34} D_{L}^{u} (V_{L}^{u})_{34}^{\dagger} (V_{L}^{u})_{35}^{\dagger} (V_{L}^{u})_{25}^{\dagger} (V_{L}^{u})_{23}^{\dagger} (V_{L}^{u})_{45}^{\dagger}
\\[1ex]
&= (V_{L}^{u})_{45} (V_{L}^{u})_{23} (V_{L}^{u})_{25} (V_{L}^{u})_{35} D_{L}^{u} (V_{L}^{u})_{35}^{\dagger} (V_{L}^{u})_{25}^{\dagger} (V_{L}^{u})_{23}^{\dagger} (V_{L}^{u})_{45}^{\dagger}
\\[1ex]
D_{R}^{u \prime} &= V_{R}^{u} D_{R}^{u} V_{R}^{u \dagger}
\\[1ex]
&= (V_{R}^{u})_{54} (V_{R}^{u})_{23} (V_{R}^{u})_{25} (V_{R}^{u})_{35} (V_{R}^{u})_{24} (V_{R}^{u})_{34} D_{R}^{u} (V_{R}^{u})_{34}^{\dagger} (V_{R}^{u})_{24}^{\dagger} (V_{R}^{u})_{35}^{\dagger} (V_{R}^{u})_{25}^{\dagger} (V_{R}^{u})_{23}^{\dagger} (V_{R}^{u})_{54}^{\dagger}
\\
&= (V_{R}^{u})_{54} (V_{R}^{u})_{23} (V_{R}^{u})_{25} (V_{R}^{u})_{35} D_{R}^{u} (V_{R}^{u})_{35}^{\dagger} (V_{R}^{u})_{25}^{\dagger} (V_{R}^{u})_{23}^{\dagger} (V_{R}^{u})_{54}^{\dagger}
\label{eqn:SMZ_uq_mb_LR_1}
\end{split}
\end{equation}
The SM $Z$ gauge coupling constants for the up-quark sector in the mass basis can be seen by (We substitute $\left( 1/2-2/3\sin^2\theta_w \right)$ by $c$ and $\left( -2/3\sin^2\theta_w \right)$ by $d$ and assume the mixing angles $\theta_{35,25,23,45(54)L,R}^{u}$ are small enough to approximate for simplicity):
\begin{equation}
\begin{split}
D_{L}^{u \prime} &\approx 
\begin{pmatrix}
c & 0 & 0 & 0 & 0 \\[1ex]
0 & c & d \theta_{25L}^{u} \theta_{35L}^{u} & d \theta_{25L}^{u} \theta_{45L}^{u} & (c-d) \theta_{25L}^{u} \\[1ex]
0 & d \theta_{25L}^{u} \theta_{35L}^{u} & c & d \theta_{35L}^{u} \theta_{45L}^{u} & (c-d) \theta_{35L}^{u} \\[1ex]
0 & d \theta_{25L}^{u} \theta_{45L}^{u} & d \theta_{35L}^{u} \theta_{45L}^{u} & c & (c-d) \theta_{45L}^{u} \\[1ex]
0 & (c-d) \theta_{25L}^{u} & (c-d) \theta_{35L}^{u} & (c-d) \theta_{45L}^{u} & d
\end{pmatrix}
\\[1ex]
D_{R}^{u \prime} &\approx 
\begin{pmatrix}
d & 0 & 0 & 0 & 0 \\[1ex]
0 & d & c \theta_{25R}^{u} \theta_{35R}^{u} & (d-c) \theta_{25R}^{u} & -c \theta_{25R}^{u} \theta_{54R}^{u} \\[1ex]
0 & c \theta_{25R}^{u} \theta_{35R}^{u} & d & (d-c) \theta_{35R}^{u}  & -c \theta_{35R}^{u} \theta_{54R}^{u} \\[1ex]
0 & (d-c) \theta_{25R}^{u} & (d-c) \theta_{35R}^{u} & c & (c-d) \theta_{54R}^{u} \\[1ex]
0 & -c \theta_{25R}^{u} \theta_{54R}^{u} & -c \theta_{35R}^{u} \theta_{54R}^{u} & (c-d) \theta_{54R}^{u} & d
\end{pmatrix}
\label{eqn:simple_DLup_DRup}
\end{split}
\end{equation}
Next, we focus on the down-type quark mass matrix and the left-handed mixing matrices for that is different when compared to other left-handed mixing matrices for the up-type or charged lepton mass matrix in that it can reach to all mixings among the three SM generations. Keeping that in mind, we start from the SM $Z$ coupling constants for the down-type quarks.
\begin{equation}
\begin{split}
D_{L}^{d}
&= 
\left(
\begin{array}{c|ccccc}
  & d_{1L} & d_{2L} & d_{3L} & d_{4L} & \widetilde{d}_{4L} \\[0.5ex]
  \hline
\overline{d}_{1L} & \left( -\frac{1}{2}+\frac{1}{3}\sin^2\theta_w \right) & 0 & 0 & 0 & 0 \\[1ex]
\overline{d}_{2L} & 0 & \left( -\frac{1}{2}+\frac{1}{3}\sin^2\theta_w \right) & 0 & 0 & 0 \\[1ex]
\overline{d}_{3L} & 0 & 0 & \left( -\frac{1}{2}+\frac{1}{3}\sin^2\theta_w \right) & 0 & 0 \\[1ex]
\overline{d}_{4L} & 0 & 0 & 0 & \left( -\frac{1}{2}+\frac{1}{3}\sin^2\theta_w \right) & 0 \\[1ex]
\overline{\widetilde{d}}_{4L} & 0 & 0 & 0 & 0 & \left( \frac{1}{3}\sin^2\theta_w \right) 
\end{array}
\right)
\\
D_{R}^{d}
&= 
\left(
\begin{array}{c|ccccc}
  & d_{1R} & d_{2R} & d_{3R} & \widetilde{d}_{4R} & d_{4R} \\[0.5ex]
  \hline
\overline{d}_{1R} & \left( \frac{1}{3}\sin^2\theta_w \right) & 0 & 0 & 0 & 0 \\[1ex]
\overline{d}_{2R} & 0 & \left( \frac{1}{3}\sin^2\theta_w \right) & 0 & 0 & 0 \\[1ex]
\overline{d}_{3R} & 0 & 0 & \left( \frac{1}{3}\sin^2\theta_w \right) & 0 & 0 \\[1ex]
\overline{\widetilde{d}}_{4R} & 0 & 0 & 0 & \left( -\frac{1}{2}+\frac{1}{3}\sin^2\theta_w \right) & 0 \\[1ex]
\overline{d}_{4R} & 0 & 0 & 0 & 0 & \left( \frac{1}{3}\sin^2\theta_w \right) 
\end{array}
\right)
\label{eqn:SMZ_DLd_DRd}
\end{split}
\end{equation}
After simplifying the whole left-handed (right-handed) mixing matrices of Equation~\ref{eqn:down_mix2}, 
\Antonio{we obtain the following matrices of $Z$ couplings with quarks
\begin{equation}
\begin{split}
D_{L}^{d \prime} = V_{L}^{d} D_{L}^{d} (V_{L}^{d})^{\dagger} \\[1ex]
D_{R}^{d \prime} = V_{R}^{d} D_{R}^{d} (V_{R}^{d})^{\dagger} 
\label{eqn:DLdp_DRdp}
\end{split}
\end{equation}}
The SM $Z$ coupling constants for the right-handed down-type quarks in the mass basis have the same form given in Equation~\ref{eqn:simple_DLep_DRep}, whereas those for the left-handed down-type quarks involves $12$ and $13$ mixings more, so it is worthwhile to look at its complete form (We again substitute $\left( -1/2+1/3 \sin^2\theta_w \right)$ by $e$ and $\left( 1/3 \sin^2\theta_w \right)$ by $f$ and assume all relevant mixing angles are small enough to approximate for simplicity)
\begin{equation}
\begin{split}
D_{L}^{d \prime} &\approx 
\begin{pmatrix}
e & 0 & 0 & 0 & 0 \\[1ex]
0 & e & e \theta_{12L}^{d} \theta_{13L}^{d} + f \theta_{25L}^{d} \theta_{35L}^{d} & -e \theta_{23L}^{d} \theta_{34L}^{d} + f \theta_{25L}^{d} \theta_{45L}^{d} & (e-f) \theta_{25L}^{d} \\[1ex]
0 & e \theta_{12L}^{d} \theta_{13L}^{d} + f \theta_{25L}^{d} \theta_{35L}^{d} & e & f \theta_{35L}^{d} \theta_{45L}^{d} & (e-f) \theta_{35L}^{d} \\[1ex]
0 & -e \theta_{23L}^{d} \theta_{34L}^{d} + f \theta_{25L}^{d} \theta_{45L}^{d} & f \theta_{35L}^{d} \theta_{45L}^{d} & e & (e-f) \theta_{45L}^{d} \\[1ex]
0 & (e-f) \theta_{25L}^{d} & (e-f) \theta_{35L}^{d} & (e-f) \theta_{45L}^{d} & f
\end{pmatrix}
\\[1ex]
D_{R}^{d \prime} &\approx 
\begin{pmatrix}
f & 0 & 0 & 0 & 0 \\[1ex]
0 & f & e \theta_{25R}^{d} \theta_{35R}^{d} & (f-e) \theta_{25R}^{d} & -e \theta_{25R}^{d} \theta_{54R}^{d} \\[1ex]
0 & e \theta_{25R}^{d} \theta_{35R}^{d} & f & (f-e) \theta_{35R}^{d}  & -e \theta_{35R}^{d} \theta_{54R}^{d} \\[1ex]
0 & (f-e) \theta_{25R}^{d} & (f-e) \theta_{35R}^{d} & e & (e-f) \theta_{54R}^{d} \\[1ex]
0 & -e \theta_{25R}^{d} \theta_{54R}^{d} & -e \theta_{35R}^{d} \theta_{54R}^{d} & (e-f) \theta_{54R}^{d} & f
\end{pmatrix},
\label{eqn:simple_DLdp_DRdp}
\end{split}
\end{equation}
where it can confirm two relations hold for the zeros appearing in the first row and column of $D_{L}^{d \prime}$: $\theta_{13L}^{d} \simeq \theta_{12L}^{d} \theta_{23L}^{d}$ and $\theta_{15L}^{d} \simeq \theta_{12L}^{d} \theta_{25L}^{d}$. Following the analytic mixings given in Equation~\ref{eqn:down_mix2}, the left-handed down type quark sector can access to all mixings among the three SM generations and this feature is also reflected on a numerical mixing matrix $V_{L}^{d}$ of Equation~\ref{eqn:a_num_VCKM}. What this implies is the SM $Z$ physics does not get affected by any specific basis we choose and this will be verified again numerically in Appendix~\ref{app:E}. 
\section{PHENOMENOLOGY IN THE CHARGED LEPTON SECTOR OF THE SM} \label{sec:V}
Now that we have defined all required mixings and coupling constants in both quark and charged lepton sectors, it is time to discuss the relevant phenomenology. As mentioned in the introduction, one of our main goals is to study the FCNC observables to constrain the possible mass range of the vector-like fermions. 
Given that the second and third generations of SM charged leptons 
acquire masses through their mixings with the fourth vector-like charged leptons, we will restrict our analysis  
to the constraints imposed on the flavor violating decays involving  
the second and third charged lepton generations such as $\tau \rightarrow \mu \gamma, \tau \rightarrow \mu \mu \mu$ and lastly $Z \rightarrow \mu \tau$ decay.
\subsection{Analytic expression for $\tau \rightarrow \mu \gamma$ decay}
\begin{figure}[H]
\centering
\includegraphics[keepaspectratio,width=\textwidth]{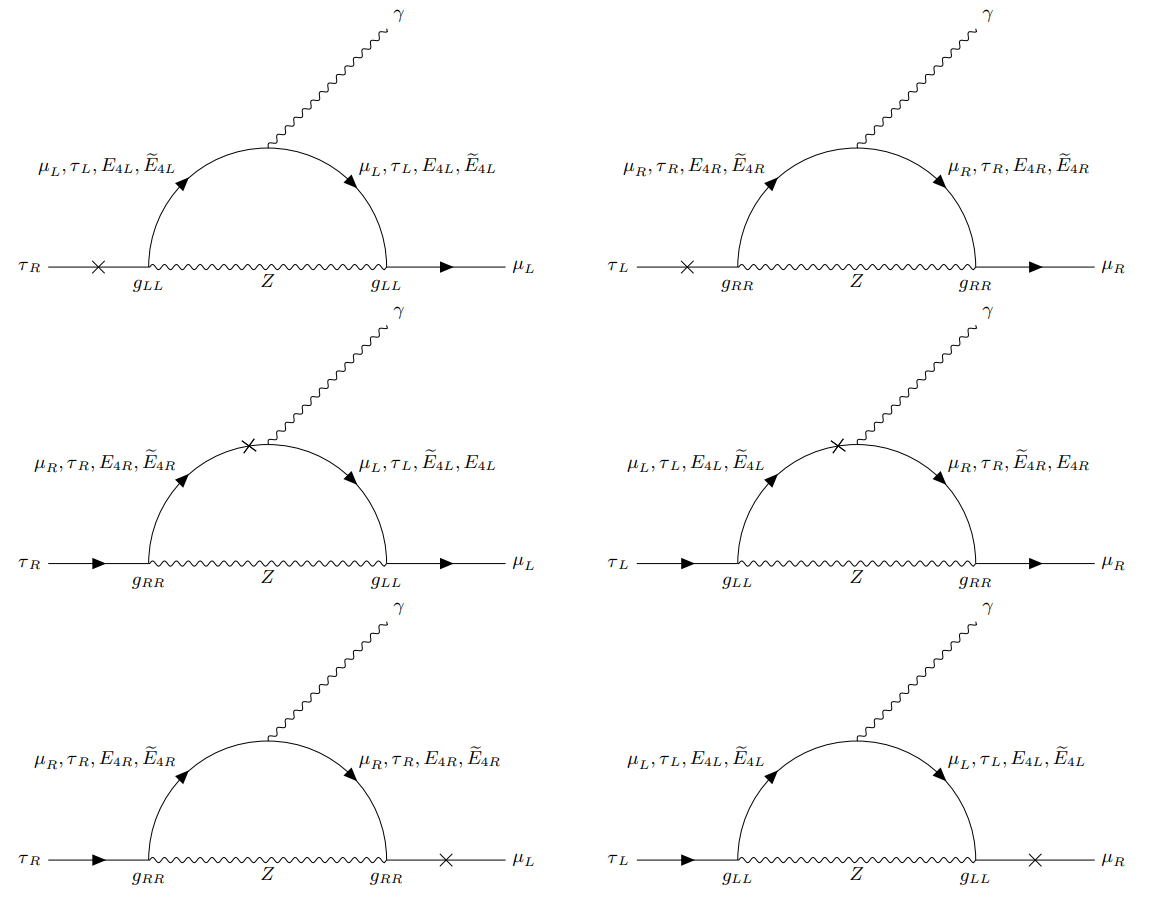}
\caption{Diagrams contributing to the charged lepton flavor violation (CLFV) $\tau \rightarrow \mu \gamma$ decay at one-loop level in the mass basis. The cross notation in each diagram means the helicity flip process.}
\label{fig:tmgdecay}
\end{figure}

The most important FCNC constraint corresponds to the charged lepton flavor violating (CLFV) $\mu \rightarrow e \gamma$ decay, however we can not make an appropriate prediction for the constraint as the electron does not acquire a mass in the model under consideration. This is due to the fact that the fermion sector of the model includes two heavy fermionic seesaw mediators. As previously mentioned, adding a fifth vector-like family to the fermion sector of the model will generate a nonvanishing electron. However in order to keep our model as economical as possible and to simplify our analysis corresponding to the FCNC constraints on vector-like masses, we restrict to the case of a fourth vector-like family in the fermionic spectrum. 
Therefore, in view of the aforementioned considerations, we first consider the CLFV decay $\tau \rightarrow \mu \gamma$ in order to determine how the model parameter space gets affected by the experimental constraint arising from this decay.
For the $\tau \rightarrow \mu \gamma$ decay, the  
leading order contribution appears in the one-loop diagrams since there is no possible contribution at tree-level. Then, all possible Feynman diagrams contributing to  
the $\tau \rightarrow \mu \gamma$ decay are given in Figure~\ref{fig:tmgdecay},
where $g_{LL(RR)}$ are the LH-LH (RH-RH) coupling constants defined in the mass basis for the charged lepton sector. These one-loop contributions mediated by the  
$Z^{\prime}$ boson were studied in one of our previous works~\cite{CarcamoHernandez:2019ydc} and their corresponding analytic prediction for the branching ratio of $\tau \rightarrow \mu \gamma$ decay is given in Equation~\ref{eqn:analytic_tmgdecay}~\cite{CarcamoHernandez:2019ydc,Raby:2017igl,Lindner:2016bgg,Lavoura:2003xp,Chiang:2011cv} shown below:
\begin{equation}
\begin{split}
\func{BR} \left( \tau \rightarrow \mu \gamma \right) &= \frac{\alpha_{\func{em}}}{1024\pi^4} \frac{m_{\tau}^5}{M_Z^4 \Gamma_{\tau}} \Big( \lvert g_{\tau \mu}^L g_{\mu \mu}^L F(x_\mu) + g_{\tau \tau}^L g_{\tau \mu}^L F(x_\tau) + g_{\tau E_4}^L g_{E_4 \mu}^L F(x_{E_4}) + g_{\tau \widetilde{E}_4}^L g_{\widetilde{E}_4 \mu}^L F(x_{\widetilde{E}_4}) \\
&+ \frac{m_{\mu}}{m_{\tau}} g_{\tau \mu}^L g_{\mu \mu}^L F(x_\mu) + \frac{m_{\mu}}{m_{\tau}} g_{\tau \tau}^L g_{\tau \mu}^L F(x_\tau) + \frac{m_{\mu}}{m_{\tau}} g_{\tau E_4}^L g_{E_4 \mu}^L F(x_{E_4}) + \frac{m_{\mu}}{m_{\tau}} g_{\tau \widetilde{E}_4}^L g_{\widetilde{E}_4 \mu}^L F(x_{\widetilde{E}_4}) \\
&+ \frac{m_{\mu}}{m_{\tau}} g_{\tau \mu}^L g_{\mu \mu}^R G(x_{\mu}) + \frac{m_{\tau}}{m_{\tau}} g_{\tau \tau}^L g_{\tau \mu}^R G(x_{\mu}) + \frac{M_{E_4}}{m_{\tau}} g_{\tau E_4}^L g_{\widetilde{E}_4 \mu}^R G(x_{E_4}) + \frac{M_{\widetilde{E}_4}}{m_{\tau}} g_{\tau \widetilde{E}_4}^L g_{E_4 \mu}^R G(x_{\widetilde{E}_4}) \rvert^2 \\
&+ \lvert g_{\tau \mu}^R g_{\mu \mu}^R F(x_\mu) + g_{\tau \tau}^R g_{\tau \mu}^R F(x_\tau) + g_{\tau E_4}^R g_{E_4 \mu}^R F(x_{E_4}) + g_{\tau \widetilde{E}_4}^R g_{\widetilde{E}_4 \mu}^R F(x_{\widetilde{E}_4}) \\
&+ \frac{m_{\mu}}{m_{\tau}} g_{\tau \mu}^R g_{\mu \mu}^R F(x_\mu) + \frac{m_{\mu}}{m_{\tau}} g_{\tau \tau}^R g_{\tau \mu}^R F(x_\tau) + \frac{m_{\mu}}{m_{\tau}} g_{\tau E_4}^R g_{E_4 \mu}^R F(x_{E_4}) + \frac{m_{\mu}}{m_{\tau}} g_{\tau \widetilde{E}_4}^R g_{\widetilde{E}_4 \mu}^R F(x_{\widetilde{E}_4}) \\
&+ \frac{m_{\mu}}{m_{\tau}} g_{\tau \mu}^R g_{\mu \mu}^L G(x_{\mu}) + \frac{m_{\tau}}{m_{\tau}} g_{\tau \tau}^R g_{\tau \mu}^L G(x_{\mu}) + \frac{M_{\widetilde{E}_4}}{m_{\tau}} g_{\tau E_4}^R g_{\widetilde{E}_4 \mu}^L G(x_{\widetilde{E}_4}) + \frac{M_{E_4}}{m_{\tau}} g_{\tau \widetilde{E}_4}^R g_{E_4 \mu}^L G(x_{E_4}) \rvert^2 \Big),
\label{eqn:analytic_tmgdecay}
\end{split}
\end{equation}
where $\alpha_{\func{em}}$ is the fine structure constant, $\Gamma_{\tau}$ is the total decay width of the tau lepton ($\Gamma_{\tau} = 5 \times \Gamma(\tau_L^- \rightarrow \nu_\tau e_L^- \overline{\nu}_e) = 2.0 \times 10^{-12}$) and $F$ and $G$ are the loop functions defined by:
\begin{equation}
\begin{split}
F(x) &= \frac{5x^4 - 14x^3 + 39x^2 - 38x - 18x^2 \func{ln}x + 8}{12(1-x)^4} \\[1ex]
G(x) &= \frac{x^3 + 3x - 6x \func{ln}x - 4}{2(1-x)^3}, \qquad x = \frac{m_{\func{loop}}^2}{M_{Z}^2}
\end{split}
\end{equation}
where $m_{\func{loop}}$ is the propagating mass of the charged leptons in the loop. 
The most dominant contributions to the $\tau \rightarrow \mu \gamma$ branching ratio correspond to the terms proportional to $M_{E_4(\widetilde{E}_4)}/m_{\tau}$ because charged vector-like leptons are heavier than $200$ GeV, thus implying that the enhancement factor $M_{E_4(\widetilde{E}_4)}/m_{\tau}$ makes those contributions much larger than the ones not involving this factor. 
However, the contributions to the $\tau \rightarrow \mu \gamma$ decay rate involving the terms having the aforementioned proportionality factor do not keep increasing 
as the vector-like fermions get heavier 
since their flavor violating coupling constants get more suppressed at the same time by the small mixing angles, defined by the ratio between Yukawa and vector-like masses.
Therefore, these compensations provide some balanced relation between the vector-like mass and the coupling of the $Z$ gauge boson with a SM charged antilepton (lepton) and heavy charged vector-like (antilepton) lepton. 
\Steve{The experimental bound for the branching ratio of $\tau \rightarrow \mu \gamma$ decay is given by~\cite{Crivellin:2020ebi,MEG:2016leq,BaBar:2009hkt}}:
\begin{equation}
\func{BR}\left( \tau \rightarrow \mu \gamma \right)_{\func{EXP}} = 4.4 \times 10^{-8}
\end{equation}
\subsection{Analytic expression for $\tau \rightarrow \mu \mu \mu$ decay}
\begin{figure}[H]
\centering
\includegraphics[keepaspectratio,width=\textwidth]{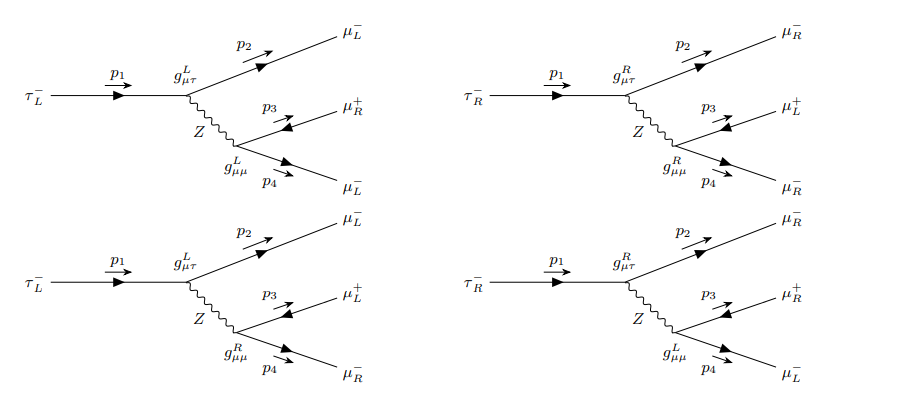}
\caption{Diagrams contributing to the charged lepton flavor violation (CLFV) $\tau \rightarrow 3\mu$ decay at tree-level. We refer to the top-left diagram as $\mathcal{M}_{LL}$ and the top-right as $\mathcal{M}_{RR}$ and similarly for the two below diagrams as $\mathcal{M}_{LR,RL}$, respectively.}
\label{fig:diagrams_tau3mu}
\end{figure}

The other interesting flavor violating decay mode is the $\tau \rightarrow \mu \mu \mu$ decay mediated by the SM $Z$ gauge boson. As the model under consideration 
 has 
 $Z$ mediated renormalizable flavor violating interactions, we can draw the Feynman diagrams for the $\tau \rightarrow \mu \mu \mu$ decay at tree-level as given in Figure~\ref{fig:diagrams_tau3mu}.
The contributions shown in Figure~\ref{fig:diagrams_tau3mu} are 
beyond Standard Model (BSM) 
effects, thus they need to be computed to set constraints on the model parameter space. 
In order to derive an analytic expression for the CLFV $\tau \rightarrow 3\mu$ decay rate mediated by the SM $Z$ gauge boson, 
 we start by writting down its definition as follows:
\begin{equation}
d\Gamma\left( \tau \rightarrow 3\mu \right) = \frac{1}{2m_{\tau}} \frac{d^3 p_2}{(2\pi)^3 2E_2} \frac{d^3 p_3}{(2\pi)^3 2E_3} \frac{d^3 p_4}{(2\pi)^3 2E_4} \lvert \mathcal{M} \rvert^2 \left( 2\pi \right)^4 \delta^{4} \left( p_1 - p_2 - p_3 - p_4 \right)
\end{equation}
Evaluating each polarized diagram in Figure~\ref{fig:diagrams_tau3mu}, it yields the following result: 
\begin{equation}
\begin{aligned}
\lvert \mathcal{M}_{LL} \rvert^2 &= \left( \frac{g_{\mu \tau}^L g_{\mu \mu}^L}{4M_Z^2} \right)^2 256\left( p_1 \cdot p_3 \right) \left( p_2 \cdot p_4 \right), \quad \lvert \mathcal{M}_{RR} \rvert^2 &= \left( \frac{g_{\mu \tau}^R g_{\mu \mu}^R}{4M_Z^2} \right)^2 256\left( p_2 \cdot p_4 \right) \left( p_1 \cdot p_3 \right) \\
\lvert \mathcal{M}_{LR} \rvert^2 &= \left( \frac{g_{\mu \tau}^L g_{\mu \mu}^R}{4M_Z^2} \right)^2 256\left( p_1 \cdot p_4 \right) \left( p_2 \cdot p_3 \right), \quad \lvert \mathcal{M}_{RL} \rvert^2 &= \left( \frac{g_{\mu \tau}^R g_{\mu \mu}^L}{4M_Z^2} \right)^2 256\left( p_1 \cdot p_4 \right) \left( p_2 \cdot p_3 \right)
\end{aligned}
\end{equation}
We are now ready to determine the squared amplitude averaged and summed over the initial and final spin states.
\begin{equation}
\begin{split}
\frac{1}{2} \sum_{\func{spin}} \lvert \mathcal{M} \rvert^2 &= \frac{1}{2} \left( \lvert \mathcal{M}_{LL} \rvert^2 + \lvert \mathcal{M}_{RR} \rvert^2 + \lvert \mathcal{M}_{LR} \rvert^2 + \lvert \mathcal{M}_{RL} \rvert^2 \right) \\
&= \frac{8}{M_Z^4} \left[ \left( g_{\mu \tau}^{2L} g_{\mu \mu}^{2L} + g_{\mu \tau}^{2R} g_{\mu \mu}^{2R} \right) \left( p_1 \cdot p_3 \right) \left( p_2 \cdot p_4 \right) + \left( g_{\mu \tau}^{2L} g_{\mu \mu}^{2R} + g_{\mu \tau}^{2R} g_{\mu \mu}^{2L} \right) \left( p_1 \cdot p_4 \right) \left( p_2 \cdot p_3 \right) \right]
\end{split}
\end{equation}
The momentum of the particles involved in the $\tau \rightarrow 3\mu$ are written in the rest frame as follows:
\begin{equation}
\begin{split}
p_1 &=  ( m_{\tau}, \vec{0} ) \\
p_2 &= \left( E_2, \vec{p}_2 \right) \\
p_3 &= \left( m_{\tau} - E_2 - E_4, -\vec{p}_2 -\vec{p}_4  \right) \\
p_4 &= \left( E_4, \vec{p}_4 \right) \\
\end{split}
\end{equation}
Then, we can carry out the inner products of momenta taking into account  
the momentum conservation ($p_1 = p_2 + p_3 + p_4$).
\begin{equation}
\begin{split}
p_1 \cdot p_3 &= m_{\tau} \left( m_{\tau} - E_2 - E_4 \right) \\
p_2 \cdot p_4 &= \frac{1}{2} \left( -m_{\tau}^2 - m_{\mu}^2 + 2m_{\tau} (E_2 + E_4) \right) \\
p_1 \cdot p_4 &= m_{\tau} E_4 \\
p_2 \cdot p_3 &= \frac{1}{2} \left( m_{\tau}^2 - m_{\mu}^2 - 2m_{\tau} E_4 \right)
\end{split}
\end{equation}
We can rewrite the squared amplitude in terms of the mass parameters after simplifying the summing over the diverse coupling constants to $g_{1,2}$, respectively ($g_1 = g_{\mu \tau}^{2L} g_{\mu \mu}^{2L} + g_{\mu \tau}^{2R} g_{\mu \mu}^{2R}, g_2 = g_{\mu \tau}^{2L} g_{\mu \mu}^{2R} + g_{\mu \tau}^{2R} g_{\mu \mu}^{2L}$).
\begin{equation}
\begin{split}
\frac{1}{2}\sum_{\func{spin}} \lvert \mathcal{M} \left( g_1, g_2, E_2, E_4 \right) \rvert^2 &= \frac{4}{M_Z^4} \Big[ g_1 \left( m_{\tau}^2 - m_{\tau} (E_2 + E_4) \right) \left( -m_{\tau}^2 - m_{\mu}^2 + 2m_{\tau} (E_2 + E_4) \right) \\
&+ g_2 \left( m_{\tau} E_4 \right) \left( m_{\tau}^2 - m_{\mu}^2 - 2m_{\tau} E_4 \right) \Big]
\end{split}
\end{equation}
Now it is time to evaluate the three body phase space integral by turning it into an effective two-body phase integral as follows (we simply drop out the prefactor $1/(2\pi)^5$ for simplicity in this derivation while keeping 
 the prefactor 
 in the computation of the aforementioned partial decay width).
\begin{equation}
\begin{split}
&\frac{d^3p_2}{2E_2} \frac{d^3p_3}{2E_3} \frac{d^3p_4}{2E_4} \delta^{4} \left( p_1 - p_2 - p_3 - p_4 \right) = \frac{d^3p_2}{2E_2} d^4p_3 \Theta \left( p_3^0 \right) \delta \left( p_3^2 \right) \frac{d^3p_4}{2E_4} \delta^{4} \left( p_1 - p_2 - p_3 - p_4 \right) \\
&= \frac{d^3p_2}{2E_2} \frac{d^3p_4}{2E_4} \Theta \left( p_1^0 - p_2^0 - p_4^0 \right) \delta \left( (p_1 - p_2 - p_4)^2 \right) \\
&=\frac{d^3p_2}{2E_2} \frac{d^3p_4}{2E_4} \Theta \left( p_1^0 - p_2^0 - p_4^0 \right) \delta \left( p_1^2 -2p_1 \cdot (p_2 + p_4) + (p_2^2 + 2p_2 \cdot p_4 + p_4^2) \right) \\
&= \frac{d^3p_2}{2E_2} \frac{d^3p_4}{2E_4} \Theta \left( p_1^0 - p_2^0 - p_4^0 \right) \delta \left( m_{\tau}^2 +2m_{\mu}^2 -2m_{\tau} (E_2 + E_4) + 2(E_2 E_4 - \lvert \vec{p}_2 \rvert \lvert \vec{p}_4 \rvert \cos\theta) \right) \\
&= \frac{d^3p_2}{2E_2} \frac{d^3p_4}{2E_4} \Theta \left( p_1^0 - p_2^0 - p_4^0 \right) \frac{1}{2\lvert \vec{p}_2 \rvert \lvert \vec{p}_4 \rvert} \delta \left( \frac{m_{\tau}^2 +2m_{\mu}^2 -2m_{\tau} (E_2 + E_4) + 2 E_2 E_4}{2\lvert \vec{p}_2 \rvert \lvert \vec{p}_4 \rvert} - \cos\theta \right)
\end{split}
\end{equation}
From the delta function, we can determine the integration range by assuming $E_2 \approx \lvert \vec{p}_2 \rvert, E_4 \approx \lvert \vec{p}_4 \rvert$. When $\cos\theta = 1$, the obtained result is
\begin{equation}
m_{\tau}^2 + 2m_{\mu}^2 - 2m_{\tau} (E_2 + E_4) = 0
\label{eqn:inside_deltafunction}
\end{equation}
From Equation~\ref{eqn:inside_deltafunction}, the integration range can be read off as follows:
\begin{equation}
\frac{m_{\mu}^2}{m_{\tau}} \leq E_2 \leq \frac{1}{2}m_{\tau}, \quad \frac{1}{2}m_{\tau} + \frac{m_{\mu}^2}{m_{\tau}} - E_2 \leq E_4 \leq \frac{1}{2}m_{\tau}
\end{equation}
It can be easily understood that once one mass parameter $E_2$ is set up by $\frac{1}{2}m_{\tau}$, the energy of the other mass parameters $E_4, E_3$ must be given by $\frac{m_{\mu}^2}{m_{\tau}}, \frac{1}{2}m_{\tau} - \frac{m_{\mu}^2}{m_{\tau}}$, respectively. Then, it remains 
to simplify the effective two body phase space integral as follows.
\begin{equation}
d^3p_2 d^3p_4 = 4\pi \lvert \vec{p}_2 \rvert^2 d\lvert \vec{p}_2 \rvert 2\pi \lvert \vec{p}_4 \rvert d\lvert \vec{p}_4 \rvert d\cos\theta, \quad \lvert \vec{p}_2 \rvert d\lvert \vec{p}_2 \rvert = E_2 dE_2, \quad \lvert \vec{p}_4 \rvert d\lvert \vec{p}_4 \rvert = E_4 dE_4
\end{equation}
Putting all pieces 
together, the decay width for the CLFV $\tau \rightarrow 3\mu$ decay at tree-level after carrying out the $\cos\theta$ integration is given by:
\begin{equation}
\Gamma\left( \tau \rightarrow 3 \mu \right) = \frac{1}{64 m_{\tau} \pi^3} \int_{m_{\mu}^2/m_{\tau}}^{\frac{1}{2}m_{\tau}} \int_{\frac{1}{2}m_{\tau}+\frac{m_{\mu}^2}{m_{\tau}}-E_2}^{\frac{1}{2}m_{\tau}} \left( \frac{1}{2}\sum_{\func{spin}} \lvert \mathcal{M} \left( g_1, g_2, E_2, E_4 \right) \rvert^2 \right) dE_4 dE_2
\end{equation}
The experimental bound for the $\tau \rightarrow 3\mu$ decay is given by~\Steve{\cite{Hayasaka:2010np}}:
\begin{equation}
\func{BR}\left( \tau \rightarrow 3\mu \right)_{\func{EXP}} = 2.1 \times 10^{-8}
\end{equation}
\subsection{Analytic expression for $Z \rightarrow \mu \tau$ decay}
\begin{figure}[H]
\centering
\includegraphics[keepaspectratio,width=\textwidth]{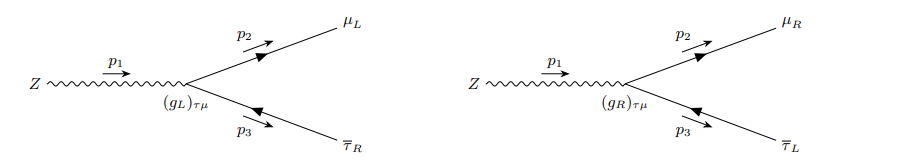}
\caption{Diagrams contributing to the charged lepton flavor violation (CLFV) $Z\tau\mu$ decay at tree-level}
\label{fig:diagrams_Zmutau}
\end{figure}

The last FCNC constraint we discuss is the $Z \rightarrow \mu \tau$ decay and diagrams contributing to the $Z \rightarrow \mu \tau$ decay are given in Figure~\ref{fig:diagrams_Zmutau}.
As in the CLFV $\tau \rightarrow 3\mu$ decay mediated by the SM $Z$ gauge boson, this CLFV $Z\rightarrow \mu\tau$ decay is also a new effect and it requires to derive its appropriate prediction from the ground. We can read off the invariant amplitude for each diagram given in Figure~\ref{fig:diagrams_Zmutau}. We refer to the left diagram as $\mathcal{M}_L$ and the right as $\mathcal{M}_R$. Then, the amplitudes are written as follows:
\begin{equation}
\begin{split}
i\mathcal{M}_L &= i (g_L)_{\tau\mu} \epsilon_\mu(p_1) \overline{u} (p_2) \gamma^\mu P_L v (p_3), \\[1ex]
i\mathcal{M}_R &= i (g_R)_{\tau\mu} \epsilon_\mu(p_1) \overline{u} (p_2) \gamma^\mu P_R v (p_3).
\label{eqn:amplitude_for_each_Zmutau}
\end{split}
\end{equation}
In order to have the squared amplitude averaged and summed, we square the amplitude given in Equation~\ref{eqn:amplitude_for_each_Zmutau} as follows:
\begin{equation}
\begin{split}
\frac{1}{3}\sum_{\func{spin}} \lvert \mathcal{M}_L \rvert^2 &= \frac{1}{3} (g_L)_{\tau\mu}^2 \left( -g_{\mu\nu} + \frac{p_{1\mu}p_{1\nu}}{M_Z^2} \right) \func{Tr} \left[ (\slashed{p}_2+m_{\mu}) \gamma^\mu P_L (\slashed{p}_3-m_{\tau}) \gamma^\nu P_L) \right] \\
&\simeq \frac{2}{3} (g_L)_{\tau\mu}^2 M_Z^2 \\
\frac{1}{3}\sum_{\func{spin}} \lvert \mathcal{M}_R \rvert^2 &= \frac{1}{3} (g_R)_{\tau\mu}^2 \left( -g_{\mu\nu} + \frac{p_{1\mu}p_{1\nu}}{M_Z^2} \right) \func{Tr} \left[ (\slashed{p}_2+m_{\mu}) \gamma^\mu P_R (\slashed{p}_3-m_{\tau}) \gamma^\nu P_R) \right] \\
&\simeq \frac{2}{3} (g_R)_{\tau\mu}^2 M_Z^2 \\
\frac{1}{3} \sum_{\func{spin}} \lvert \mathcal{M} \rvert^2 &= \frac{1}{3} \sum_{\func{spin}} \left( \lvert \mathcal{M}_L \rvert^2 + \lvert \mathcal{M}_R \rvert^2 \right) = \frac{2}{3}\left( (g_L)_{\tau\mu}^2 + (g_R)_{\tau\mu}^2 \right) M_Z^2 
\end{split}
\end{equation}
Then, the decay rate equation is given by:
\begin{equation}
\begin{split}
d\Gamma\left( Z\rightarrow\mu\tau \right) &= \frac{1}{2M_Z} \frac{d^3p_2}{(2\pi)^3 2E_2} \frac{d^3p_3}{(2\pi)^3 2E_3} \lvert \mathcal{M} \rvert^2 (2\pi)^4 \delta^{(4)} \left( p_1 - p_2 - p_3 \right) \\
\Gamma\left( Z\rightarrow\mu\tau \right) &= \frac{\lvert p^* \rvert}{32\pi^2 M_Z^2} \int \lvert \mathcal{M} \rvert^2 d\Omega \\
&= \frac{M_Z}{24\pi} \left( (g_L)_{\tau\mu}^2 + (g_R)_{\tau\mu}^2 \right)
\end{split}
\end{equation}
where $p^* \simeq M_Z/2$. Then, we are ready to write our prediction for the branching ratio of $Z\mu\tau$ decay at tree-level
\begin{equation}
\func{BR}\left(Z \rightarrow \mu\tau \right) = \frac{\Gamma\left( Z \rightarrow \mu\tau \right)}{\Gamma_{Z}} \simeq \frac{1}{2.5} \frac{M_Z}{24\pi} \left( (g_L)_{\tau\mu}^2 + (g_R)_{\tau\mu}^2 \right),
\end{equation}
where $\Gamma_{Z}$ is the total decay width of the SM $Z$ gauge boson ($\Gamma_{Z} \simeq 2.5\func{GeV}$). The experimental bound of the CLFV $Z \rightarrow \mu\tau$ decay is known as~\Steve{\cite{DELPHI:1996iox}}:
\begin{equation}
\func{BR}\left( Z \rightarrow \mu\tau \right)_{\func{EXP}} = 1.2 \times 10^{-5}
\end{equation}
\subsection{Numerical analysis for each prediction in the charged lepton sector}
We have discussed some 
relevant CLFV decay modes such as the $\tau \rightarrow \mu \gamma, \tau \rightarrow 3\mu$ and $Z \rightarrow \mu\tau$ from a theoretical point of view. 
By defining the renormalizable flavor violating interactions 
we showed that it is possible for the new physics to arise in a simple scenario thanks to the presence of vector-like charged leptons, which play a crucial role for these CLFV decay modes to happen. It is an encouraging feature that the mass range of the vector-like charged leptons can be constrained by the experimental bound of the CLFV decays and numerical scans for this feature will be discussed 
in detail in the following subsection.
\subsubsection{Free parameter setup}
For the numerical scan for the charged lepton sector, we first proceed to 
set up a possible mass range of the mass parameters of Equation~\ref{eqn:cl_2} \Steve{in Table~\ref{tab:parameter_region_initial_scan}}.

\begin{center}
{\renewcommand{\arraystretch}{1.5} 
\begin{tabular}{cc}
\toprule
\toprule
\textbf{Mass parameter} & \textbf{Scanned Region($\func{GeV}$)} \\ 
\midrule
$y_{24}^{e} v_{d} = m_{24}$ & $\pm [1,10]$ \\
$y_{34}^{e} v_{d} = m_{34}$ & $\pm [1,10]$ \\
$y_{43}^{e} v_{d} = m_{43}$ & $\pm [1,10]$ \\
$x_{34}^{L} v_{\phi} = m_{35}$ & $\pm [50,200]$ \\
$x_{42}^{e} v_{\phi} = m_{52}$ & $\pm [50,200]$ \\
$x_{43}^{e} v_{\phi} = m_{53}$ & $\pm [50,200]$ \\
$M_{45}^{L}$ & $\pm [150,2000]$ \\
$M_{54}^{e}$ & $\pm [150,2000]$ \\
\bottomrule
\bottomrule 
\end{tabular}
\captionof{table}{Initial parameter setup for scanning the mass of the vector-like charged leptons} \label%
{tab:parameter_region_initial_scan}}
\end{center}
There are a few of features to be noticed before we start the numerical scan.
\begin{enumerate}
\item We assumed a vev for the SM up-type Higgs $H_u$ 
very close to $246\func{GeV}$, whereas the one 
of the SM down-type Higgs $H_d$ is assumed to be very small compared to the $v_u = \langle H_u \rangle$ and is ranged from $1$ to $10\func{GeV}$. The two vevs hold the relation $v_u^2 + v_d^2 = (246\func{GeV})^2$. We made that assumption since we are considering an scenario close to the decoupling limit where the neutral CP even part of $H_u$ is mostly identified with the $126$ GeV SM like Higgs boson. 
\item As the vev of the singlet flavon $\phi$ is a free parameter, we varied it in the 
range $[50,200]\func{GeV}$ whereas the mass parameters $m_{35,52,53}$ were varied in a range of values consistent 
with the observed hierarchical structure of the charged lepton masses. Furthermore,  
the vector-like masses are also other free parameters assumed to be larger or equal than 
$150\func{GeV}$ in order to successfully fulfill the 
experimental bounds on exotic charged lepton masses.
\item What we need to constrain in this numerical scan is the predicted muon and tau masses as well as the $23$ mixing angle. For the muon and tau masses, we required that the 
obtained values of the muon and tau masses to be in the range 
$[1 \pm 0.1] \times m_{\mu,\tau}$. Considering that the sizeable off-diagonal elements of the PMNS mixing matrix mainly arise from the neutrino sector, all mixing angles in the charged lepton sector are required to be  
as small as possible and 
thus we limit them to be lower than $0.2$.
\end{enumerate}
\subsubsection{Numerical scan result for the charged lepton sector}
\begin{figure}[H]
\centering
\begin{subfigure}{0.48\textwidth}
\includegraphics[keepaspectratio,width=0.95\textwidth]{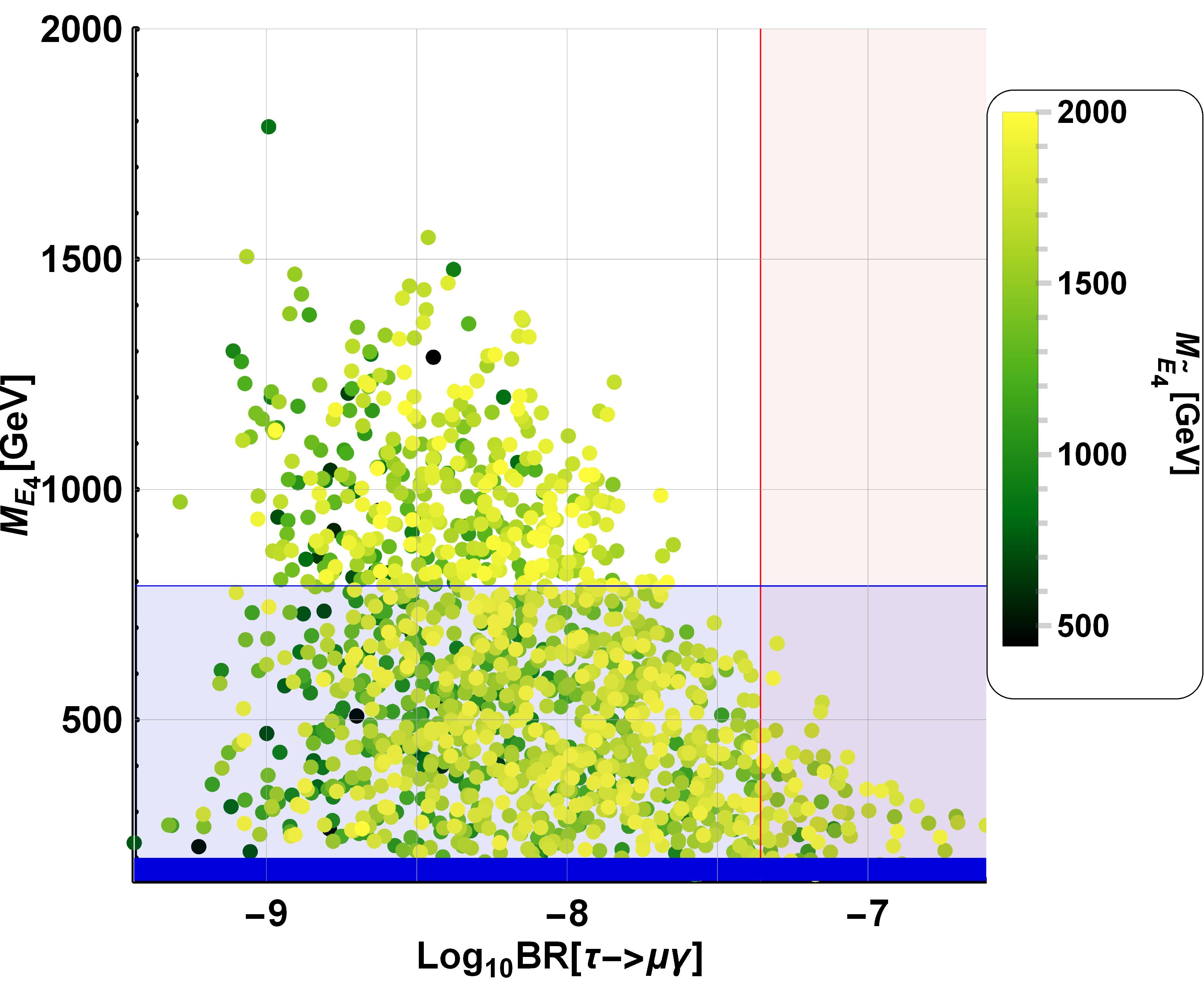}
\end{subfigure}
\begin{subfigure}{0.48\textwidth}
\includegraphics[keepaspectratio,width=0.95\textwidth]{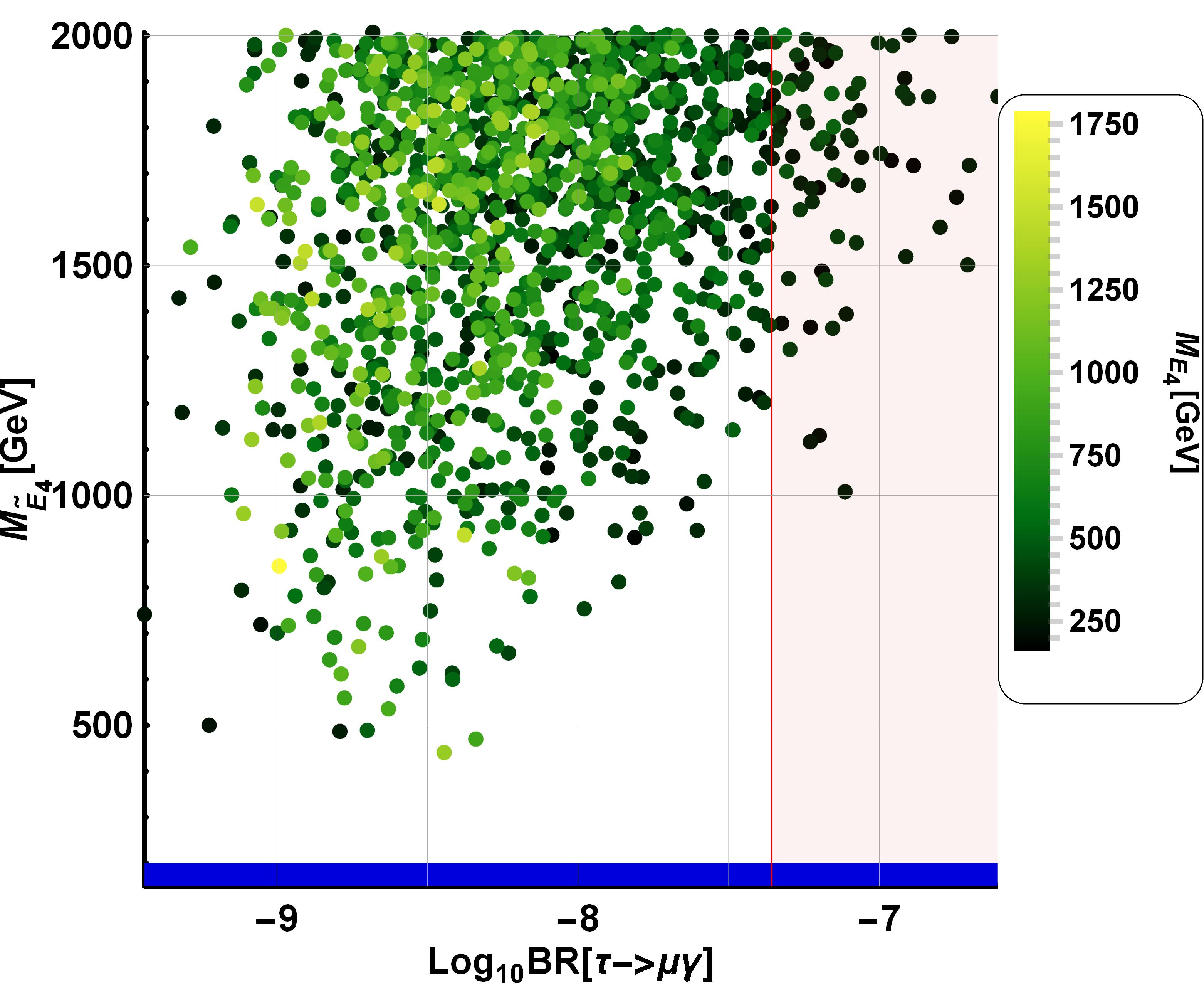}
\end{subfigure}
\begin{subfigure}{0.48\textwidth}
\includegraphics[keepaspectratio,width=0.95\textwidth]{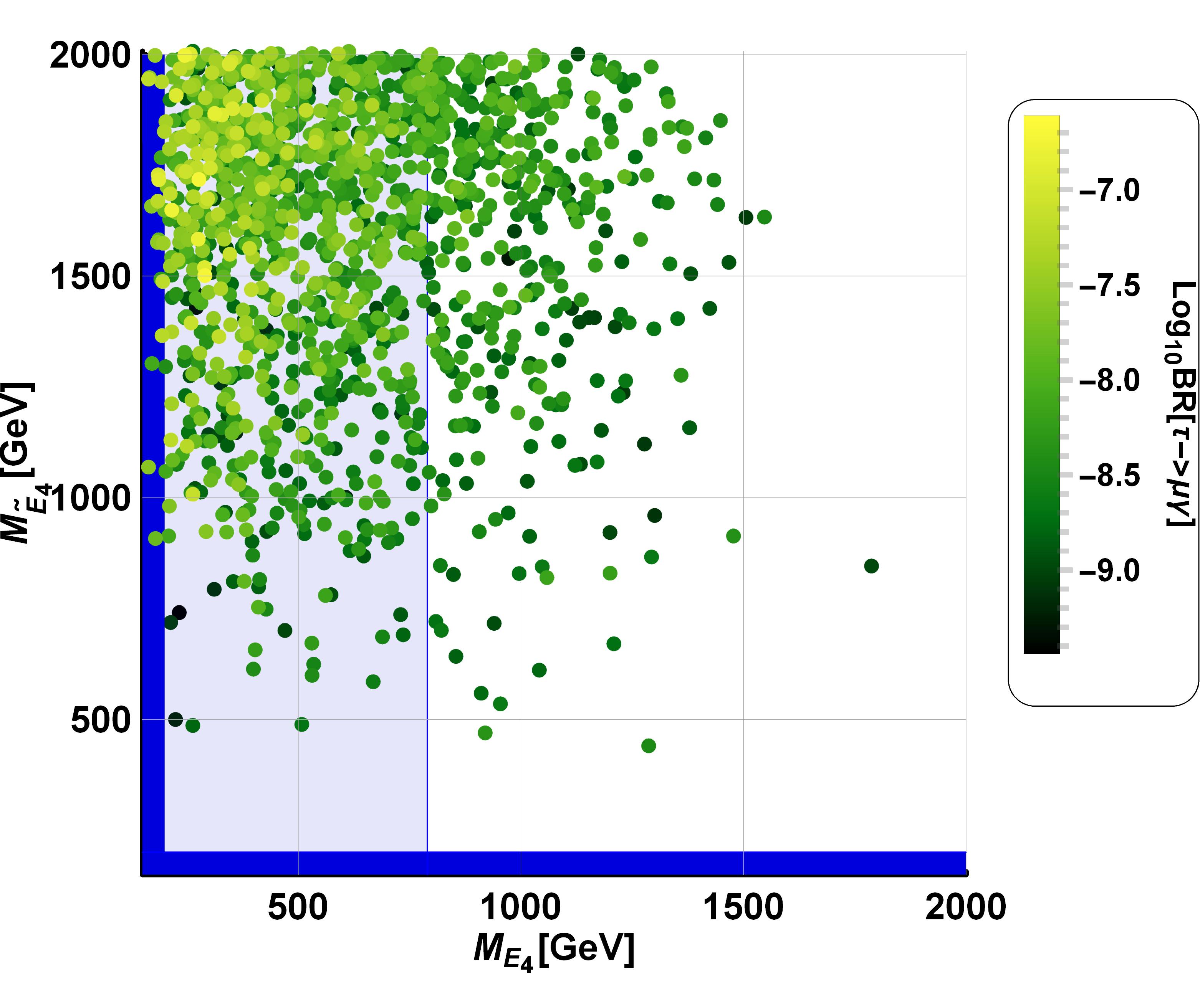}
\end{subfigure}
\begin{subfigure}{0.48\textwidth}
\includegraphics[keepaspectratio,width=0.95\textwidth]{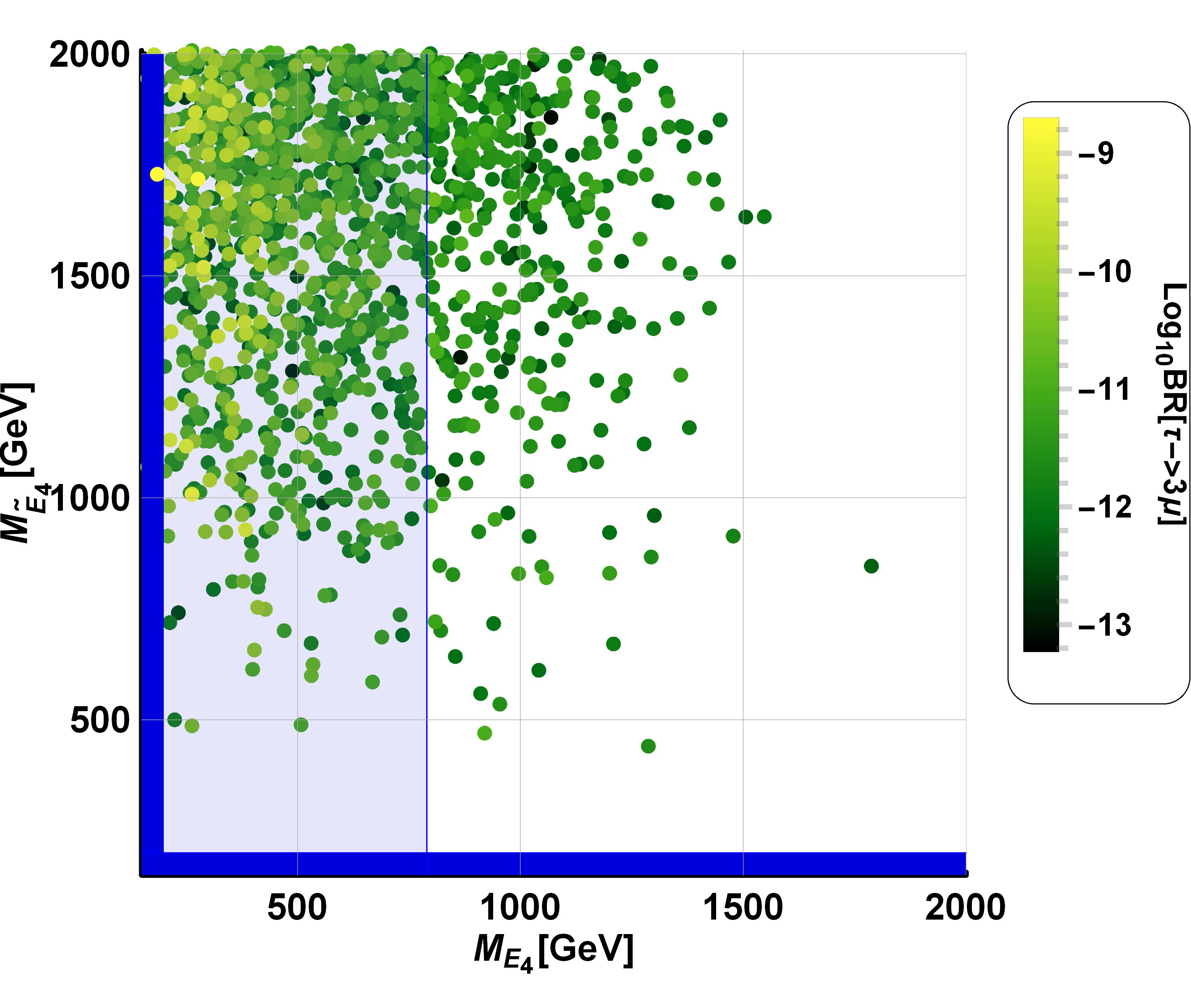}
\end{subfigure}
\begin{subfigure}{0.48\textwidth}
\includegraphics[keepaspectratio,width=0.95\textwidth]{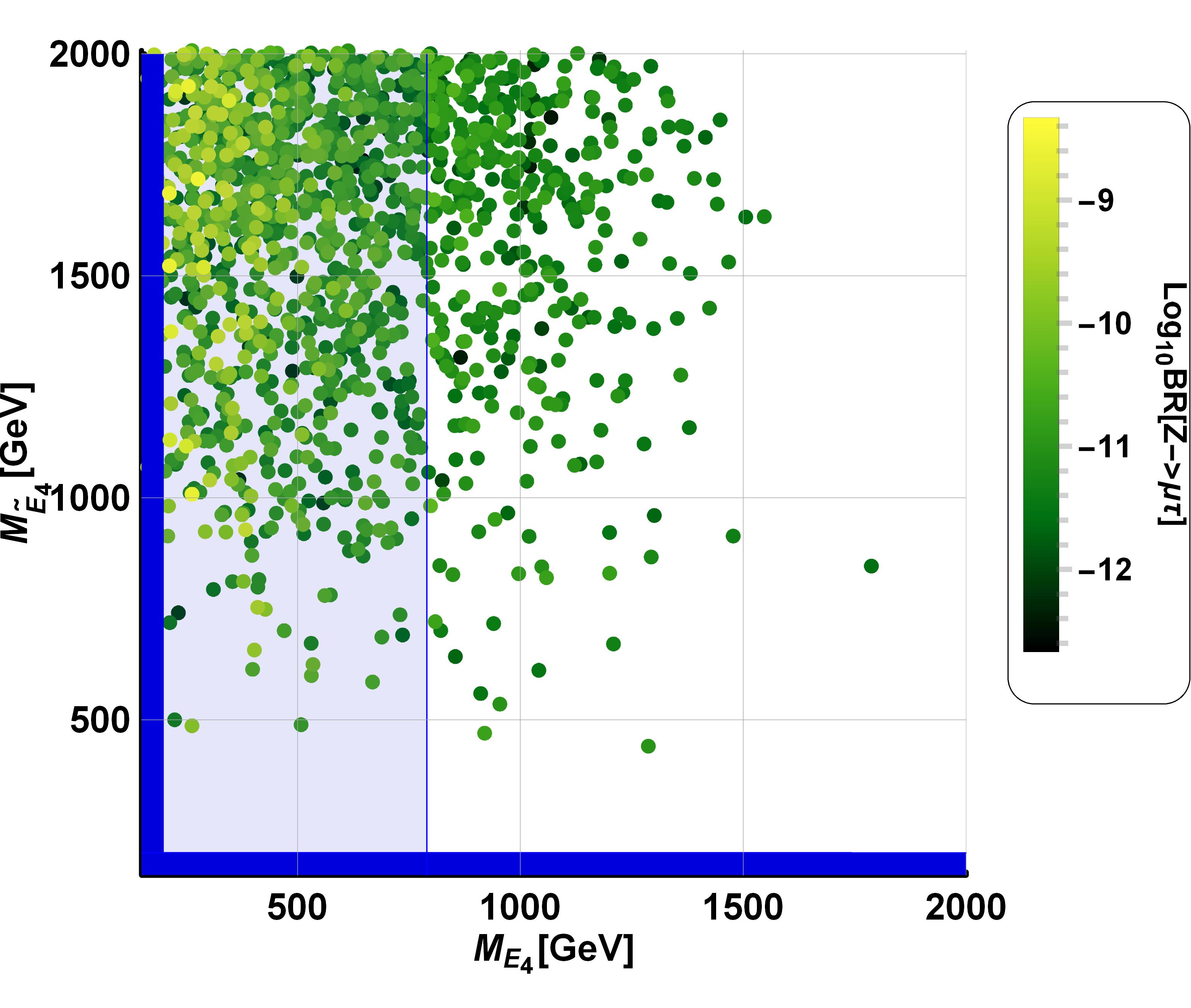}
\end{subfigure}
\caption{Scanned mass region of the vector-like charged leptons and contributions of the flavor violating interactions with the SM $Z$ gauge boson to the diverse CLFV decays $\tau \rightarrow \mu \gamma, \tau \rightarrow 3\mu$ and $Z \rightarrow \mu \tau$. The used constraints are the predicted muon and tau mass to be put between $[1 \pm 0.1] \times m_{\mu,\tau}$ and the $23$ mixing angle to be less than $0.2$. The darker blue region appearing in each diagram means either the singlet or doublet vector-like masses $M_{\widetilde{E}_4}, M_{E_4}$ are excluded up to $200\func{GeV}$ by \Steve{reference~\cite{Xu:2018pnq,Guedes:2021oqx}}. The brighter blue region means the doublet vector-like mass $M_{E_4}$ is excluded up to $790\func{GeV}$ by the CMS~\cite{Bhattiprolu:2019vdu,CMS:2019hsm}. The brighter red region appearing in above two plots is the region excluded by the experimental bound for the $\func{BR}\left( \tau \rightarrow \mu \gamma \right)_{\func{EXP}} = 4.4 \times 10^{-8}$.}
\label{fig:cl_CLFV_decays}
\end{figure}

\begin{figure}[H]
\centering 
\begin{subfigure}{0.48\textwidth}
\includegraphics[keepaspectratio,width=0.95\textwidth]{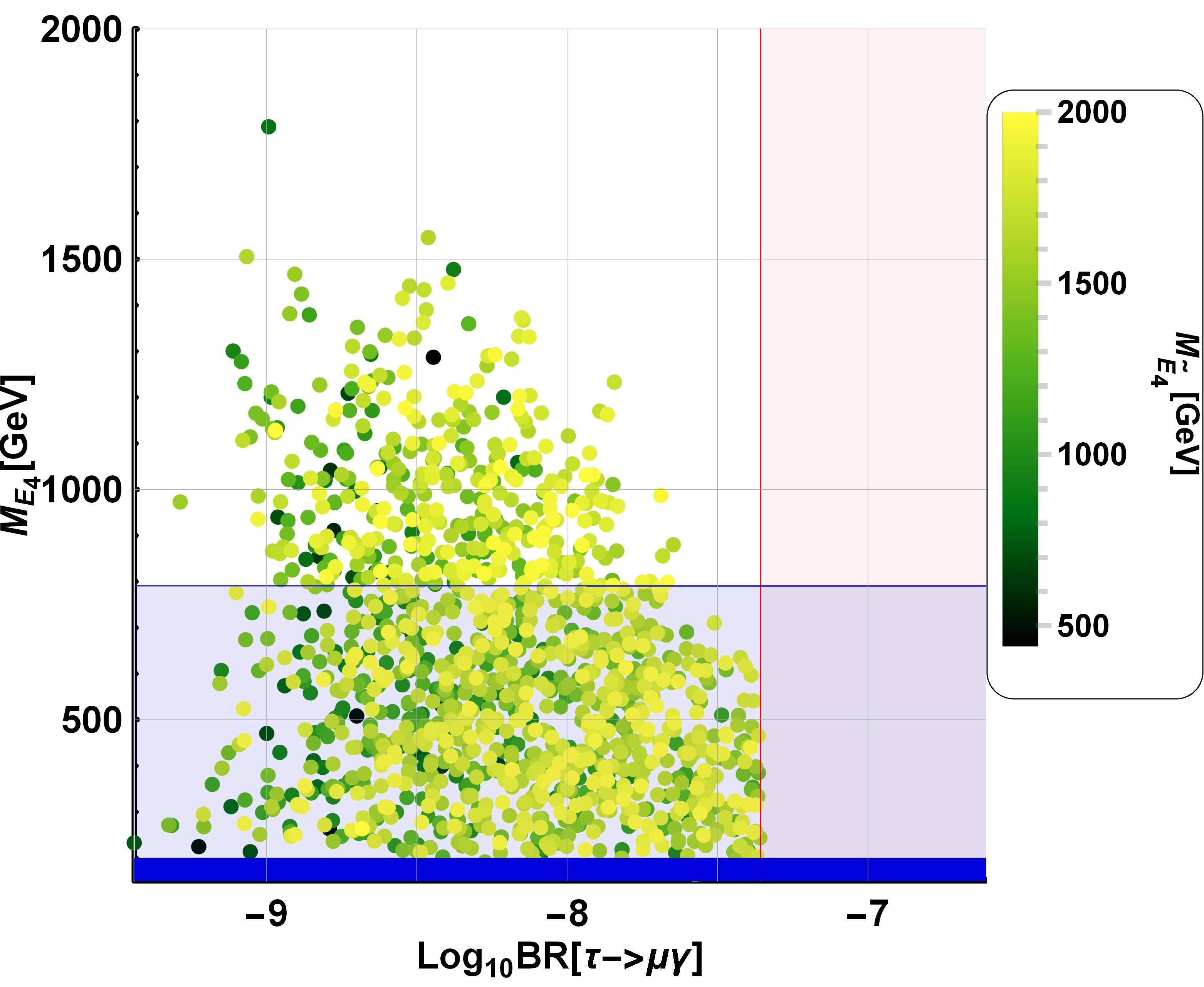}
\end{subfigure}
\begin{subfigure}{0.48\textwidth}
\includegraphics[keepaspectratio,width=0.95\textwidth]{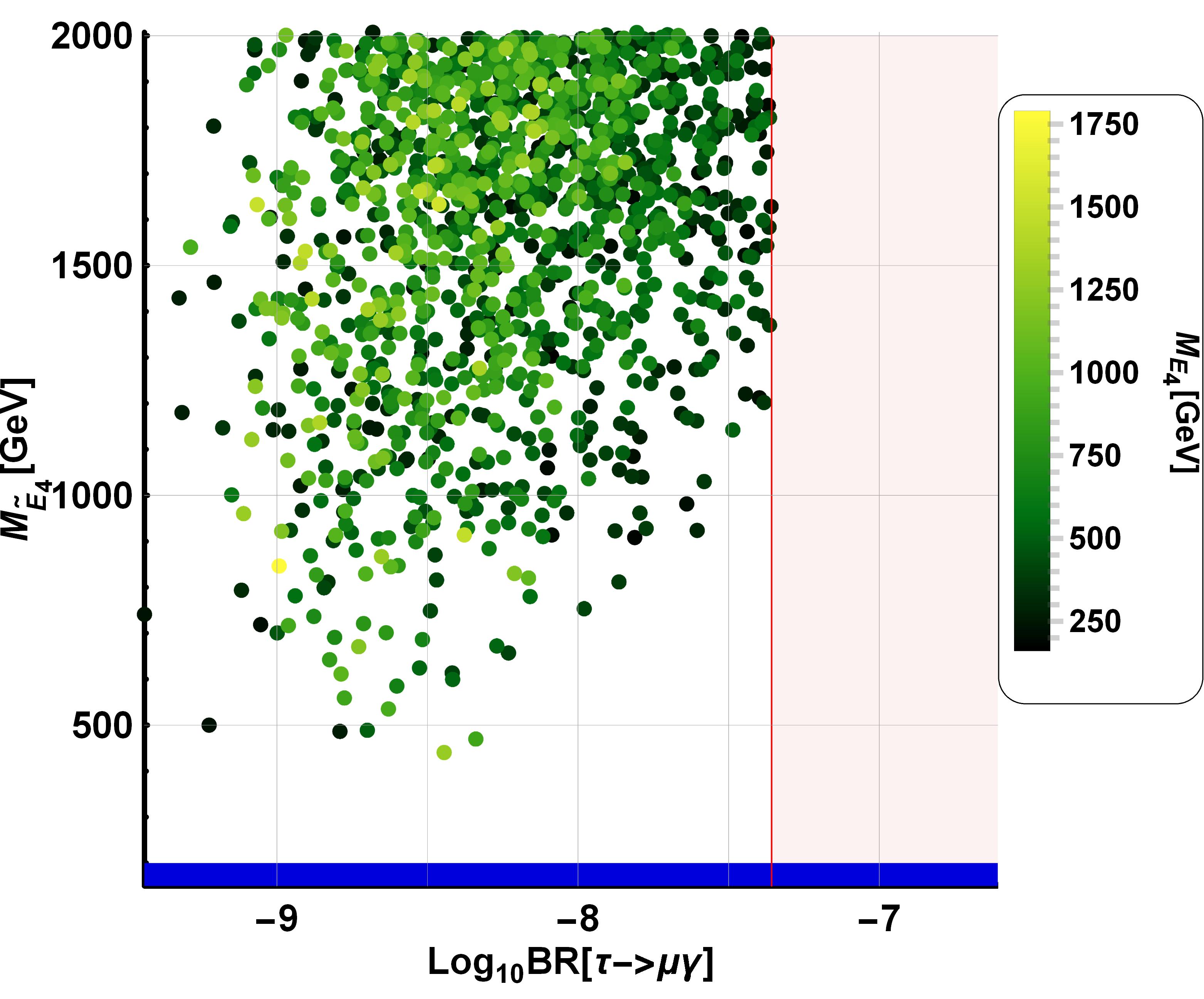}
\end{subfigure}
\begin{subfigure}{0.48\textwidth}
\includegraphics[keepaspectratio,width=0.95\textwidth]{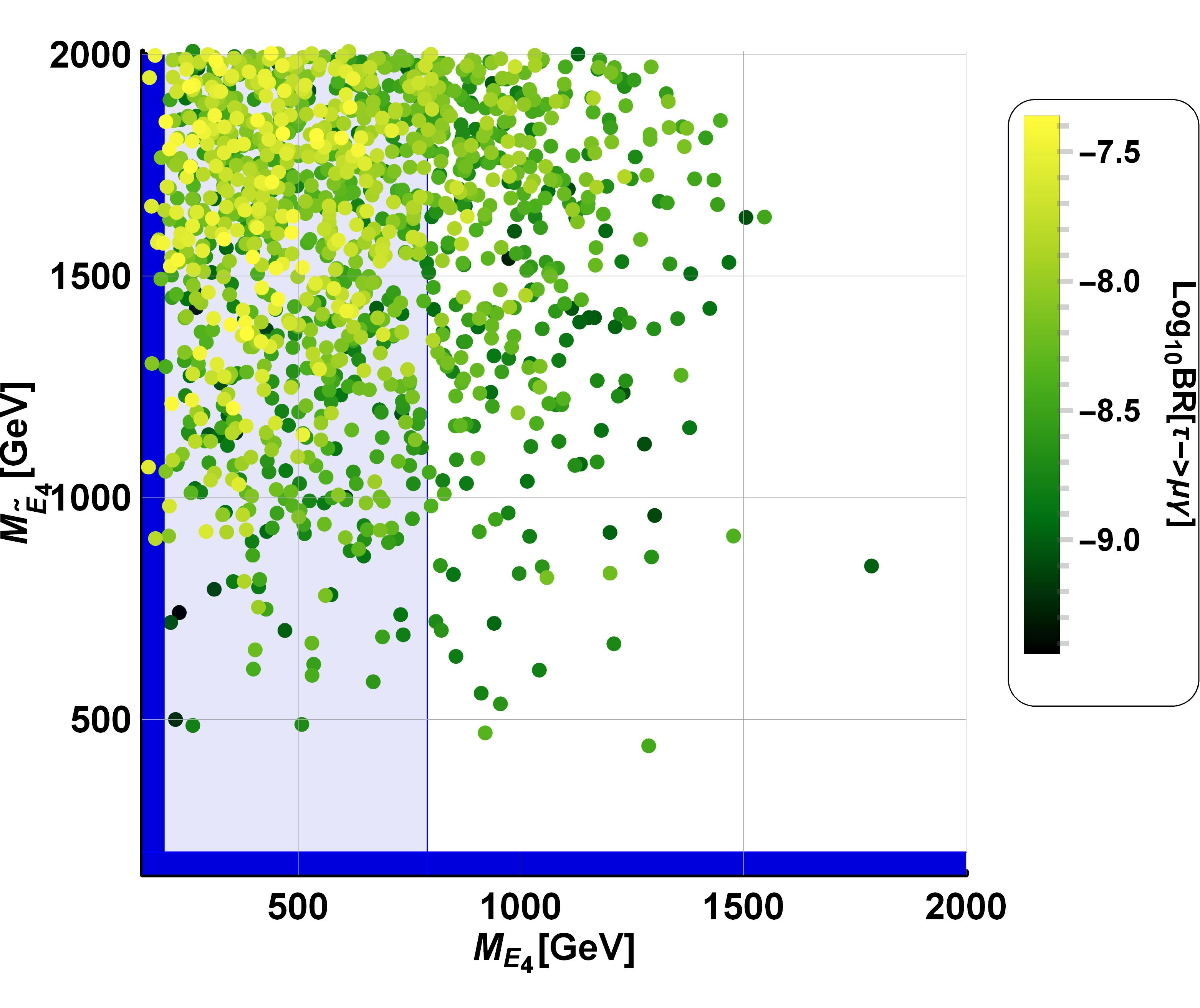}
\end{subfigure}
\begin{subfigure}{0.48\textwidth}
\includegraphics[keepaspectratio,width=0.95\textwidth]{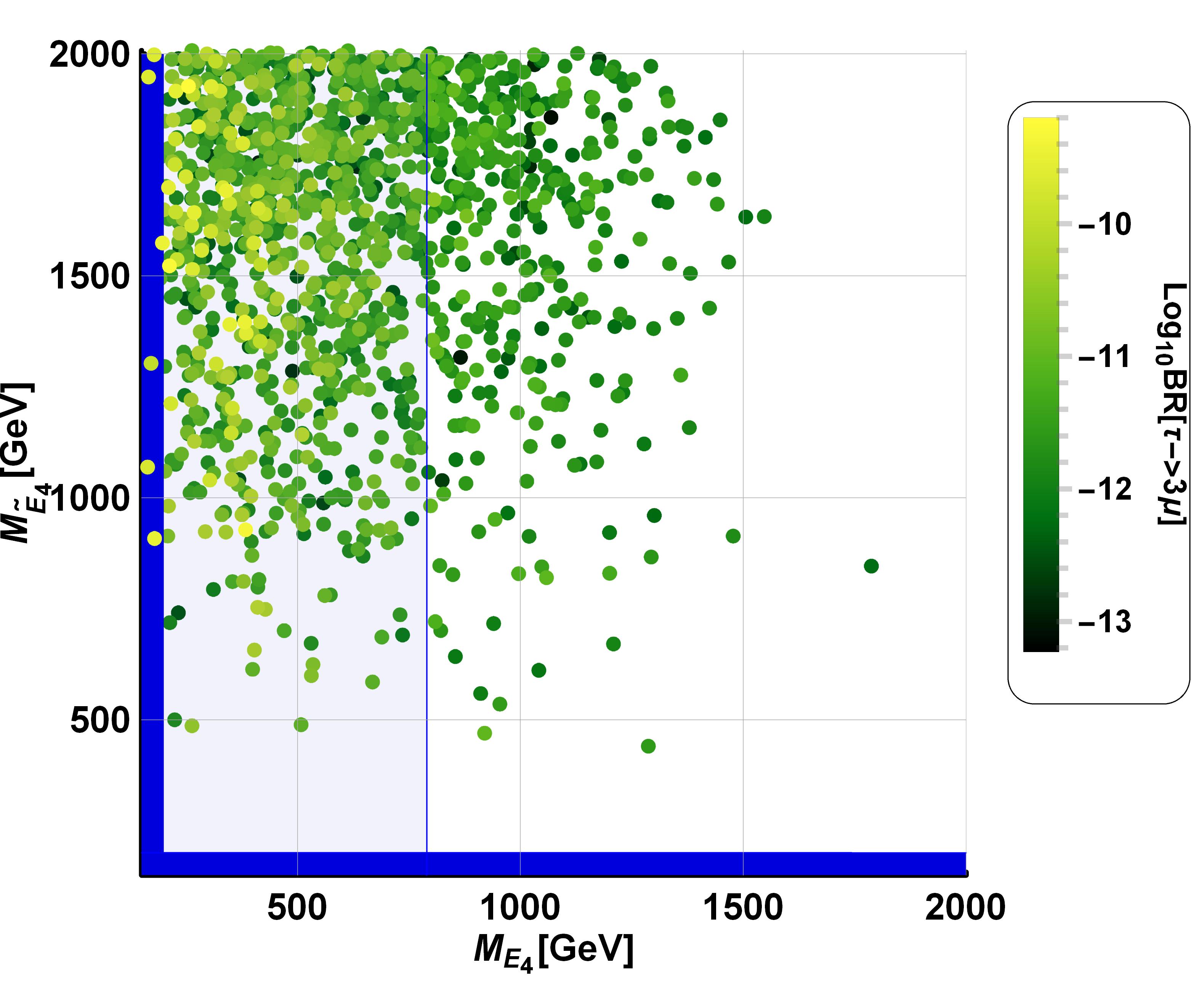}
\end{subfigure}
\begin{subfigure}{0.48\textwidth}
\includegraphics[keepaspectratio,width=0.95\textwidth]{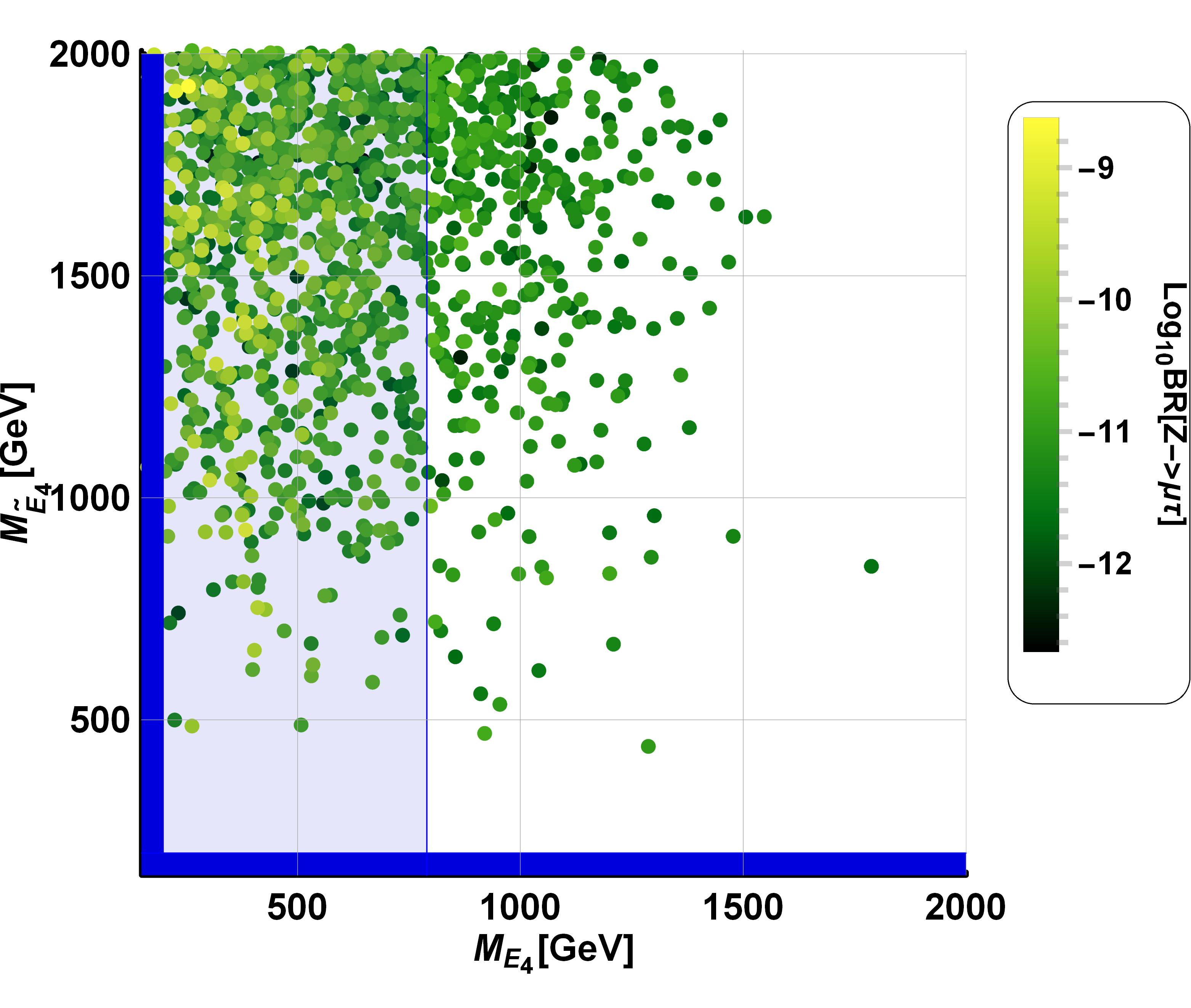}
\end{subfigure}
\caption{Reduced number of numerical predictions. The numerical predictions are constrained by the experimental value of the branching ratio of $\tau \rightarrow \mu \gamma$ decay. None of them are constrained by the branching ratio of $\tau \rightarrow 3\mu$ and $Z \rightarrow \mu \tau$ experimental bounds.}
\label{fig:cl_CLFV_second}
\end{figure}

The scanned mass range of the vector-like charged leptons are shown 
in Figure~\ref{fig:cl_CLFV_decays}.
The first we need to discuss is the experimental bounds for the vector-like charged leptons appearing in Figure~\ref{fig:cl_CLFV_decays}. The darker blue region is the excluded region for the vector-like charged leptons by $200\func{GeV}$~\Steve{\cite{Xu:2018pnq,Guedes:2021oqx}}. The vector-like mass $M_{E_4}$ consists of the doublet vector-like charged leptons $E_{4L},\widetilde{E}_{4R}$, whereas the other vector-like mass $M_{\widetilde{E}_4}$ consists of the singlet vector-like charged leptons $\widetilde{E}_{4L}, E_{4R}$. Therefore, $M_{E_4}$ is the doublet vector-like mass, whereas $M_{\widetilde{E}_4}$ is the singlet vector-like mass, and the doublet vector-like mass is excluded by CMS up to $790\func{GeV}$~\cite{Bhattiprolu:2019vdu,CMS:2019hsm}, expressed by the brighter blue region of Figure~\ref{fig:cl_CLFV_decays}. The second is our predictions for the branching ratio of $\tau \rightarrow \mu \gamma$ in Figure~\ref{fig:cl_CLFV_decays}. The relevant experimental bound for each CLFV decay is given by:
\begin{equation}
\begin{split}
&\func{BR}\left( \tau \rightarrow \mu \gamma \right)_{\func{EXP}} = 4.4 \times 10^{-8} 
\\[1ex]
&-
\func{Log}_{10} \func{BR} \left( \tau \rightarrow \mu \gamma \right)_{\func{EXP}} \simeq -7.4 
\\[1ex] 
&\func{BR}\left( \tau \rightarrow 3\mu \right)_{\func{EXP}} = 2.1 \times 10^{-8} 
\\[1ex]
&- \func{Log}_{10} \func{BR} \left( \tau \rightarrow 3\mu \right)_{\func{EXP}} \simeq -7.7 
\\[1ex]
&\func{BR}\left( Z \rightarrow \mu \tau \right)_{\func{EXP}} = 1.2 \times 10^{-5} 
\\[1ex]
&- \func{Log}_{10} \func{BR} \left( Z \rightarrow \mu \tau \right)_{\func{EXP}} \simeq -4.9 
\end{split}
\end{equation}
Our predictions for the CLFV $\tau \rightarrow 3\mu$ and $Z \rightarrow \mu \tau$ decays are not excluded by the experimental bound, however those are not the case for the CLFV decay $\tau \rightarrow \mu \gamma$, 
which exceed its upper experimental bound in some parts of the parameter space. This is due to, in some parts of the parameter space, the dominant contributions to the $\tau \rightarrow \mu \gamma$ decay involving a charged exotic lepton as well as chirality flip in the internal line and proportional to $M/m_{\tau}$ because of the sizeable large value of the charged exotic lepton - SM charged lepton mass ratio. 
After removing all excluded points by the experimental bound of the $\tau \rightarrow \mu \gamma$ decay, 
we obtain Figure~\ref{fig:cl_CLFV_second}.
Looking at our numerical predictions for the branching ratio of the $\tau \rightarrow \mu \gamma$ decay shown in Figure~\ref{fig:cl_CLFV_second}, some 
of them are constrained by the experimental limit of this 
branching ratio, however most of them survive, which implies that  
our numerical predictions for the branching ratio of $\tau \rightarrow \mu \gamma$ are not significantly constrained by its experimental bound. Furthermore, 
none of the numerical predictions for the branching ratios of the $\tau \rightarrow 3\mu$ and $Z \rightarrow \mu \tau$ decays  
are constrained by its experimental bound. However, our numerical predictions for the different 
CLFV decays can significantly be constrained by the future LHC upgrades having higher center of mass energy and luminosity than the ones of the current LHC, which will allow to set tighest constraints on charged exotic vector-like masses, thus leading to stronger constraints on the model parameter space.  
Regarding the CLFV $Z \rightarrow \mu \tau$ decay, the FCC-ee experiment has planned to generate $10^{12}$ SM $Z$ gauge bosons, which will allow to probe our model since  
 the branching ratio of the $Z \rightarrow \mu \tau$ decay can reach values of the order of $10^{-10}$ in the allowed region of the parameter space. Thus, the $Z \rightarrow \mu \tau$ decay is within the reach of the FCC-ee experiment, whose  
$Z$ factory 
\cite{DeRomeri:2016gum,Calibbi:2021pyh} will be crucial to verify or rule out this model. Concluding this subsection, our numerical predictions are not significantly constrained by any of the CLFV $\tau \rightarrow \mu \gamma, \tau \rightarrow 3\mu$ and $Z \rightarrow \mu \tau$ decays and might be able to be seen from the $Z$ factory for the first time, predicting the doublet vector-like charged lepton mass which is ranged from $790$ to $1600\func{GeV}$ whereas the singlet vector-like charged lepton mass is ranged from $500$ to $2000\func{GeV}$ or above than that.
\section{QUARK SECTOR PHENOMENOLOGY}
\label{sec:VI}
\begin{figure}[H]
\centering
\includegraphics[keepaspectratio,width=\textwidth]{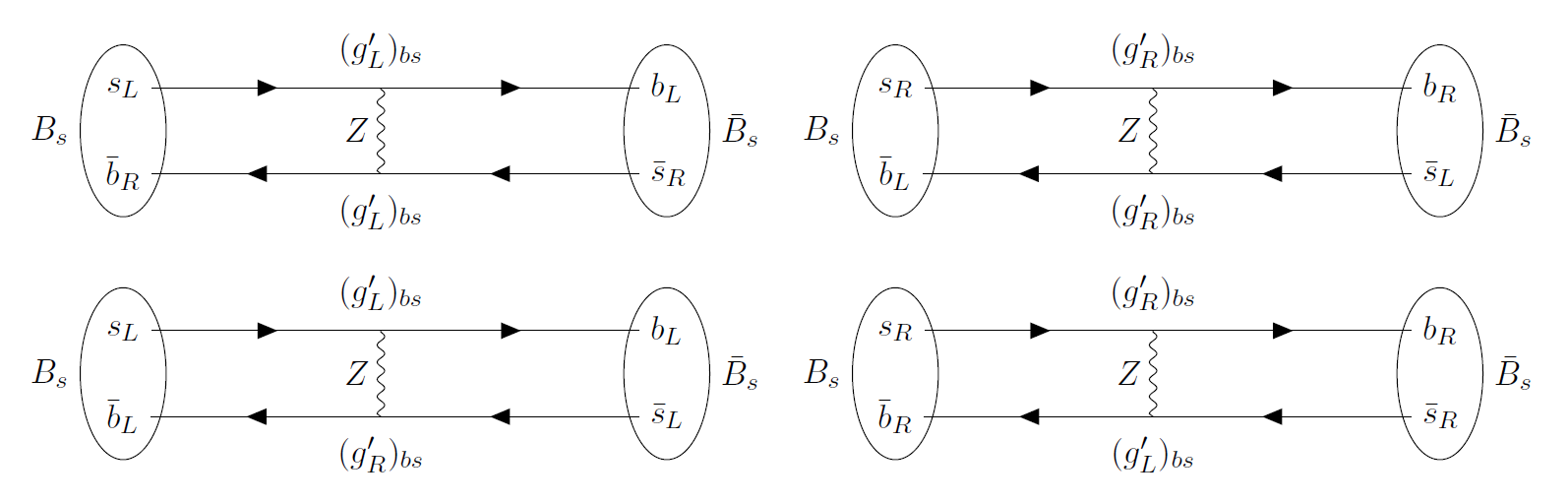}
\caption{Feynman diagrams contributing to the 
the $B_s-\bar{B}_s$ meson mixing involving the 
 tree-level exchange of the $Z$ gauge boson in the polarized basis}
\label{fig:BsBbars_mixing} 
\end{figure}

We have discussed the up- and down-type quark mass matrices pointed out that they have a different form, since the quark doublet rotation used in the up-type quark sector can not remove the down-type Yukawa term. This difference between up- and down-type quark mass matrices cause a distinct feature for each sector as follows:
\begin{itemize}
\item The up-type quark mass matrix can reach to the $23$ left (right)-handed mixing.
\item The down-type quark mass matrix can access to all left-handed mixings among the three SM generations, whereas the right-handed mixing can only have the $23$ mixing.
\end{itemize}
The interesting feature of the down-type quark mass matrix allows for flavor changing $Z$ interactions with down type quarks which yield 
neutral meson oscillations such as $K, B_d$ and $B_s$. Furthermore, an important feature to be mentioned is that the 
first generation of SM charged fermions do not acquire masses.  
Due to this property, our predictions for the neutral meson oscillations including the $d$ quark give a very suppressed energy difference corresponding to $10^{-40} \func{GeV}$, which is impossible to reach with the current experimental sensitivity. Then, the rest of the neutral meson oscillation $B_s$ is possible and an encouraging feature of the $B_s$ meson oscillation in our proposed model 
is the $B_s$ meson oscillation mediated by the SM $Z$ gauge boson can be calculated at tree-level as given in Figure~\ref{fig:BsBbars_mixing}.
From our numerical analysis we have found that the BSM contributions to the  
$B_s$ meson oscillation arising from 
the tree-level exchange of the  
$Z$ gauge boson yield the meson mass splitting of the order of 
$10^{-15}$ GeV or even less than that value, which is quite negligible compared to its corresponding   
experimental bound of $10^{-11} \func{GeV}$. The very suppressed  
new physics effect for the $B_s$ meson oscillation can be explained by considering   
the flavor violating coupling constants at each vertex of each diagram, whose value is about $10^{-6,-7}$ and this values are determined by 
the ratio between Yukawa and vector-like masses. For this reason, in the study of the phenomenological implications of our model in the flavor changing neutral interactions in the quark sector, we do not consider 
 the neutral meson oscillations as well as the $B_s \rightarrow \mu^+\mu^-$ decay. It is worth mentioning that the $B_s \rightarrow \mu^+\mu^-$ decay gives weaker effects than the neutral meson oscillations. Considering these facts, we conclude that the rare $t \rightarrow c Z$ decay and the CKM mixing matrix can constrain the quark sector of our  
 model, and thus  
 we discuss these two phenomenological aspects in the following subsections. 
\subsection{Analytic expression for the $t \rightarrow c Z$ decay}
\begin{figure}[H]
\centering
\includegraphics[keepaspectratio,width=\textwidth]{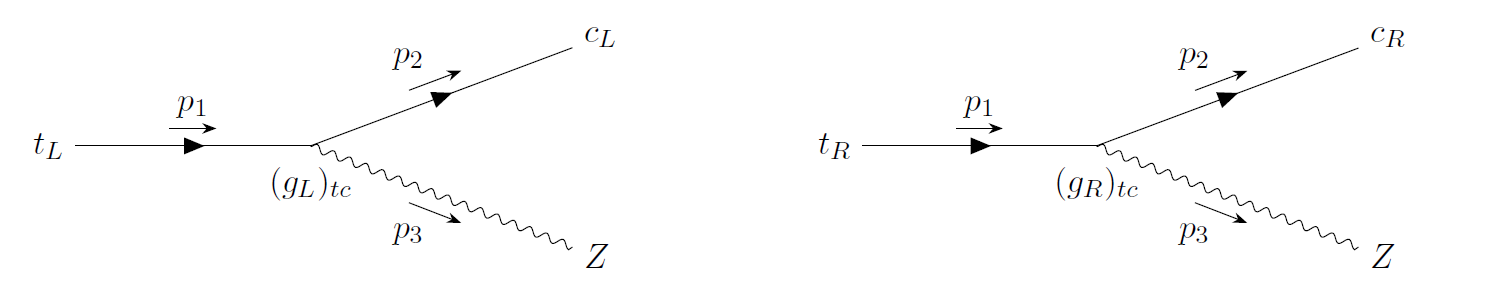}
\caption{tree-level Feynman diagrams contributing to the rare $t \rightarrow c Z$ decay}
\label{fig:tcZ}
\end{figure}

The $t \rightarrow c Z$ decay, which only appears at one-loop level in the SM,  
can take place at tree-level in our proposed model, thanks to the $Z$ mediated flavor changing neutral current interactions in the quark sector. In our proposed model, the $t \rightarrow cZ$ receives tree-level contributions which are depicted in 
Figure~\ref{fig:tcZ}.
Denoting the invariant amplitudes for the Feynman diagrams of the left and right panels of Figure~\ref{fig:tcZ} as $\mathcal{M}_L$ and $\mathcal{M}_R$, respectively, we find that they can be written as: 
\begin{equation}
\begin{split}
i\mathcal{M}_L &= i (g_L)_{tc} \epsilon_\mu^* (p_3) \overline{u} (p_2) \gamma^\mu P_L u (p_1) \\
i\mathcal{M}_R &= i (g_R)_{tc} \epsilon_\mu^* (p_3) \overline{u} (p_2) \gamma^\mu P_R u (p_1)
\label{eqn:amplitude_for_each_tcZ}
\end{split}
\end{equation}
In order to have the squared amplitude averaged and summed, we square the amplitudes given in Equation~\ref{eqn:amplitude_for_each_tcZ} and then sum over the different spin states, as follows:
\begin{equation}
\begin{gathered}
\frac{1}{2}\sum_{\func{spin}} \lvert \mathcal{M}_L \rvert^2 = (g_L)_{ct}^2 \left( (p_2 \cdot p_1) + \frac{2}{M_Z^2} (p_2 \cdot p_3)(p_1 \cdot p_3) \right) \\
\frac{1}{2}\sum_{\func{spin}} \lvert \mathcal{M}_R \rvert^2 = (g_R)_{ct}^2 \left( (p_2 \cdot p_1) + \frac{2}{M_Z^2} (p_2 \cdot p_3)(p_1 \cdot p_3) \right) \\
\frac{1}{2} \sum_{\func{spin}} \lvert \mathcal{M} \rvert^2 = \frac{1}{2} \sum_{\func{spin}} \left( \lvert \mathcal{M}_L \rvert^2 + \lvert \mathcal{M}_R \rvert^2 \right) = \frac{1}{2}\left( (g_L)_{ct}^2 + (g_R)_{ct}^2 \right) \left( (p_2 \cdot p_1) + \frac{2}{M_Z^2} (p_2 \cdot p_3)(p_1 \cdot p_3) \right) 
\end{gathered}
\end{equation}
Then, the decay rate equation is given by:
\begin{equation}
\begin{split}
d\Gamma\left( t \rightarrow c Z \right) &= \frac{1}{2m_t} \frac{d^3p_2}{(2\pi)^3 2E_2} \frac{d^3p_3}{(2\pi)^3 2E_3} \lvert \mathcal{M} \rvert^2 (2\pi)^4 \delta^{(4)} \left( p_1 - p_2 - p_3 \right) \\
\Gamma\left( t \rightarrow c Z \right) &= \frac{\lvert p^* \rvert}{32\pi^2 m_t^2} \int \lvert \mathcal{M} \rvert^2 d\Omega \\
&= \frac{1}{8\pi m_t^2} \frac{1}{2m_t}(m_t^2 - M_Z^2) ( (g_L)_{ct}^2 + (g_R)_{ct}^2 ) \\
&\times \left[ \frac{m_t^2 + m_c^2 - M_Z^2}{2} + \frac{2}{M_Z^2} (\frac{m_t^2 - m_c^2 - M_Z^2}{2}) (\frac{m_t^2 - m_c^2 + M_Z^2}{2}) \right]
\end{split}
\end{equation}
where $p^* \simeq \frac{1}{2m_t} \left( m_t^2 - M_Z^2 \right)$.
Then, we are ready to write down our prediction for the branching ratio of the tree-level $t \rightarrow c Z$ decay~\Steve{\cite{ATLAS:2018zsq}}: 
\begin{equation}
\func{BR}\left(t \rightarrow c Z \right) = \frac{\Gamma\left( t \rightarrow c Z \right)}{\Gamma_{t}} < \func{BR} \left( t \rightarrow c Z \right)_{\func{EXP}} = 2.4 \times 10^{-4} \quad (95\% \func{CL})
\label{eqn:BRtcZ_exp}
\end{equation}
where $\Gamma_t = 1.32 \func{GeV}$.
\subsection{Analytic expression for the CKM mixing matrix}
In order to discuss the CKM mixing matrix, the first task we need to investigate is the $W$ current of the SM in order to see how the CKM mixing matrix can take place (we only consider the three SM generations at the moment).
\begin{equation}
\begin{split}
\mathcal{L}_{\func{SM}}^{W} &= g j_{\mu}^{W+} W^{\mu +} = \frac{g}{\sqrt{2}} \left( \overline{u}_{L}^{i} \gamma_{\mu} d_{L}^{i} \right) W^{\mu +} 
\\
&= \frac{g}{\sqrt{2}} \left( \overline{u}_{L}^{i} (V_{L}^{u\dagger} V_{L}^{u}) \gamma_{\mu} (V_{L}^{d\dagger} V_{L}^{d}) d_{L}^{i} \right) W^{\mu +} = \frac{g}{\sqrt{2}} \left( \overline{u}_{L}^{i \prime} \gamma_{\mu} (V_{\func{CKM}}) d_{L}^{i \prime} \right) W^{\mu +}  
\end{split}
\end{equation}
where $u_{L}^{i \prime}, d_{L}^{i \prime}$ are the up- and down-type quarks of the SM in the mass basis and $V_{\func{CKM}}$ is the CKM mixing matrix defined as $V_{L}^{u} V_{L}^{d \dagger}$. Now we extend the quark spectrum by considering the vector-like quarks, thus implying that 
the $W$ current takes the form:
\begin{equation}
\begin{split}
\mathcal{L}^{W} &= 
\frac{g}{\sqrt{2}}
\begin{pmatrix}
\overline{u}_{1L} & \overline{u}_{2L} & \overline{u}_{3L} & \overline{u}_{4L} & \overline{\widetilde{u}}_{4L}
\end{pmatrix}
\gamma_{\mu}
\begin{pmatrix}
1 & 0 & 0 & 0 & 0 \\
0 & 1 & 0 & 0 & 0 \\
0 & 0 & 1 & 0 & 0 \\
0 & 0 & 0 & 1 & 0 \\
0 & 0 & 0 & 0 & 0 
\end{pmatrix}
\begin{pmatrix}
d_{1L} \\
d_{2L} \\
d_{3L} \\
d_{4L} \\
\widetilde{d}_{4L}
\end{pmatrix}
W^{\mu +}
\\
&= 
\frac{g}{\sqrt{2}}
\begin{pmatrix}
\overline{u}_{1L} & \overline{u}_{2L} & \overline{u}_{3L} & \overline{u}_{4L} & \overline{\widetilde{u}}_{4L}
\end{pmatrix}
V_{L}^{u \dagger} V_{L}^{u}
\gamma_{\mu}
\begin{pmatrix}
1 & 0 & 0 & 0 & 0 \\
0 & 1 & 0 & 0 & 0 \\
0 & 0 & 1 & 0 & 0 \\
0 & 0 & 0 & 1 & 0 \\
0 & 0 & 0 & 0 & 0 
\end{pmatrix}
V_{L}^{d \dagger} V_{L}^{d}
\begin{pmatrix}
d_{1L} \\
d_{2L} \\
d_{3L} \\
d_{4L} \\
\widetilde{d}_{4L}
\end{pmatrix}
W^{\mu +}
\\
&= 
\frac{g}{\sqrt{2}}
\begin{pmatrix}
\overline{u}_{L} & \overline{c}_{L} & \overline{t}_{L} & \overline{U}_{4L} & \overline{\widetilde{U}}_{4L}
\end{pmatrix}
\gamma_{\mu}
V_{L}^{u}
\begin{pmatrix}
1 & 0 & 0 & 0 & 0 \\
0 & 1 & 0 & 0 & 0 \\
0 & 0 & 1 & 0 & 0 \\
0 & 0 & 0 & 1 & 0 \\
0 & 0 & 0 & 0 & 0 
\end{pmatrix}
V_{L}^{d \dagger}
\begin{pmatrix}
d_{L} \\
s_{L} \\
b_{L} \\
D_{4L} \\
\widetilde{D}_{4L}
\end{pmatrix}
W^{\mu +}
\end{split}
\label{eqn:CKM_analytic}
\end{equation}
where the CKM mixing prediction in our model is given by
\begin{equation}
V_{\func{CKM}} = 
V_{L}^{u}
\begin{pmatrix}
1 & 0 & 0 & 0 & 0 \\
0 & 1 & 0 & 0 & 0 \\
0 & 0 & 1 & 0 & 0 \\
0 & 0 & 0 & 1 & 0 \\
0 & 0 & 0 & 0 & 0 
\end{pmatrix}
V_{L}^{d \dagger}, \quad \text{for the upper-left $3 \times 3$ block}
\label{eqn:CKM_prediction}
\end{equation}
where $V_{L}^{u}$ is the mixing matrix for the up-type quarks defined in Equation~\ref{eqn:up_mixing} and $V_{L}^{d}$ is the one corresponding to  
the down-type quarks given in Equation~\ref{eqn:down_mixing}. The zero appearing in the middle matrix between $V_{L}^{u}$ and $V_{L}^{d \dagger}$ arises from the left-handed vector-like quark singlets $\widetilde{U}_{4L}$ and $\widetilde{D}_{4L}$ which do not interact with the $W$ currents, so our prediction for the CKM mixing matrix does not feature the unitarity requirement and this leads to the need of relaxing the unitarity constraint of the CKM quark mixing matrix. That deviation of unitarity of the CKM quark mixing matrix is due to the presence of heavy vector-like quarks and this aspect was studied in 
\cite{Branco:2021vhs} in the context of theory with different particle spectrum and symmetry than ours. 
Few sigma of SM deviations from the first row of the CKM mixing matrix without unitarity were analyzed in \cite{Branco:2021vhs}. 
The deviation from the Unitarity of the CKM will also be discussed in our numerical result and the experimental CKM mixing matrix without unitarity is given by~\cite{Branco:2021vhs,ParticleDataGroup:2020ssz}:
\begingroup
\setlength\arraycolsep{5pt}
\begin{equation}
\lvert K_{\func{CKM}} \rvert
=
\begin{pmatrix}
0.97370 \pm 0.00014 & 0.22450 \pm 0.00080 & 0.00382 \pm 0.00024 \\[1.5ex]
0.22100 \pm 0.00400 & 0.98700 \pm 0.01100 & 0.04100 \pm 0.00140 \\[1.5ex]
0.00800 \pm 0.00030 & 0.03880 \pm 0.00110 & 1.01300 \pm 0.03000 
\end{pmatrix}
\end{equation}
\endgroup
\subsection{Numerical analysis for each prediction in the quark sector}
When compared to the charged lepton sector simulation, the quark sector becomes much more complicated  
since we need 
to fit the masses of the $c,t,s$ and $b$ quarks simultaneously as well as the CKM mixing matrix without imposing the unitarity requirement. Therefore, we fit the masses of the four quarks first by using a fitting function $\chi_{\func{mass}}^{2}$ and then we start a second fitting procedure by using another fitting function $\chi_{\func{CKM}}^{2}$ and this will be discussed in detail  
in the following subsections.
\subsubsection{The fitting function $\chi^2$ and free parameter setup}
We set up our parameter region as follows:
\begin{center}
{\renewcommand{\arraystretch}{1.5} 
\begin{tabular}{cc}
\toprule
\toprule
\textbf{Mass parameter} & \textbf{Scanned Region($\func{GeV}$)} \\ 
\midrule
$y_{24}^{u} v_{u} = m_{24}^{u}$ & $\pm [10,50]$ \\[0.5ex]
$y_{34}^{u} v_{u} = m_{34}^{u}$ & $\pm [200,400]$ \\[0.5ex]
$y_{43}^{u} v_{u} = m_{43}^{u}$ & $\pm [200,400]$ \\[0.5ex]
$x_{34}^{Q} v_{\phi} = m_{35}^{Q}$ & $m_{35}^{Q}$ \\[0.5ex]
$x_{42}^{u} v_{\phi} = m_{52}^{u}$ & $\pm [500,700]$ \\[0.5ex]
$x_{43}^{u} v_{\phi} = m_{53}^{u}$ & $\pm [50,500]$ \\[0.5ex]
$M_{45}^{Q}$ & $M_{45}^{Q}$ \\[0.5ex]
$M_{54}^{u}$ & $\pm [1000,3000]$ \\[0.5ex]
\midrule
$y_{14}^{d} v_{d} = m_{14}^{d}$ & $\pm [1,10]$ \\[0.5ex]
$y_{24}^{d} v_{d} = m_{24}^{d}$ & $\pm [5,20]$ \\[0.5ex]
$y_{34}^{d} v_{d} = m_{34}^{d}$ & $\pm [10,30]$ \\[0.5ex]
$y_{43}^{d} v_{d} = m_{43}^{d}$ & $\pm [5,10]$ \\[0.5ex]
$x_{34}^{Q} v_{\phi} = m_{35}^{Q}$ & $\pm [10,100]$ \\[0.5ex]
$x_{42}^{d} v_{\phi} = m_{52}^{d}$ & $\pm [10,100]$ \\[0.5ex]
$x_{43}^{d} v_{\phi} = m_{53}^{d}$ & $\pm [10,100]$ \\[0.5ex]
$M_{45}^{Q}$ & $\pm [1000,3000]$ \\[0.5ex]
$M_{54}^{d}$ & $\pm [1000,3000]$ \\[0.5ex]
\bottomrule
\bottomrule 
\end{tabular}
\captionof{table}{Initial parameter setup for scanning mass of the vector-like quarks} 
\label{tab:parameter_region_initial_quarks}}
\end{center}
There are a few things to be noticed as in the charged lepton case.
\begin{enumerate}
\item The relation $v_{u}^{2} + v_{d}^{2} = (246\func{GeV})^2$ still holds and the mass parameters $m_{24,34,43}^{u}$ can not exceed the upper perturbative limit on 
the Yukawa constant $\sqrt{4\pi} \simeq 3.54$ multiplied by the vev 
$\approx 240\func{GeV}$,  
of the $H_{u}$ Higgs, thus yielding the bound of $850\func{GeV}$ for these mass parameters. These restrictions have been taken into account through the whole fitting process.
\item The down-type Higgs $H_{d}$ has a very small vev, which is about order of $10\func{GeV}$, based on our previous analysis 
~\cite{Hernandez:2021tii}, and the range of values of the mass parameters $m_{14,24,34,43}^{d}$ are considered under this assumption.
\item Since we do not know the correct scale of $v_{\phi}$, we considered  
$m_{35}^{Q}, m_{52,53}^{u}$ and $m_{52,53}^{d}$ as free parameters. For the same reason, the vector-like masses $M_{45}^{Q}$ and $M_{54}^{u,d}$ are considered free parameters as well.
\item The mass parameters $m_{35}^{Q}$ and $M_{45}^{Q}$ appear in a  
common term shared by both up- and down-type quark sector mass matrices and this feature has been discussed in the paragraph below  
Equation~\ref{eqn:diff_up_down}.
\end{enumerate}
The next thing to do is to set up the two fitting functions $\chi_{\func{mass}}^{2}$ and $\chi_{\func{CKM}}^{2}$ as follows:
\begin{equation}
\chi_{\func{mass}}^2 = \sum_{f=c,t,s,b} = \frac{(m_{f}^{\func{pred}}-m_{f}^{\func{EXP}})^2}{(\delta m_{f}^{\func{EXP}})^2}, \quad \chi_{\func{CKM}}^2 = \sum_{i,j=1,2,3} \frac{((V_{\func{CKM}}^{\func{pred}})_{ij} - (V_{\func{CKM}}^{\func{EXP}})_{ij})^2}{((\delta V_{\func{CKM}}^{\func{EXP}})_{ij})^2},
\end{equation}
where the superscript $\func{pred}$ means our prediction to its experimental value and the delta means error bar of the physical quantity at $1\sigma$. Our first goal is to fit the masses of the four quarks simultaneously. For the charged lepton case, we require that our obtained muon and tau masses to be in the range 
$[1 \pm 0.1] \times m_{\mu,\tau}$ and this requirement is also imposed for  
the $c,s$ and $b$ quarks excepting for the $t$ quark since the $t$ quark is too heavy. Besides that, we require that the obtained top quark mass 
to be in the range 
 $[1 \pm 0.01] \times m_{t}$ instead of $[1 \pm 0.1] \times m_{t}$. After the mass parameters have been converged to be put between the arranged range of each quark mass, 
 we use the other fitting function $\chi_{\func{CKM}}^{2}$ to fit our prediction for the CKM mixing matrix. Using one of the defined fitting functions, we need to vary the mass parameters of Table~\ref{tab:parameter_region_initial_quarks} by a factor of $\left[ 1 \pm \kappa \right]$ where $\kappa = 0.1$ in order to find better mass parameters. We rename the given parameters from the initial parameter setup by adding an subscript $r$ to the mass parameters. The varied mass parameters are given in Table~\ref{tab:parameter_region_initial_q_second}.
\begin{center}
{\renewcommand{\arraystretch}{1.5} 
\begin{tabular}{cc}
\toprule
\toprule
\textbf{Mass parameter} & \textbf{Scanned Region($\func{GeV}$)} \\ 
\midrule
$y_{24}^{u} v_{u} = m_{24}^{u}$ & $\left[ 1 \pm \kappa \right] \times m_{24r}^{u}$ \\[0.5ex]
$y_{34}^{u} v_{u} = m_{34}^{u}$ & $\left[ 1 \pm \kappa \right] \times m_{34r}^{u}$ \\[0.5ex]
$y_{43}^{u} v_{u} = m_{43}^{u}$ & $\left[ 1 \pm \kappa \right] \times m_{43r}^{u}$ \\[0.5ex]
$x_{34}^{Q} v_{\phi} = m_{35}^{Q}$ & $m_{35}^{Q}$ \\[0.5ex]
$x_{42}^{u} v_{\phi} = m_{52}^{u}$ & $\left[ 1 \pm \kappa \right] \times m_{42r}^{u}$ \\[0.5ex]
$x_{43}^{u} v_{\phi} = m_{53}^{u}$ & $\left[ 1 \pm \kappa \right] \times m_{43r}^{u}$ \\[0.5ex]
$M_{45}^{Q}$ & $M_{45}^{Q}$ \\[0.5ex]
$M_{54}^{u}$ & $\left[ 1 \pm \kappa \right] \times M_{54r}^{u}$ \\[0.5ex]
\midrule
$y_{14}^{d} v_{d} = m_{14}^{d}$ & $\left[ 1 \pm \kappa \right] \times m_{14r}^{d}$ \\[0.5ex]
$y_{24}^{d} v_{d} = m_{24}^{d}$ & $\left[ 1 \pm \kappa \right] \times m_{24r}^{d}$ \\[0.5ex]
$y_{34}^{d} v_{d} = m_{34}^{d}$ & $\left[ 1 \pm \kappa \right] \times m_{34r}^{d}$ \\[0.5ex]
$y_{43}^{d} v_{d} = m_{43}^{d}$ & $\left[ 1 \pm \kappa \right] \times m_{43r}^{d}$ \\[0.5ex]
$x_{34}^{Q} v_{\phi} = m_{35}^{Q}$ & $\left[ 1 \pm \kappa \right] \times m_{35r}^{Q}$ \\[0.5ex]
$x_{42}^{d} v_{\phi} = m_{52}^{d}$ & $\left[ 1 \pm \kappa \right] \times m_{42r}^{d}$ \\[0.5ex]
$x_{43}^{d} v_{\phi} = m_{53}^{d}$ & $\left[ 1 \pm \kappa \right] \times m_{43r}^{d}$ \\[0.5ex]
$M_{45}^{Q}$ & $\left[ 1 \pm \kappa \right] \times M_{45r}^{Q}$ \\[0.5ex]
$M_{54}^{d}$ & $\left[ 1 \pm \kappa \right] \times M_{54r}^{d}$ \\[0.5ex]
\midrule
$\kappa$ & $0.1$ \\
\bottomrule
\bottomrule 
\end{tabular}
\captionof{table}{Next parameter setup after the initial parameter setup to find better mass parameters} 
\label{tab:parameter_region_initial_q_second}}
\end{center}
We vary the parameter space given in Table~\ref{tab:parameter_region_initial_q_second} by first using the fitting function $\chi_{\func{mass}}^{2}$  
in order to find a suitable mass prediction for the four quarks $t,b,c$ and $b$. Once the obtained masses of these quarks are allocated in the ranges 
($\left[ 1 \pm 0.1 \right] \times m_{c,s,b}$ and $\left[ 1 \pm 0.01 \right] \times m_{t}$), we proceed
to fit the CKM quark mixing matrix once more by using the other fitting function $\chi_{\func{CKM}}^{2}$ and it is worth mentioning that  
fitting the CKM mixing matrix is much more challenging due to the very small experimental errors of the CKM matrix elements. We display a benchmark point most converged for the CKM mixing matrix at the next subsection to discuss the possible deviation from the SM result arising from the first row of the CKM mixing matrix.
\subsubsection{Numerical scan result for the quark sector}
We start with the most converged benchmark point ($\chi_{\func{CKM}}^{2} = 956.828$) after repeating the varying many times
\begin{equation}
\begin{split}
M^{u} &= 
\begin{pmatrix}
0 & 0 & 0 & 0 & 0 \\[0.5ex]
0 & 0 & 0 & 0 & 14.474 \\[0.5ex]
0 & 0 & 0 & 1206.340 & 277.563 \\[0.5ex]
0 & 0 & 273.503 & -1775.200 & 0 \\[0.5ex]
0 & 550.990 & 434.462 & 0 & -5624.050 
\end{pmatrix}
\quad
M^{d} = 
\begin{pmatrix}
0 & 0 & 0 & 0 & -0.938 \\[0.5ex]
0 & 0 & 0 & 0 & -4.041 \\[0.5ex]
0 & 0 & 0 & 1206.340 & -27.427 \\[0.5ex]
0 & 0 & -5.636 & -1775.200 & 0 \\[0.5ex]
0 & 72.915 & -75.760 & 0 & 2623.620
\end{pmatrix}
\\[2ex]
M_{\func{diag}}^{u} &= 
\begin{pmatrix}
0 & 0 & 0 & 0 & 0 \\[0.5ex]
0 & 1.255 & 0 & 0 & 0 \\[0.5ex]
0 & 0 & 171.303 & 0 & 0 \\[0.5ex]
0 & 0 & 0 & 2155.890 & 0 \\[0.5ex]
0 & 0 & 0 & 0 & 5674.840 
\end{pmatrix}
\quad
M_{\func{diag}}^{d} = 
\begin{pmatrix}
0 & 0 & 0 & 0 & 0 \\[0.5ex]
0 & 0.094 & 0 & 0 & 0 \\[0.5ex]
0 & 0 & 3.875 & 0 & 0 \\[0.5ex]
0 & 0 & 0 & 2146.190 & 0 \\[0.5ex]
0 & 0 & 0 & 0 & 2625.960 
\end{pmatrix},
\label{eqn:MuMd_diagMuMd}
\end{split}
\end{equation}
where the above two mass matrices of Equation~\ref{eqn:MuMd_diagMuMd} are the mass matrices for the up- and down-type quarks in the flavor basis, whereas the below two mass matrices are ones fully diagonalized, so revealing all propagating quark mass. From the mass matrices of Equation~\ref{eqn:MuMd_diagMuMd}, we have the mixing matrices $V_{L}^{u}$ and $V_{L}^{d}$ and arrive to our CKM prediction using the formula of Equation~\ref{eqn:CKM_prediction}.
\begin{equation}
V_{\func{CKM}}^{\func{pred}} 
=
\begin{pmatrix}
0.97409 & 0.22602 & 0.00799 & -6.38471 \times 10^{-6} & -0.00036 \\[0.5ex]
0.22615 & -0.97372 & -0.02697 & 3.80102 \times 10^{-5} & 0.00147 \\[0.5ex]
0.00166 & 0.02815 & -0.99880 & -0.00766 & 0.00874 \\[0.5ex]
0.00003 & 0.00019 & -0.00916 & 0.99919 & -0.01773 \\[0.5ex]
-0.00057 & 0.00112 & 0.03812 & 0.03539 & -0.00096 
\label{eqn:CKM_bestfit}
\end{pmatrix}
\end{equation} 
where a feature we should remember is the left-handed down-quark sector are able to reach to all mixings among the three SM generations, whereas the only left-handed $23$ mixing is allowed for the up-quark sector in this BSM model. The experimental CKM mixing matrix without unitarity is given in Equation~\ref{eqn:CKM_exp_wx_uni}.
\begingroup
\setlength\arraycolsep{5pt}
\begin{equation}
\lvert K_{\func{CKM}} \rvert
=
\begin{pmatrix}
0.97370 \pm 0.00014 & 0.22450 \pm 0.00080 & 0.00382 \pm 0.00024 \\[1.5ex]
0.22100 \pm 0.00400 & 0.98700 \pm 0.01100 & 0.04100 \pm 0.00140 \\[1.5ex]
0.00800 \pm 0.00030 & 0.03880 \pm 0.00110 & 1.01300 \pm 0.03000 
\end{pmatrix}
\label{eqn:CKM_exp_wx_uni}
\end{equation}
\endgroup
Restricting our attention to the upper-left $3\times 3$ block of Equation~\ref{eqn:CKM_bestfit}, it can be compared to its experimental bound given in Equation~\ref{eqn:CKM_exp_wx_uni}. In order to confirm that our prediction for the CKM mixing matrix is consistent with the experimental data, 
it requires for the upper-left $3 \times 3$ block of Equation~\ref{eqn:CKM_bestfit} to be inside the $3\sigma$ experimentally allowed range  
as follows:
\begin{equation}
\lvert (K_{\func{CKM}})_{ij} \rvert - 3\lvert (\delta K_{\func{CKM}})_{ij} \rvert < \lvert (V_{\func{CKM}}^{\func{pred}})_{ij} \rvert < \lvert (K_{\func{CKM}})_{ij} \rvert + 3\lvert (\delta K_{\func{CKM}} )_{ij}, \rvert, \quad \text{for $i,j=1,2,3$}
\end{equation}
and we confirm that the $13, 23, 31, 32$ elements in the CKM prediction of Equation~\ref{eqn:CKM_bestfit} cannot be fitted within the $3\sigma$ range with a small difference. 
From our numerical analysis we find that in our model the  
CKM quark mixing matrix mainly arises from the down type quark sector and has a subleading correction coming from the up type quark sector.
It is worth mentioning that the inclusion of an additional
vector-like family in our proposed model to provide masses for the 
first generation of SM charged fermions 
will lead to an improvement of our predictions related to the CKM quark mixing matrix.
However, that approach of having a fifth vector-like fermion family goes beyond the 
scope in this work and is deferred for a future publication.  
Furthermore, in this section we also discuss the possible deviation of the first row of the CKM mixing matrix without unitarity and this study is also covered in \Steve{references~\cite{Branco:2021vhs,Belfatto:2021jhf}} with an isosinglet vector-like quark in a model different than the one considered in this paper. According to 
\Steve{\cite{Branco:2021vhs,Belfatto:2021jhf}}, 
the deviation $\Delta$ of unitarity is defined as follows:
\begin{equation}
\Delta = 1 - \lvert V_{ud} \rvert^2 - \lvert V_{us} \rvert^2 - \lvert V_{ub} \rvert^2,
\end{equation}
and its experimental value is given by 
\cite{Belfatto:2019swo}.
\begin{equation}
\sqrt{\Delta} \sim 0.04
\label{eqn:dev_exp}
\end{equation}
Calculating the deviation of unitarity $\Delta$ from the best fitted CKM prediction of Equation~\ref{eqn:CKM_bestfit}, the result is
\begin{equation}
\begin{split}
\sqrt{\Delta} &\simeq 0.00035 
\end{split}
\end{equation}
Therefore, the deviation of unitarity derived from \Antonio{the model under consideration} 
is too small to be observed compared to its experimental bound given in Equation~\ref{eqn:dev_exp}. Lastly, we discuss the rare $t \rightarrow c Z$ decay and collect all benchmark points satisfying $\chi_{\func{CKM}}^{2} < 980$ (notice that the most converged point reports $\chi^2_{\func{CKM}} = 956.828$).
\begin{figure}[H]
\centering
\begin{subfigure}{0.48\textwidth}
\includegraphics[keepaspectratio,width=0.9\textwidth]{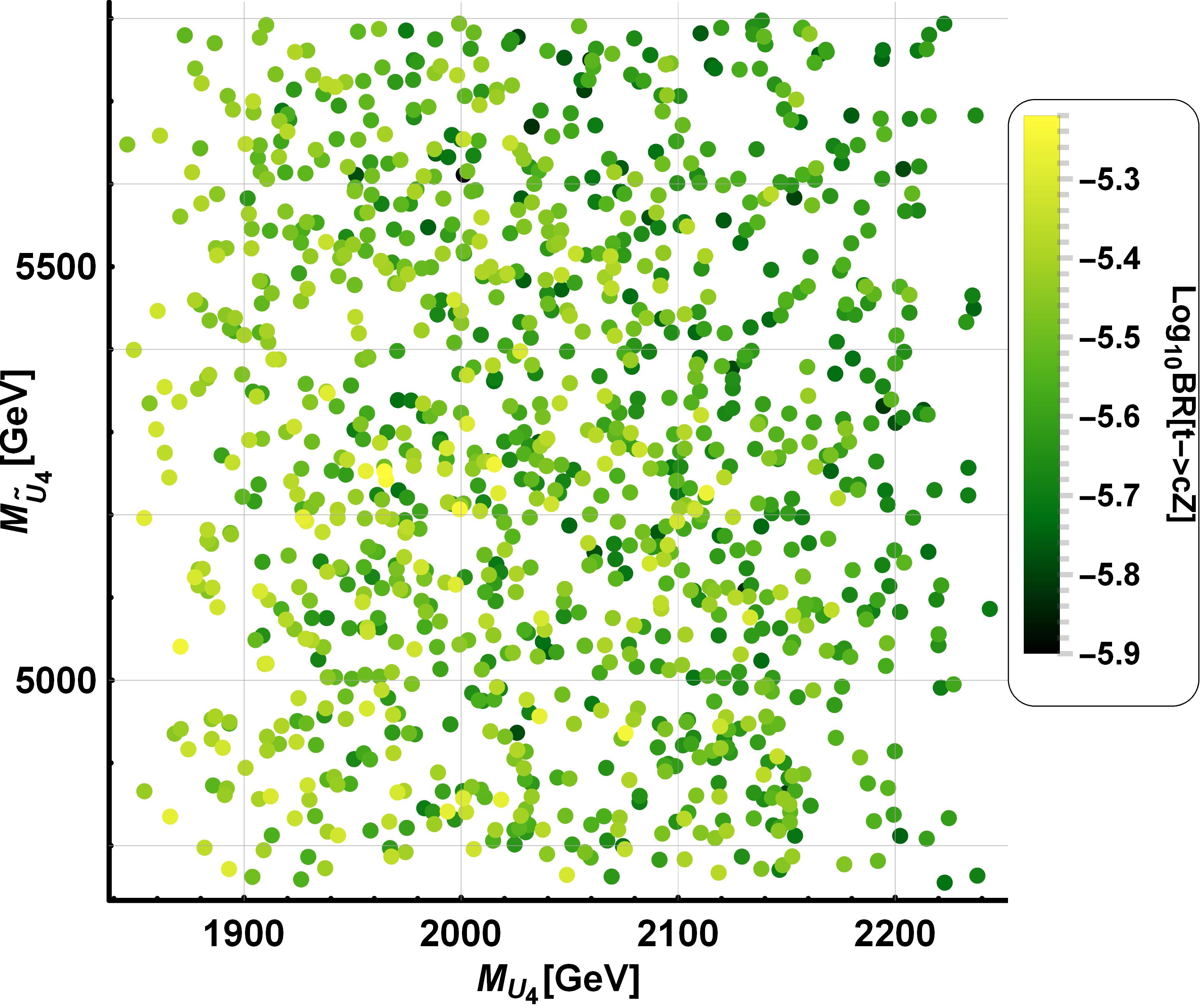}
\end{subfigure}
\hspace{0.1cm}
\begin{subfigure}{0.48\textwidth}
\includegraphics[keepaspectratio,width=0.9\textwidth]{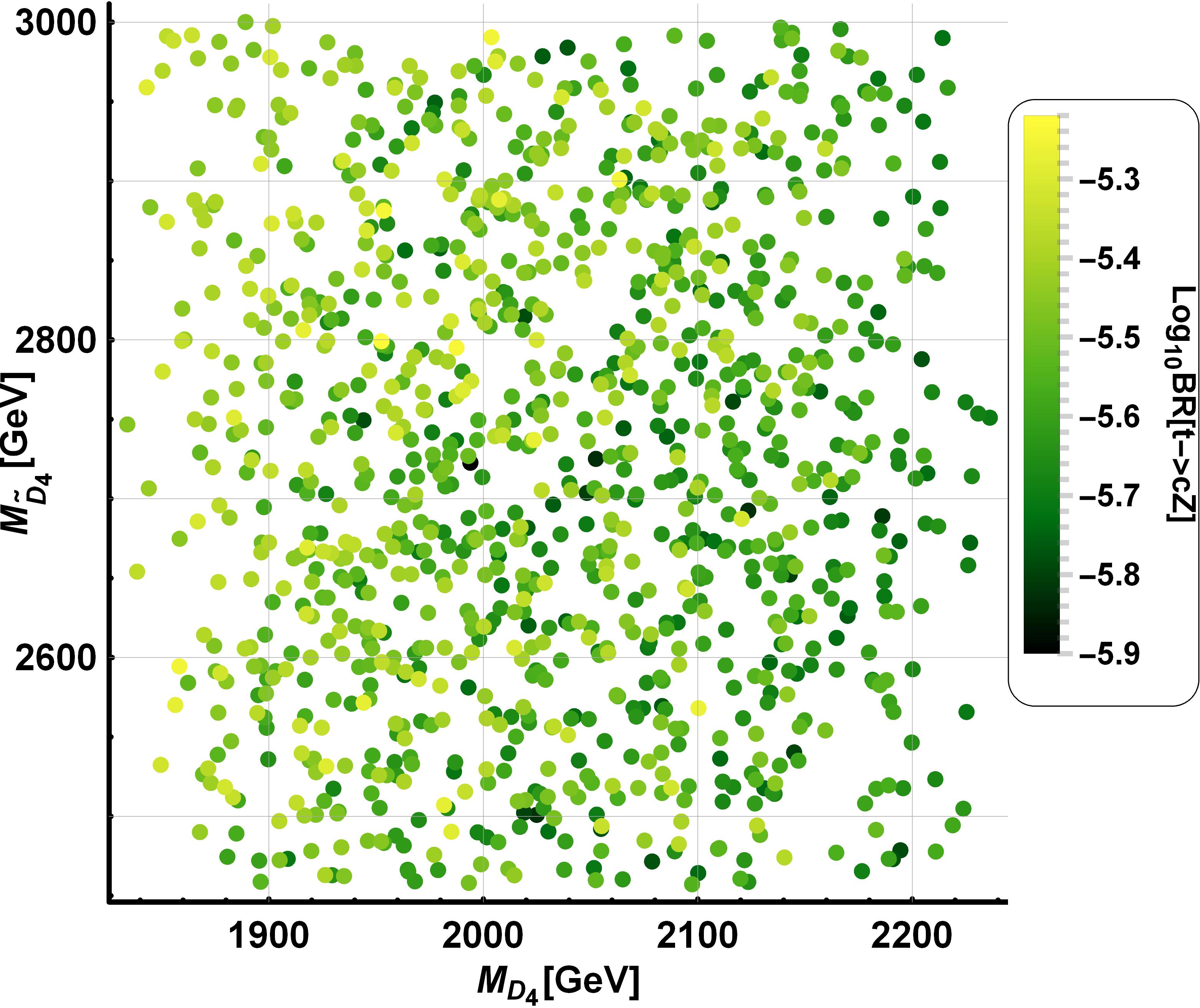}
\end{subfigure}
\caption{Scanned mass region of the vector-like quarks and contributions of the flavor violating interactions with the SM $Z$ gauge boson to the rare  $t \rightarrow c Z$ decay. The used constraints are the predicted $c,s,b$ and $t$ quark mass to be put between $[1 \pm 0.1] \times m_{c,s,b}$ and $[1 \pm 0.01] \times m_{t}$ and the CKM mixing matrix.}
\label{fig:tcZ_decay}
\end{figure}
Figure~\ref{fig:tcZ_decay} displays the allowed values of vector-like quark masses consistent with the constraints arising from the rare  $t \rightarrow c Z$ decay. Our obtained values for the vector-like quark masses are consistent with their lower experimental bound of $1000\func{GeV}$ arising from collider searches. In our numerical analysis the vector-like doublet up-type quark mass $M_{U_4}$ is ranged from $1850\func{GeV}$ up to about 
$2250\func{GeV}$ and the vector-like singlet up-type quark mass $M_{\widetilde{U}_4}$ is varied from $4750\func{GeV}$ to $5800\func{GeV}$. Regarding the exotic down type quark sector, we have varied  
the vector-like doublet down-type quark mass $M_{D_4}$ 
from $1850\func{GeV}$ up to $2250\func{GeV}$ and the vector-like singlet down-type quark mass $M_{\widetilde{D}_4}$ 
from $2450\func{GeV}$ up to $3000\func{GeV}$. As seen from Figure~\ref{fig:tcZ_decay} the order of magnitude of the obtained values for the branching ratio of the rare $t \rightarrow c Z$ decay range from $10^{-6}$ up to $10^{-5}$, which is consistent with 
its current experimental bound, whose logarithmic value is about $-3.6$ as indicated by 
Equation~\ref{eqn:BRtcZ_exp}.

\section{CONCLUSION} \label{sec:VII}
In this work we have considered a model where the SM fermion sector is extended by the inclusion of a 
fourth vector-like family and the scalar sector is augmented by the incorporation of an extra scalar doublet and a gauge singlet scalar. In addition, we have assumed a global $U(1)^{\prime}$ symmetry under which all particles are charged except the SM chiral quark and lepton fields. 
The model explains the hierarchical structure of the SM quark and lepton masses 
by assuming that the SM Yukawa interactions are forbidden by the $U(1)^{\prime}$ symmetry and arise effectively after it is spontaneously broken, due to 
induced mixing with the fourth vector-like family. This mixing also results in non-standard couplings of the $W$ and $Z$ gauge bosons which have been studied here for the first time.

This setup leads to sizeable \Antonio{branching fractions for the FCNC decays such as} $\mu \rightarrow e \gamma$, $Z\to\mu\tau$ and $t\rightarrow cZ$, within the reach of the future experimental sensitivity. These FCNC decays are studied in detail in this work, in order to set constraints on the model parameter space. 
A great advantage of the approach taken in this work with respect to the ones considered in extensions of the SM having a $Z^{\prime}$ gauge boson 
is that it makes the study of the FCNC observables simpler than in the latter since in the former \Antonio{we can avoid assuming specific values for the unknown $U(1)^{\prime}$ coupling and  
$Z^{\prime}$ gauge boson mass.} 
This makes the present phenomenology based on $W$ and $Z$ gauge boson couplings
more predictive than if the $U(1)^{\prime}$ were a spontaneously broken gauge symmetry, leading to a massive $Z'$.

Given that the hierarchical structure of the SM is implemented in our proposed model, the extended mass matrices for the charged lepton and quark sectors need to be completely and accurately diagonalised, as 
the starting point of our analytical and numerical analysis.
Since we only consider a fourth vector-like family, the model
cannot provide masses for the first generation of SM charged fermions, 
nevertheless this is a good approximation given that the first generation of the SM fermions are very light. For this reason, we mainly focus on the study of FCNC observables 
involving the second and third generations of SM fermions in both quark and lepton sectors. 

In the chosen convenient basis, the different shape of the down-type quark mass matrix allows all left-handed mixings between the three SM generations, whereas the up type quark sector can have only the $23$ left-handed mixing, while all quarks and charged lepton have the only $23$ right-handed mixing, and we have checked that the results are basis independent.
This feature implies that we can obtain
 a prediction for the CKM mixing matrix and this is one of main phenomenological aspects analyzed in this work.
 In order to diagonalize the fermionic mass matrices, in an analytic approximation,
 we have defined the $SU(2)$ conserving and $SU(2)$ violating mixings and we have shown that the $SU(2)$ violating mixing plays a crucial role for generating the \Antonio{$Z$ mediated flavor violating interactions. 
 Furthermore, the
 extension of the SM fermion sector by the inclusion of a vector-like family makes the matrices of $Z$ couplings with fermions 
 different than the identity matrix 
 due to the appearance of non-zero off-diagonal matrix elements of the $Z$ coupling matrices which will give rise to flavor violating $Z$ decays. The non-zero off-diagonal SM $Z$ gauge coupling constants are generally proportional to two of the small mixing angles, which are defined by the ratio between the SM fermion and vector-like masses, thus leading to small values.}
\\~
\Antonio{Defining all the required $Z$ gauge coupling constants with fermions in the mass basis, as discussed above, we began by 
analyzing the FCNC processes of the charged lepton sector.} 
We have found that in the lepton sector, 
the following three 
FCNC decays are allowed: 
$\tau \rightarrow \mu \gamma, \tau \rightarrow 3 \mu$ and $Z \rightarrow \mu \tau$. Regarding 
the $\tau \rightarrow \mu \gamma$ decay, we discussed its leading contribution, which  
arises from the Feynman diagrams having a chirality flip in the internal fermionic lines and being proportional to 
$M/m_{\tau}$, where $M$ is the mass scale of the heavy charged vector-like leptons. However, the dominant terms cannot be as big as the vector-like masses get heavier since their coupling constants get suppressed at the same time, thus providing a 
balanced relation between the vector-like masses and their coupling constants. We have found that 
our predictions for the vector-like charged lepton masses are not severely constrained by the $\tau \rightarrow \mu \gamma$ decay since most of the obtained values for the $\tau \rightarrow \mu \gamma$ decay are consistent with its experimental upper bound. In the concerning to the  
$\tau \rightarrow 3\mu$ and $Z \rightarrow \mu \tau$ decays, we have derived an analytic expression for their corresponding rates at tree-level finding that  
none of  
our predictions is constrained by the experimental bounds of these decays. Considering the FCC-ee experiment which have planned to generate $10^{12}$ the $Z$ gauge bosons and our numerical prediction for the $Z \rightarrow \mu \tau$ branching ratio is of the  
order of $10^{-9}$ at most, thus implying that our model can be tested at the $Z$ factory via the $Z \rightarrow \mu \tau$ decay. 
 However, the CMS provided that the doublet vector-like mass can be constrained up to $790\func{GeV}$~\cite{Hernandez:2021tii,Xu:2018pnq} and our numerical predictions for the vector-like charged leptons are severely constrained by the CMS result. Therefore, we can expect that the vector-like charged lepton doublet mass is ranged from $790$ to nearly $1600 \func{GeV}$, whereas the vector-like charged lepton singlet mass is ranged from $500$ to $2000 \func{GeV}$ or above than that. 

Turning to the quark sector phenomenology, we have analyzed the rare $t \rightarrow c Z$ decay as well as the CKM mixing 
to set constraints on the quark sector parameters. It is worth mentioning that the
neutral meson $K, B_{d}, B_{s}$ oscillations do not set constraints on the quark sector parameters of our model since their new physics effects are quite negligible compared to the SM expectation.  
We have derived analytic expressions for the rare $t \rightarrow c Z$ decay as well as for the CKM mixing matrix. \Antonio{Due to the mixings between SM fermions and vector-like fermions, the CKM quark mixing matrix is not unitary, thus implying that the unitarity requirement has to be relaxed} 
~\cite{Branco:2021vhs,ParticleDataGroup:2020ssz}. \Antonio{Using the most converged benchmark point, we showed how dominant the down-type quark mixing matrix plays a crucial role in the CKM mixing matrix and we discussed the deviation of unitarity arisen from the first row of the CKM mixing matrix, whose value is too small to be experimentally measured.
Finally, we investigate the branching ratio for the $t \rightarrow c Z$ decay and found that our numerical predictions are not excluded by its experimental bound, for 
 vector-like doublet up-type and down type quark masses $M_{U_4}$ and $M_{D_4}$ in the window \mbox{$1850$\text{GeV}$\leqslant M_{U_4},M_{D_4}\leqslant 2250\text{GeV}$} as well as 
 vector-like singlet up and down type quark masses $M_{\widetilde{U}_4}$ and $M_{\widetilde{D}_4}$ in the ranges \mbox{$4750$\text{GeV}$\leqslant M_{\widetilde{U}_4}\leqslant 5800\text{GeV}$}, 
and \mbox{$2450$\text{GeV}$\leqslant M_{\widetilde{D}_4}\leqslant 3000\text{GeV}$}, respectively.} 

In conclusion, we have analysed a range of FCNCs arising from non-standard $W$ and $Z$ gauge boson couplings in an extension of the SM with a fourth vector-like family, which can also address the hierarchy of quark and lepton masses, leading to several interesting rare decays which may be probed in future high luminosity experiments.

\section*{Acknowledgements}
This research has received funding from Fondecyt (Chile), Grant No.~1210378, ANID PIA/APOYO AFB180002 and Milenio-ANID-ICN2019\_044. SFK acknowledges the STFC Consolidated Grant ST/L000296/1 and the European Union's Horizon 2020 Research and Innovation programme under Marie Sk\l {}odowska-Curie grant agreement HIDDeN European ITN project (H2020-MSCA-ITN-2019//860881-HIDDeN).

\appendix

\section{Analytic approximated step-by-step diagonalization for the charged lepton sector} \label{app:A}
In order to diagonalize the mass matrix of Equation~\ref{eqn:cl_1} in an analytic way, we employ the method of 
mixing formalism and define intermediate mass basis.
The flavor basis is used when writing 
the initial mass matrix of Equation~\ref{eqn:cl_1}, whereas 
the true mass basis corresponds to 
 the fully diagonalized mass matrix, which reveals the masses of all propagating charged leptons. The intermediate mass basis is a basis where the heavy particles appearing in all terms generating the entries proportional to $v_{\phi}$ are integrated out remaining other terms unrotated. This separation makes the difference between $SU(2)$ conserving and $SU(2)$ violating mixings, which will be defined later, clear, which will become important when we consider the flavor violating interactions mediated by the SM $Z$ gauge boson. Before we carry out the digonalization step-by-step, it is convenient to rearrange the mass matrix of Equation~\ref{eqn:cl_1} by switching the Yukawa terms by mass parameters and by swapping the fourth and fifth column in order to make the heavy vector-like masses locate in the diagonal positions as given in Equation~\ref{eqn:cl_2s}
\begin{equation}
M^{e }
=
\left( 
\begin{array}{c|ccccc}
& e _{1R} & e _{2R} & e _{3R} & e _{4R} & \widetilde{L }_{4R} \\[0.5ex] \hline
\overline{L }_{1L} & 0 & 0 & 0 & 0 & 0 \\[1ex]
\overline{L }_{2L} & 0 & 0 & 0 & m_{24} & 0 \\[1ex] 
\overline{L }_{3L} & 0 & 0 & 0 & m_{34} & m_{35} \\[1ex]
\overline{L }_{4L} & 0 & 0 & m_{43} & 0 & M_{45}^{L } \\[1ex]
\overline{\widetilde{e }}_{4L} & 0 & m_{52} & m_{53} & M_{54}^{e} & 0 \\ 
\end{array}%
\right)
=
\left( 
\begin{array}{c|ccccc}
& e _{1R} & e _{2R} & e _{3R} & \widetilde{L }_{4R} & e _{4R} \\[0.5ex] \hline
\overline{L }_{1L} & 0 & 0 & 0 & 0 & 0 \\[1ex]
\overline{L }_{2L} & 0 & 0 & 0 & 0 & m_{24} \\[1ex] 
\overline{L }_{3L} & 0 & 0 & 0 & m_{35} & m_{34} \\[1ex]
\overline{L }_{4L} & 0 & 0 & m_{43} & M_{45}^{L } & 0 \\[1ex]
\overline{\widetilde{e }}_{4L} & 0 & m_{52} & m_{53} & 0 & M_{54}^{e} \\ 
\end{array}%
\right), 
\label{eqn:cl_2s}
\end{equation}
where the indices running from $1$ to $3$ correspond to the three SM families, the index $4$ labels the  
fourth vector-like particles 
and lastly the index $5$ denotes 
the tilde particles, which are a partner of the vector-like particles. Now we are ready to diagonalize the mass matrix of Equation~\ref{eqn:cl_2s} and the first step is to get the intermediate mass basis and to integrate out the particles generating the entries proportional to 
$v_{\phi}$ 
(equally, all mass terms involving index $5$). For this task, we first consider $34$ rotation in the left-handed fields to turn off the mass term $m_{35}$.
\begin{gather}
V_{34}^{L} M^{e }
=
\left( 
\begin{array}{c|ccccc}
& e _{1R} & e _{2R} & e _{3R} & \widetilde{L }_{4R} & e _{4R} \\[0.5ex] \hline
\overline{L }_{1L} & 0 & 0 & 0 & 0 & 0 \\[1ex]
\overline{L }_{2L} & 0 & 0 & 0 & 0 & m_{24} \\[1ex] 
\overline{L }_{3L}^{\prime} & 0 & 0 & -\frac{m_{35} m_{43}}{M_{45}^{L \prime}} & 0 & \frac{m_{34} M_{45}^{L}}{M_{45}^{L \prime}} \\[1ex]
\overline{L }_{4L}^{\prime} & 0 & 0 & \frac{m_{43} M_{45}^{L}}{M_{45}^{L \prime}} & M_{45}^{L \prime} & \frac{m_{34} m_{35}^{L}}{M_{45}^{L \prime}} \\[1ex]
\overline{\widetilde{e }}_{4L} & 0 & m_{52} & m_{53} & 0 & M_{54}^{e} \\ 
\end{array}%
\right), \\[2ex]
M_{45}^{L \prime} = \sqrt{M_{45}^{L 2} + m_{35}^2}, \quad s_{34}^{L} = \frac{m_{35}}{M_{45}^{L \prime}}, \quad c_{34}^{L} = \frac{M_{45}^{L}}{M_{45}^{L \prime}}, \quad V_{34}^{L} = \begin{pmatrix}
1 & 0 & 0 & 0 & 0 \\[1ex]
0 & 1 & 0 & 0 & 0 \\[1ex]
0 & 0 & c_{34}^{L} & -s_{34}^{L} & 0 \\[1ex]
0 & 0 & s_{34}^{L} & c_{34}^{L} & 0 \\[1ex]
0 & 0 & 0 & 0 & 1
\end{pmatrix},
\label{eqn:mm_cl_LH_34}
\end{gather}
where the primed fields correspond to the rotated fields. 
Throughout this whole work, the fields characterized by a capital letter are the ones belonging to a 
$SU(2)$ doublet under the SM gauge symmetry, whereas those ones denoted by a small letter are $SU(2)$ singlets. Then, the next rotation is $34$ rotation in the right-handed leptonic fields to turn off the $m_{53}$ entry. It is worth mentioning that 
the order used in the rotation of the left-handed fields is $12345$, whereas for the rotation of the right-handed fields the corresponding order is $12354$ since for consistency reasons we assigned the index $5$ for the tilde particle, 

\begin{gather}
V_{34}^{L} M^{e } (V_{34}^{e})^{\dagger}
=
\left( 
\begin{array}{c|ccccc}
& e _{1R} & e _{2R} & e _{3R}^{\prime} & \widetilde{L }_{4R} & e _{4R}^{\prime} \\[0.5ex] \hline
\overline{L }_{1L} & 0 & 0 & 0 & 0 & 0 \\[1ex]
\overline{L }_{2L} & 0 & 0 & -\frac{m_{24} m_{53}}{M_{54}^{e \prime}} & 0 & \frac{m_{24} M_{54}^{e}}{M_{54}^{e \prime}} \\[1ex] 
\overline{L }_{3L}^{\prime} & 0 & 0 & \frac{-m_{34} m_{53} M_{45}^{L} -m_{35} m_{43} M_{54}^{e}}{M_{45}^{L \prime} M_{54}^{e \prime}} & 0 & \frac{-m_{35} m_{43} m_{53} +m_{34} M_{45}^{L} M_{54}^{e}}{M_{45}^{L \prime} M_{54}^{e \prime}} \\[1ex]
\overline{L }_{4L}^{\prime} & 0 & 0 & \frac{-m_{34} m_{35} m_{53} +m_{43} M_{45}^{L} M_{54}^{e}}{M_{45}^{L \prime} M_{54}^{e \prime}} & M_{45}^{L \prime} & \frac{m_{43} m_{53} M_{45}^{L} +m_{34} m_{35} M_{54}^{e}}{M_{45}^{L \prime} M_{54}^{e \prime}} \\[1ex]
\overline{\widetilde{e }}_{4L} & 0 & m_{52} & 0 & 0 & M_{54}^{e \prime} \\ 
\end{array}%
\right), \\[2ex]
M_{54}^{e \prime} = \sqrt{M_{54}^{e 2} + m_{53}^2}, \quad s_{34}^{e} = \frac{m_{53}}{M_{54}^{e \prime}}, \quad c_{34}^{e} = \frac{M_{54}^{e}}{M_{54}^{e \prime}}, \quad V_{34}^{e} = \begin{pmatrix}
1 & 0 & 0 & 0 & 0 \\[1ex]
0 & 1 & 0 & 0 & 0 \\[1ex]
0 & 0 & c_{34}^{e} & 0 & -s_{34}^{e} \\[1ex]
0 & 0 & 0 & 1 & 0 \\[1ex]
0 & 0 & s_{34}^{e} & 0 & c_{34}^{e}
\end{pmatrix}.
\label{eqn:mm_cl_RH_34}
\end{gather}
The last step to arrive at the intermediate mass basis is the $24$ rotation in the right-handed fields.
\begin{gather}
V_{34}^{L} M^{e } (V_{34}^{e})^{\dagger} (V_{24}^{e})^{\dagger}
=
\\[2ex]
\left( 
\begin{array}{c|ccccc}
& e _{1R} & e _{2R}^{\prime} & e _{3R}^{\prime} & \widetilde{L }_{4R} & e _{4R}^{\prime \prime} \\[0.5ex] \hline
\overline{L }_{1L} & 0 & 0 & 0 & 0 & 0 \\[1ex]
\overline{L }_{2L} & 0 & -\frac{m_{24} m_{52} M_{54}^{e}}{M_{54}^{e \prime} M_{54}^{e \prime \prime}} & -\frac{m_{24} m_{53}}{M_{54}^{e \prime}} & 0 & \frac{m_{24} M_{54}^{e}}{M_{54}^{e \prime \prime}} \\[1ex] 
\overline{L }_{3L}^{\prime} & 0 & \frac{m_{52} (m_{35} m_{43} m_{53} -m_{34} M_{45}^{L} M_{54}^{e})}{M_{45}^{L \prime} M_{54}^{e \prime} M_{54}^{e \prime \prime}} & \frac{-m_{34} m_{53} M_{45}^{L} -m_{35} m_{43} M_{54}^{e}}{M_{45}^{L \prime} M_{54}^{e \prime}} & 0 & \frac{-m_{35} m_{43} m_{53} +m_{34} M_{45}^{L} M_{54}^{e}}{M_{45}^{L \prime} M_{54}^{e \prime \prime}} \\[1ex]
\overline{L }_{4L}^{\prime} & 0 & -\frac{m_{52} (m_{43} M_{45}^{L} m_{53} +m_{34} m_{35} M_{54}^{e})}{M_{45}^{L \prime} M_{54}^{e \prime} M_{54}^{e \prime \prime}} & \frac{-m_{34} m_{35} m_{53} +m_{43} M_{45}^{L} M_{54}^{e}}{M_{45}^{L \prime} M_{54}^{e \prime}} & M_{45}^{L \prime} & \frac{m_{43} m_{53} M_{45}^{L} +m_{34} m_{35} M_{54}^{e}}{M_{45}^{L \prime} M_{54}^{e \prime \prime}} \\[1ex]
\overline{\widetilde{e }}_{4L} & 0 & 0 & 0 & 0 & M_{54}^{e \prime \prime} \label{eqn:mm_intermediate} \\ 
\end{array}%
\right), \\[2ex]
M_{54}^{e \prime \prime} = \sqrt{M_{54}^{e \prime 2} + m_{52}^2}, \quad s_{24}^{e} = \frac{m_{52}}{M_{54}^{e \prime \prime}}, \quad c_{24}^{e} = \frac{M_{54}^{e \prime}}{M_{54}^{e \prime \prime}}, \quad V_{34}^{e} = \begin{pmatrix}
1 & 0 & 0 & 0 & 0 \\[1ex]
0 & c_{24}^{e} & 0 & 0 & -s_{24}^{e} \\[1ex]
0 & 0 & 1 & 0 & 0 \\[1ex]
0 & 0 & 0 & 1 & 0 \\[1ex]
0 & s_{24}^{e} & 0 & 0 & c_{24}^{e}
\end{pmatrix}.
\label{eqn:mm_cl_RH_24}
\end{gather}
The mass matrix given in Equation~\ref{eqn:mm_intermediate} is the intermediate mass basis and this diagonalization is exactly consistent with the one for the SM charged lepton sector in one of our works~\cite{Hernandez:2021tii}. When we diagonalized the charged lepton sector in \cite{Hernandez:2021tii}, we assumed all off-diagonal elements to be zero and this is actually a quite suitable assumption since the differences between the Yukawa induced mass terms and the vector-like masses are quite large. However, we consider all small mixings in order to get the fully diagonalized mass matrix in this work rather than setting 
them to zero, since we are interested in studying diverse FCNC constraints by scanning all possible and allowed mass ranges of the vector-like fermions in both SM quark and lepton sectors and the FCNC constraints are sensitive to the small mixings as it will be shown below.  
One more feature to be mentioned 
 in this diagonalization is that all the mixings have been made between 
  the same $SU(2)$ multiplets. In other words, the $SU(2)$ doublet left-handed fields are mixed with the another $SU(2)$ doublet left-handed fields, whereas the $SU(2)$ singlet right-handed fields are mixed with the another $SU(2)$ singlet right-handed fields, so we call this mixing ``$SU(2)$ conserving mixing". This $SU(2)$ conserving mixing can not cause the flavor violating interactions with the SM $Z$ gauge boson since they involve 
  an identity matrix resulting 
  from the SM $Z$ gauge interactions. Therefore, the next diagonalization process becomes especially important when we start exploring the FCNC constraints. Before we start the next diagonalization, it is convenient to reparameterize all elements of Equation~\ref{eqn:mm_intermediate} by a simpler one as given in Equation~\ref{eqn:mm_cl_rp}.
\begin{equation}
M^{e \prime} = V_{34}^{L} M^{e } (V_{34}^{e})^{\dagger} (V_{24}^{e})^{\dagger}
=
\left( 
\begin{array}{c|ccccc}
& e _{1R} & e _{2R}^{\prime} & e _{3R}^{\prime} & \widetilde{L }_{4R} & e _{4R}^{\prime \prime} \\[0.5ex] \hline
\overline{L }_{1L} & 0 & 0 & 0 & 0 & 0 \\[1ex]
\overline{L }_{2L} & 0 & m_{22}^{\prime} & m_{23}^{\prime} & 0 & m_{24}^{\prime} \\[1ex] 
\overline{L }_{3L}^{\prime} & 0 & m_{32}^{\prime} & m_{33}^{\prime} & 0 & m_{34}^{\prime} \\[1ex]
\overline{L }_{4L}^{\prime} & 0 & m_{42}^{\prime} & m_{43}^{\prime} & M_{45}^{L \prime} & m_{44}^{\prime} \\[1ex]
\overline{\widetilde{e }}_{4L} & 0 & 0 & 0 & 0 & M_{54}^{e \prime \prime} \\ 
\end{array}%
\right) 
\label{eqn:mm_cl_rp}
\end{equation}
We carry out first the $35$ rotation in the left-handed fields of mass matrix~\ref{eqn:mm_cl_rp} and this is a start of ``$SU(2)$ violating mixing". As already mentioned, since the difference between the Yukawa mass and vector-like mass is significantly sizeable, it is possible to simplify the mixing matrices in terms of the relevant small mixing angle $\theta_{35}^{L}$.
\begin{gather}
V_{35}^{L} M^{e \prime} 
=
\left( 
\begin{array}{c|ccccc}
& e _{1R} & e _{2R}^{\prime} & e _{3R}^{\prime} & \widetilde{L }_{4R} & e _{4R}^{\prime \prime} \\[0.5ex] \hline
\overline{L }_{1L} & 0 & 0 & 0 & 0 & 0 \\[1ex]
\overline{L }_{2L} & 0 & m_{22}^{\prime} & m_{23}^{\prime} & 0 & m_{24}^{\prime} \\[1ex] 
\overline{L }_{3L}^{\prime \prime} & 0 & m_{32}^{\prime} & m_{33}^{\prime} & 0 & m_{34}^{\prime} - M_{54}^{e \prime \prime} \theta_{35}^{L} \\[1ex]
\overline{L }_{4L}^{\prime} & 0 & m_{42}^{\prime} & m_{43}^{\prime} & M_{45}^{L \prime} & m_{44}^{\prime} \\[1ex]
\overline{\widetilde{e }}_{4L}^{\prime} & 0 & m_{32}^{\prime} \theta_{35}^{L} & m_{33}^{\prime} \theta_{35}^{L} & 0 & M_{54}^{e \prime \prime} + m_{34}^{\prime} \theta_{35}^{L} \\ 
\end{array}%
\right)
\approx
\left( 
\begin{array}{c|ccccc}
& e _{1R} & e _{2R}^{\prime} & e _{3R}^{\prime} & \widetilde{L }_{4R} & e _{4R}^{\prime \prime} \\[0.5ex] \hline
\overline{L }_{1L} & 0 & 0 & 0 & 0 & 0 \\[1ex]
\overline{L }_{2L} & 0 & m_{22}^{\prime} & m_{23}^{\prime} & 0 & m_{24}^{\prime} \\[1ex] 
\overline{L }_{3L}^{\prime \prime} & 0 & m_{32}^{\prime} & m_{33}^{\prime} & 0 & 0 \\[1ex]
\overline{L }_{4L}^{\prime} & 0 & m_{42}^{\prime} & m_{43}^{\prime} & M_{45}^{L \prime} & m_{44}^{\prime} \\[1ex]
\overline{\widetilde{e }}_{4L}^{\prime} & 0 & 0 & 0 & 0 & M_{54}^{e \prime \prime} \\ 
\end{array}%
\right)
\\[2ex]
\theta_{35}^{L} = \frac{m_{34}^{\prime}}{M_{54}^{e \prime \prime}}, \quad
V_{35}^{L} = \begin{pmatrix}
1 & 0 & 0 & 0 & 0 \\[1ex]
0 & 1 & 0 & 0 & 0 \\[1ex]
0 & 0 & 1 & 0 & -\theta_{35}^{L} \\[1ex]
0 & 0 & 0 & 1 & 0 \\[1ex]
0 & 0 & \theta_{35}^{L} & 0 & 1 
\end{pmatrix}
\end{gather}
The next step is the $25$ rotation in the left-handed fields to turn off the mass term $m_{24}^{\prime}$.
\begin{gather}
V_{25}^{L} V_{35}^{L} M^{e \prime}
=
\left( 
\begin{array}{c|ccccc}
& e _{1R} & e _{2R}^{\prime} & e _{3R}^{\prime} & \widetilde{L }_{4R} & e _{4R}^{\prime \prime} \\[0.5ex] \hline
\overline{L }_{1L} & 0 & 0 & 0 & 0 & 0 \\[1ex]
\overline{L }_{2L}^{\prime} & 0 & m_{22}^{\prime} & m_{23}^{\prime} & 0 & m_{24}^{\prime} - M_{54}^{e \prime \prime} \theta_{25}^{L} \\[1ex] 
\overline{L }_{3L}^{\prime \prime} & 0 & m_{32}^{\prime} & m_{33}^{\prime} & 0 & 0 \\[1ex]
\overline{L }_{4L}^{\prime} & 0 & m_{42}^{\prime} & m_{43}^{\prime} & M_{45}^{L \prime} & m_{44}^{\prime} \\[1ex]
\overline{\widetilde{e }}_{4L}^{\prime \prime} & 0 & m_{22}^{\prime} \theta_{25}^{L} & m_{23}^{\prime} \theta_{25}^{L} & 0 & M_{54}^{e \prime \prime} + m_{24}^{\prime} \theta_{25}^{L} \\ 
\end{array}%
\right)
\approx
\left( 
\begin{array}{c|ccccc}
& e _{1R} & e _{2R}^{\prime} & e _{3R}^{\prime} & \widetilde{L }_{4R} & e _{4R}^{\prime \prime} \\[0.5ex] \hline
\overline{L }_{1L} & 0 & 0 & 0 & 0 & 0 \\[1ex]
\overline{L }_{2L}^{\prime} & 0 & m_{22}^{\prime} & m_{23}^{\prime} & 0 & 0 \\[1ex] 
\overline{L }_{3L}^{\prime \prime} & 0 & m_{32}^{\prime} & m_{33}^{\prime} & 0 & 0 \\[1ex]
\overline{L }_{4L}^{\prime} & 0 & m_{42}^{\prime} & m_{43}^{\prime} & M_{45}^{L \prime} & m_{44}^{\prime} \\[1ex]
\overline{\widetilde{e }}_{4L}^{\prime \prime} & 0 & 0 & 0 & 0 & M_{54}^{e \prime \prime} \\ 
\end{array}%
\right)
\\[2ex]
\theta_{25}^{L} = \frac{m_{24}^{\prime}}{M_{54}^{e \prime \prime}}, \quad
V_{25}^{L} = \begin{pmatrix}
1 & 0 & 0 & 0 & 0 \\[1ex]
0 & 1 & 0 & 0 & -\theta_{25}^{L} \\[1ex]
0 & 0 & 1 & 0 & 0 \\[1ex]
0 & 0 & 0 & 1 & 0 \\[1ex]
0 & \theta_{25}^{L} & 0 & 0 & 1 
\end{pmatrix}
\end{gather}
The next is the $35$ rotation in the right-handed fields to turn off the mass term $m_{42}^{\prime}$.
\begin{gather}
V_{25}^{L} V_{35}^{L} M^{e \prime} (V_{35}^{e})^{\dagger} 
=
\left( 
\begin{array}{c|ccccc}
& e _{1R} & e _{2R}^{\prime} & e _{3R}^{\prime \prime} & \widetilde{L }_{4R}^{\prime} & e _{4R}^{\prime \prime} \\[0.5ex] \hline
\overline{L }_{1L} & 0 & 0 & 0 & 0 & 0 \\[1ex]
\overline{L }_{2L}^{\prime} & 0 & m_{22}^{\prime} & m_{23}^{\prime} & m_{23}^{\prime} \theta_{35}^{e} & 0 \\[1ex] 
\overline{L }_{3L}^{\prime \prime} & 0 & m_{32}^{\prime} & m_{33}^{\prime} & m_{33}^{\prime} \theta_{35}^{e} & 0 \\[1ex]
\overline{L }_{4L}^{\prime} & 0 & m_{42}^{\prime} & m_{43}^{\prime} - M_{45}^{L \prime} \theta_{35}^{e} & M_{45}^{L \prime} + m_{43}^{\prime} \theta_{35}^{e} & m_{44}^{\prime} \\[1ex]
\overline{\widetilde{e }}_{4L}^{\prime \prime} & 0 & 0 & 0 & 0 & M_{54}^{e \prime \prime} \\ 
\end{array}%
\right)
\approx
\left( 
\begin{array}{c|ccccc}
& e _{1R} & e _{2R}^{\prime} & e _{3R}^{\prime \prime} & \widetilde{L }_{4R}^{\prime} & e _{4R}^{\prime \prime} \\[0.5ex] \hline
\overline{L }_{1L} & 0 & 0 & 0 & 0 & 0 \\[1ex]
\overline{L }_{2L}^{\prime} & 0 & m_{22}^{\prime} & m_{23}^{\prime} & 0 & 0 \\[1ex] 
\overline{L }_{3L}^{\prime \prime} & 0 & m_{32}^{\prime} & m_{33}^{\prime} & 0 & 0 \\[1ex]
\overline{L }_{4L}^{\prime} & 0 & m_{42}^{\prime} & 0 & M_{45}^{L \prime} & m_{44}^{\prime} \\[1ex]
\overline{\widetilde{e }}_{4L}^{\prime \prime} & 0 & 0 & 0 & 0 & M_{54}^{e \prime \prime} \\ 
\end{array}%
\right)
\\[2ex]
\theta_{35}^{e} = \frac{m_{43}^{\prime}}{M_{45}^{L \prime}}, \quad
V_{35}^{e} = \begin{pmatrix}
1 & 0 & 0 & 0 & 0 \\[1ex]
0 & 1 & 0 & 0 & 0 \\[1ex]
0 & 0 & 1 & -\theta_{35}^{e} & 0 \\[1ex]
0 & 0 & \theta_{35}^{e} & 1 & 0 \\[1ex]
0 & 0 & 0 & 0 & 1 
\end{pmatrix}
\end{gather}
After performing 
the right-handed $25$ rotation, we have a mass matrix, whose form is block diagonal.
\begin{gather}
V_{25}^{L} V_{35}^{L} M^{e \prime} (V_{35}^{e})^{\dagger} (V_{25}^{e})^{\dagger} =
\\[2ex]
\left( 
\begin{array}{c|ccccc}
& e _{1R} & e _{2R}^{\prime \prime} & e _{3R}^{\prime \prime} & \widetilde{L }_{4R}^{\prime \prime} & e _{4R}^{\prime \prime} \\[0.5ex] \hline
\overline{L }_{1L} & 0 & 0 & 0 & 0 & 0 \\[1ex]
\overline{L }_{2L}^{\prime} & 0 & m_{22}^{\prime} & m_{23}^{\prime} & m_{22}^{\prime} \theta_{25}^{e} & 0 \\[1ex] 
\overline{L }_{3L}^{\prime \prime} & 0 & m_{32}^{\prime} & m_{33}^{\prime} & m_{32}^{\prime} \theta_{25}^{e} & 0 \\[1ex]
\overline{L }_{4L}^{\prime} & 0 & m_{42}^{\prime} - M_{45}^{L \prime} \theta_{25}^{e} & 0 & M_{45}^{L \prime} + m_{42}^{\prime} \theta_{25}^{e} & m_{44}^{\prime} \\[1ex]
\overline{\widetilde{e }}_{4L}^{\prime \prime} & 0 & 0 & 0 & 0 & M_{54}^{e \prime \prime} \\ 
\end{array}%
\right)
\approx
\left( 
\begin{array}{c|ccccc}
& e _{1R} & e _{2R}^{\prime \prime} & e _{3R}^{\prime \prime} & \widetilde{L }_{4R}^{\prime \prime} & e _{4R}^{\prime \prime} \\[0.5ex] \hline
\overline{L }_{1L} & 0 & 0 & 0 & 0 & 0 \\[1ex]
\overline{L }_{2L}^{\prime} & 0 & m_{22}^{\prime} & m_{23}^{\prime} & 0 & 0 \\[1ex] 
\overline{L }_{3L}^{\prime \prime} & 0 & m_{32}^{\prime} & m_{33}^{\prime} & 0 & 0 \\[1ex]
\overline{L }_{4L}^{\prime} & 0 & 0 & 0 & M_{45}^{L \prime} & m_{44}^{\prime} \\[1ex]
\overline{\widetilde{e }}_{4L}^{\prime \prime} & 0 & 0 & 0 & 0 & M_{54}^{e \prime \prime} \\ 
\end{array}%
\right)
\\[2ex]
\theta_{25}^{e} = \frac{m_{42}^{\prime}}{M_{45}^{L \prime}}, \quad
V_{25}^{e} = \begin{pmatrix}
1 & 0 & 0 & 0 & 0 \\[1ex]
0 & 1 & 0 & -\theta_{25}^{e} & 0 \\[1ex]
0 & 0 & 1 & 0 & 0 \\[1ex]
0 & \theta_{25}^{e} & 0 & 1 & 0 \\[1ex]
0 & 0 & 0 & 0 & 1 
\end{pmatrix}
\end{gather}
We arrive at the fully diagonalized mass matrix by diagonalizing the upper-left $3 \times 3$ block as well as the lower-right $2 \times 2$ block as shown below in Equation~\ref{eqn:mass_pro_cl}.
\begin{equation}
\begin{split}
&V_{45}^{L} V_{23}^{L} V_{35}^{L} V_{25}^{L} M^{e \prime} (V_{35}^{e})^{\dagger} (V_{25}^{e})^{\dagger} (V_{23}^{e})^{\dagger} (V_{54}^{e})^{\dagger} = \func{diag}\left( 0, m_{\mu}, m_{\tau}, M_{E_4}, M_{\widetilde{E}_4} \right) \\
&V_{45}^{L} V_{23}^{L} V_{35}^{L} V_{25}^{L} V_{34}^{L} M^{e} (V_{34}^{e})^{\dagger} (V_{24}^{e})^{\dagger} (V_{35}^{e})^{\dagger} (V_{25}^{e})^{\dagger} (V_{23}^{e})^{\dagger} (V_{54}^{e})^{\dagger} = \func{diag}\left( 0, m_{\mu}, m_{\tau}, M_{E_4}, M_{\widetilde{E}_4} \right) 
\label{eqn:mass_pro_cl}
\end{split}
\end{equation}
As mentioned in the introduction, the SM charged lepton belonging to the first family, namely, the electron does not acquire a mass 
with one vector-like family as seen in Equation~\ref{eqn:mass_pro_cl}. This is due to the fact that the model under consideration has two leptonic seesaw mediators, which provide tree-level masses to the muon and tau leptons. It is worth mentioning that the number of seesaw mediators has to be larger or equal than the number of SM fermion families in order to provide masses to the SM fermions. A non vanishing electron mass can be generated by introducing one extra vector-like family as done in the reference~\cite{Hernandez:2021tii}. Then, we can easily confirm how the SM charged leptons in the flavor basis are connected with those ones in the mass basis via the following unitary mixing matrices.
\begin{equation}
\begin{split}
\begin{pmatrix}
e_{L} \\[0.5ex]
\mu_{L} \\[0.5ex]
\tau_{L} \\[0.5ex]
E_{4L} \\[0.5ex]
\widetilde{E}_{4L}
\end{pmatrix}
&=
\begin{pmatrix}
e_{1L} \\[0.5ex]
e_{2L}^{\prime} \\[0.5ex]
e_{3L}^{\prime \prime} \\[0.5ex]
e_{4L}^{\prime} \\[0.5ex]
\widetilde{e}_{4L}^{\prime \prime}
\end{pmatrix}
=
V^L  
\begin{pmatrix}
e_{1L} \\[0.5ex]
e_{2L} \\[0.5ex]
e_{3L} \\[0.5ex]
e_{4L} \\[0.5ex]
\widetilde{e}_{4L}
\end{pmatrix}
=
V_{45}^L V_{23}^L V_{25}^L V_{35}^L V_{34}^L  
\begin{pmatrix}
e_{1L} \\[0.5ex]
e_{2L} \\[0.5ex]
e_{3L} \\[0.5ex]
e_{4L} \\[0.5ex]
\widetilde{e}_{4L}
\end{pmatrix}
,
\\
\begin{pmatrix}
e_{R} \\[0.5ex]
\mu_{R} \\[0.5ex]
\tau_{R} \\[0.5ex]
\widetilde{E}_{4R} \\[0.5ex]
E_{4R}
\end{pmatrix}
&=
\begin{pmatrix}
e_{1R} \\[0.5ex]
e_{2R}^{\prime \prime} \\[0.5ex]
e_{3R}^{\prime \prime} \\[0.5ex]
\widetilde{e}_{4R}^{\prime \prime} \\[0.5ex]
e_{4R}^{\prime \prime}
\end{pmatrix}
=
V^e
\begin{pmatrix}
e_{1R} \\[0.5ex]
e_{2R} \\[0.5ex]
e_{3R} \\[0.5ex]
\widetilde{e}_{4R} \\[0.5ex]
e_{4R}
\end{pmatrix}
=
V_{54}^e V_{23}^e V_{25}^e V_{35}^e V_{24}^e V_{34}^e
\begin{pmatrix}
e_{1R} \\[0.5ex]
e_{2R} \\[0.5ex]
e_{3R} \\[0.5ex]
\widetilde{e}_{4R} \\[0.5ex]
e_{4R}
\end{pmatrix}.
\label{eqn:cl_umm_LH_RH}
\end{split}
\end{equation}
The left-handed $34$ mixing $V_{34}^{L}$ and right-handed $24,34$ mixings $V_{24,34}^{e}$ are the $SU(2)$ conserving mixings, whereas the left-handed $25,35$ mixings $V_{25,35}^{L}$ and right-handed $25,35$ mixings $V_{25,35}^{e}$ are the $SU(2)$ violating mixings. We will see that these $SU(2)$ violating mixings play a crucial role in generating the renormalizable flavor violating mixings mediated by the SM $Z$ gauge boson in section~\ref{sec:IV}. \Huchan{\Antonio{It is worth mentioning that 
this step-by-step diagonalization is 
a quite good approximation to the corresponding 
numerical diagonalization carried out by the singular value decomposition (SVD) method since the former yields similar results to the ones obtained from the latter, with very small differences due to the fact that 
all off-diagonal elements resulting from the step-by-step diagonalization are quite negligible and thus they can be approximated to zero, as discussed in detail  
 in Appendix~\ref{app:D}.}}
\section{Analytic approximated step-by-step diagonalization for the up-quark sector} \label{app:B}
The initial mass matrix for the up-type quark sector in the flavor basis is given by:
\begin{equation}
M^{u }=\left( 
\begin{array}{c|ccccc}
& u _{1R} & u _{2R} & u _{3R} & u _{4R} & \widetilde{Q }_{4R} \\[0.5ex] \hline
\overline{Q }_{1L} & 0 & 0 & 0 & 0 & 0 \\[1ex]
\overline{Q }_{2L} & 0 & 0 & 0 & y_{24}^{u } v_{u} & 0 \\[1ex] 
\overline{Q }_{3L} & 0 & 0 & 0 & y_{34}^{u } v_{u} & x_{34}^{Q } v_{\phi} \\[1ex]
\overline{Q }_{4L} & 0 & 0 & y_{43}^{u } v_{u} & 0 & M_{44}^{Q } \\[1ex]
\overline{\widetilde{u }}_{4L} & 0 & x_{42}^{u} v_{\phi}
& x_{43}^{u} v_{\phi} & M_{44}^{u} & 0 \\ 
\end{array}%
\right)
=
\left( 
\begin{array}{c|ccccc}
& u _{1R} & u _{2R} & u _{3R} & \widetilde{Q }_{4R} & u _{4R} \\[0.5ex] \hline
\overline{Q }_{1L} & 0 & 0 & 0 & 0 & 0 \\[1ex]
\overline{Q }_{2L} & 0 & 0 & 0 & 0 & m_{24}^{u} \\[1ex] 
\overline{Q }_{3L} & 0 & 0 & 0 & m_{35}^{u} & m_{34}^{u} \\[1ex]
\overline{Q }_{4L} & 0 & 0 & m_{43}^{u} & M_{44}^{Q } & 0 \\[1ex]
\overline{\widetilde{u }}_{4L} & 0 & m_{52}^{u}
& m_{53}^{u} & 0 & M_{44}^{u} \\ 
\end{array}%
\right), \label{eqn:uq_1s}
\end{equation}
The mass matrix of Equation~\ref{eqn:uq_1s} in the flavor basis is exactly consistent with the one corresponding to the charged lepton sector excepting for a few substitutions $y^{e} \rightarrow y^{u}$, $v_{d} \rightarrow v_{u}$, $x^{L} \rightarrow x^{Q}$ and $x^{e} \rightarrow x^{u}$. However, these substitutions do not change the whole structure of the mass matrix, so we do not need to derive all the required mixings from the initial mass matrix, instead the given mixings in the charged lepton sector can be reused as follows (For the charged lepton sector, it is enough to notice the symbol $L$ means left-handed doublet and $e$ means right-handed singlet. However, it becomes complicated in the quark sector since the mass matrices in the up- and down-type quark have a different form, so we change the mixing notation by $V_{L(R)}^{u,d}$ instead of 
$V^{Q}$.):
\begin{equation}
\begin{split}
\begin{pmatrix}
u_{L} \\[0.5ex]
c_{L} \\[0.5ex]
t_{L} \\[0.5ex]
U_{4L} \\[0.5ex]
\widetilde{U}_{4L}
\end{pmatrix}
=
\begin{pmatrix}
u_{1L} \\[0.5ex]
u_{2L}^{\prime} \\[0.5ex]
u_{3L}^{\prime \prime} \\[0.5ex]
u_{4L}^{\prime} \\[0.5ex]
\widetilde{u}_{4L}^{\prime \prime}
\end{pmatrix}
&=
V_{L}^{u}  
\begin{pmatrix}
u_{1L} \\[0.5ex]
u_{2L} \\[0.5ex]
u_{3L} \\[0.5ex]
u_{4L} \\[0.5ex]
\widetilde{u}_{4L}
\end{pmatrix}
=
(V_{L}^{u})_{45} (V_{L}^{u})_{23} (V_{L}^{u})_{25} (V_{L}^{u})_{35} (V_{L}^{u})_{34}  
\begin{pmatrix}
u_{1L} \\[0.5ex]
u_{2L} \\[0.5ex]
u_{3L} \\[0.5ex]
u_{4L} \\[0.5ex]
\widetilde{u}_{4L}
\end{pmatrix}
,
\\
\begin{pmatrix}
u_{R} \\[0.5ex]
c_{R} \\[0.5ex]
t_{R} \\[0.5ex]
\widetilde{U}_{4R} \\[0.5ex]
U_{4R}
\end{pmatrix}
=
\begin{pmatrix}
u_{1R} \\[0.5ex]
u_{2R}^{\prime \prime} \\[0.5ex]
u_{3R}^{\prime \prime} \\[0.5ex]
\widetilde{u}_{4R}^{\prime \prime} \\[0.5ex]
u_{4R}^{\prime \prime}
\end{pmatrix}
&=
V_{R}^{u}
\begin{pmatrix}
u_{1R} \\[0.5ex]
u_{2R} \\[0.5ex]
u_{3R} \\[0.5ex]
\widetilde{u}_{4R} \\[0.5ex]
u_{4R}
\end{pmatrix}
=
(V_{R}^{u})_{54} (V_{R}^{u})_{23} (V_{R}^{u})_{25} (V_{R}^{u})_{35} (V_{R}^{u})_{24} (V_{R}^{u})_{34}
\begin{pmatrix}
u_{1R} \\[0.5ex]
u_{2R} \\[0.5ex]
u_{3R} \\[0.5ex]
\widetilde{u}_{4R} \\[0.5ex]
u_{4R}
\end{pmatrix}.
\label{eqn:ana_up_mixing}
\end{split}
\end{equation}
\Huchan{As mentioned in subsection~\ref{sec:III_2}, this approximated step-by-step diagonalization for the up-quark sector requires more caution since some of the off-diagonal elements 
\Antonio{being of order unity} and appearing as a result of mixings can be sizeable due to the heavy top quark mass and \Antonio{the heavy exotic up type quark masses thus requiring  
the use of the numerical SVD technique} for the correct diagonalization and the SVD diagonalization will be used \Antonio{in our numerical scans in the main body of this work. 
The comparison} between a numerical mixing matrix derived \Antonio{from the SVD method and the one obtained from the analytic perturbative diagonalization will be} discussed in Appendix~\ref{app:E}.}
\section{Analytic approximated step-by-step diagonalization for the down-quark sector} \label{app:C}
We start from the initial down-type mass matrix given in Equation~\ref{eqn:diff_up_down} in the flavor basis.
\begin{equation}
M^{d }=\left( 
\begin{array}{c|ccccc}
& d _{1R} & d _{2R} & d _{3R} & d _{4R} & \widetilde{Q }_{4R} \\[0.5ex] \hline
\overline{Q }_{1L} & 0 & 0 & 0 & y_{14}^{d } v_{d} & 0 \\[1ex]
\overline{Q }_{2L} & 0 & 0 & 0 & y_{24}^{d } v_{d} & 0 \\[1ex] 
\overline{Q }_{3L} & 0 & 0 & 0 & y_{34}^{d } v_{d} & x_{34}^{Q } v_{\phi} \\[1ex]
\overline{Q }_{4L} & 0 & 0 & y_{43}^{d } v_{d} & 0 & M_{44}^{Q } \\[1ex]
\overline{\widetilde{d }}_{4L} & 0 & x_{42}^{d} v_{\phi}
& x_{43}^{d} v_{\phi} & M_{44}^{d} & 0 \\ 
\end{array}%
\right)
\end{equation}
As in the charged lepton case, it is convenient to rearrange the Yukawa mass terms by mass parameters and to swap the fourth and fifth column.
\begin{equation}
M^{d }
=
\left( 
\begin{array}{c|ccccc}
& d _{1R} & d _{2R} & d _{3R} & d _{4R} & \widetilde{Q }_{4R} \\[0.5ex] \hline
\overline{Q }_{1L} & 0 & 0 & 0 & m_{14}^{d} & 0 \\[1ex]
\overline{Q }_{2L} & 0 & 0 & 0 & m_{24}^{d} & 0 \\[1ex] 
\overline{Q }_{3L} & 0 & 0 & 0 & m_{34}^{d} & m_{35}^{d} \\[1ex]
\overline{Q }_{4L} & 0 & 0 & m_{43}^{d} & 0 & M_{44}^{Q } \\[1ex]
\overline{\widetilde{d }}_{4L} & 0 & m_{52}^{d}
& m_{53}^{d} & M_{44}^{d} & 0 \\ 
\end{array}%
\right)
=
\left( 
\begin{array}{c|ccccc}
& d _{1R} & d _{2R} & d _{3R} & \widetilde{Q }_{4R} & d _{4R} \\[0.5ex] \hline
\overline{Q }_{1L} & 0 & 0 & 0 & 0 & m_{14}^{d} \\[1ex]
\overline{Q }_{2L} & 0 & 0 & 0 & 0 & m_{24}^{d} \\[1ex] 
\overline{Q }_{3L} & 0 & 0 & 0 & m_{35}^{d} & m_{34}^{d} \\[1ex]
\overline{Q }_{4L} & 0 & 0 & m_{43}^{d} & M_{44}^{Q } & 0 \\[1ex]
\overline{\widetilde{d }}_{4L} & 0 & m_{52}^{d}
& m_{53}^{d} & 0 & M_{44}^{d} \\ 
\end{array}%
\right)
\end{equation}
In order to proceed from the flavor basis to the intermediate mass basis, the first thing to do is to carry out the $SU(2)$ conserving mixings $\theta_{34L}^{d}$ and $\theta_{24,34R}^{d}$ and we display the intermediate mass matrix for the down-type quarks without middle steps since the process is exactly same as the charged lepton case (After calculating all mixings required, we simplified the calculated mass parameters by $m^{\prime}$).
\begin{equation}
M^{d \prime}
=
V_{34L}^{d} M^{d} (V_{34R}^{d})^{\dagger} (V_{24R}^{d})^{\dagger} 
= 
\left(
\begin{array}{c|ccccc}
 & d_{1R} & d_{2R}^{\prime} & d_{3R}^{\prime} & \widetilde{d}_{4R} & d_{4R}^{\prime \prime} \\
 \hline
\overline{d}_{1L} & 0 & m_{12}^{d \prime} & m_{13}^{d \prime} & 0 & m_{14}^{d \prime} \\
\overline{d}_{2L} & 0 & m_{22}^{d \prime} & m_{23}^{d \prime} & 0 & m_{24}^{d \prime} \\
\overline{d}_{3L}^{d \prime} & 0 & m_{32}^{d \prime} & m_{33}^{d \prime} & 0 & m_{34}^{d \prime} \\
\overline{d}_{4L}^{\prime} & 0 & m_{42}^{d \prime} & m_{43}^{d \prime} & M_{45}^{Q \prime}
   & m_{44}^{d \prime} \\
\overline{\widetilde{d}}_{4L} & 0 & 0 & 0 & 0 & M_{54}^{d \prime \prime} \\
\end{array}
\right)
\end{equation}
We should carry out the $SU(2)$ violating mixings to turn off the mass parameters $m_{14,24,34,42,43}^{d \prime}$ and the mixing angles are very suppressed by the ratio between Yukawa and vector-like masses. Then the block diagonal form of this mass matrix appears as follows:
\begin{equation}
M^{d \prime \prime} = V_{15L}^{d} V_{25L}^{d} V_{35L}^{d} M^{d \prime} (V_{35R}^{d})^{\dagger} (V_{25R}^{d})^{\dagger}
= 
\left(
\begin{array}{c|ccccc}
 & d_{1R} & d_{2R}^{\prime \prime} & d_{3R}^{\prime \prime} & \widetilde{d}_{4R}^{\prime \prime} & d_{4R}^{\prime \prime} \\
 \hline
\overline{d}_{1L}^{\prime} & 0 & m_{12}^{d \prime} & m_{13}^{d \prime} & 0 & 0 \\
\overline{d}_{2L}^{\prime} & 0 & m_{22}^{d \prime} & m_{23}^{d \prime} & 0 & 0 \\
\overline{d}_{3L}^{\prime \prime} & 0 & m_{32}^{d \prime} & m_{33}^{d \prime} & 0 & 0 \\
\overline{d}_{4L}^{\prime} & 0 & 0 & 0 & M_{45}^{Q \prime}
   & m_{44}^{d \prime} \\
\overline{\widetilde{d}}_{4L}^{\prime \prime \prime} & 0 & 0 & 0 & 0 & M_{54}^{d \prime \prime} \\
\end{array}
\right)
\label{eqn:md_block_diagonal}
\end{equation}
An important feature of the mass matrix of Equation~\ref{eqn:md_block_diagonal} is the mass parameters of the first row is proportional to those of the second row by a factor (In other words, $m_{12}^{d \prime} / m_{22}^{d \prime} = m_{13}^{d \prime} / m_{23}^{d \prime}$. We follow the convention to diagonalize the upper-left $3 \times 3$ block~\cite{King:2002nf} rather than simply rotating the upper-left block. As the mass matrix of Equation~\ref{eqn:md_block_diagonal} consists of only real numbers, we can exclude the complex numbers in the convention and the convention is given by:
\begin{equation}
V_{12L}^{d} V_{13L}^{d} V_{23L}^{d} M^{d \prime \prime} (V_{23R}^{d})^{\dagger} (V_{13R}^{d})^{\dagger} (V_{12R}^{d})^{\dagger} = \func{diag} \left( 0 , m_{s}, m_{b}, M_{D_4}, M_{\widetilde{D}_4} \right)
\end{equation}
and then we arrive to the fully diagonalized mass matrix, which reveals all propagating mass for the down-type quarks. Then, the connection from the flavor to mass basis for the down-type quarks can be seen via the unitary mixing matrices as follows (notice that the right-handed down-type quark mixing matrices $(V_{R}^{d})_{12,13}$ remain as an identity matrix as the relevant mass matrix has the form of 
$\begin{pmatrix}
0 & m_a \\
0 & m_b
\end{pmatrix}$ 
and this form generally induces only left-handed mixing matrices).
\begin{equation}
\begin{split}
\begin{pmatrix}
d_{L} \\
s_{L} \\
b_{L} \\
D_{4L} \\
\widetilde{D}_{4L}
\end{pmatrix}
=
\begin{pmatrix}
d_{1L}^{\prime \prime \prime} \\
d_{2L}^{\prime \prime \prime} \\
d_{3L}^{\prime \prime \prime \prime} \\
d_{4L}^{\prime \prime} \\
\widetilde{d}_{4L}^{\prime \prime \prime \prime}
\end{pmatrix}
&=
V_{L}^{d}
\begin{pmatrix}
d_{1L} \\
d_{2L} \\
d_{3L} \\
d_{4L} \\
\widetilde{d}_{4L}
\end{pmatrix}
=
(V_{L}^{d})_{45} (V_{L}^{d})_{12} (V_{L}^{d})_{13} (V_{L}^{d})_{23} (V_{L}^{d})_{15} (V_{L}^{d})_{25} (V_{L}^{d})_{35} (V_{L}^{d})_{34}  
\begin{pmatrix}
d_{1L} \\
d_{2L} \\
d_{3L} \\
d_{4L} \\
\widetilde{d}_{4L}
\end{pmatrix}
\\
\begin{pmatrix}
d_{R} \\
s_{R} \\
b_{R} \\
D_{4R} \\
\widetilde{D}_{4R}
\end{pmatrix}
=
\begin{pmatrix}
d_{1R} \\
d_{2R}^{\prime \prime \prime} \\
d_{3R}^{\prime \prime \prime} \\
\widetilde{d}_{4R}^{\prime \prime \prime} \\
d_{4R}^{\prime \prime \prime}
\end{pmatrix}
&=
V_{R}^{d}
\begin{pmatrix}
d_{1R} \\
d_{2R} \\
d_{3R} \\
\widetilde{d}_{4R} \\
d_{4R}
\end{pmatrix}
=
(V_{R}^{d})_{54} (V_{R}^{d})_{23} (V_{R}^{d})_{25} (V_{R}^{d})_{35} (V_{R}^{d})_{24} (V_{R}^{d})_{34}
\begin{pmatrix}
d_{1R} \\
d_{2R} \\
d_{3R} \\
\widetilde{d}_{4R} \\
d_{4R}
\end{pmatrix}
\label{eqn:down_mix2}
\end{split}
\end{equation}
\section{Numerical comparison for the charged lepton sector} \label{app:D}
\Antonio{We have previouly stressed that the analytical charged lepton mixing matrix 
is quite close to the numerical one 
and we will compare them 
in this Appendix. For this comparison, we start from the 
charged lepton mass matrix in the flavor basis, evaluated in one of the benchmark points used in our numerical scans:} 
\begin{equation}
M^{e}
=
\begin{pmatrix}
0 & 0 & 0 & 0 & 0 \\
0 & 0 & 0 & 0 & -2.151 \\
0 & 0 & 0 & 161.657 & 3.955 \\
0 & 0 & 4.600 & 536.050 & 0 \\
0 & 51.135 & 97.915 & 0 & 696.178
\end{pmatrix}
\end{equation}
Firstly, we evaluate the mixing matrices $V^{L,e}$ using the analytic mixings \Antonio{of} Equation~\ref{eqn:cl_umm_LH_RH}. The analytic mixing matrices $V_{\func{ana}}^{L,e}$ are given by: 
\begin{equation}
\begin{split}
V_{\func{ana}}^{L} 
&=
\left(
\begin{array}{ccccc}
 1. & 0. & 0. & 0. & 0. \\
 0. & 0.985598 & 0.161888 & -0.0488209 & 0.00211728 \\
 0. & 0.169076 & -0.943613 & 0.284567 & 0.00548067 \\
 0. & 0.00002015 & 0.288689 & 0.957399 & -0.00668649 \\
 0. & -0.00301343 & 0.00675946 & 0.00494558 & 0.99996 \\
\end{array}
\right)
\\
V_{\func{ana}}^{e}
&=
\left(
\begin{array}{ccccc}
 1. & 0. & 0. & 0. & 0. \\
 0. & 0.986254 & -0.157386 & 0.00134068 & -0.050305 \\
 0. & 0.14846 & 0.97769 & -0.00738828 & -0.148413 \\
 0. & -0.000610678 & 0.00669679 & 0.999958 & -0.00627468 \\
 0. & 0.0725407 & 0.138946 & 0.00531106 & 0.987625 \\
\end{array}
\right)
\end{split}
\end{equation}
Notice that the mixing matrices $V_{L,R}^{u}$ and $V_{R}^{d}$ have exactly the same structure than \Antonio{the charged lepton mixing matrix since} 
 all off-diagonal elements in the first row and column are zero, however the mixing matrix $V_{L}^{d}$ is different as it can have mixings with the down-type first generation as seen in Equation~\ref{eqn:down_mixing})

The mixing matrices $V_{\func{num}}^{L,e}$ derived by the numerical SVD are given by:
\begin{equation}
\begin{split}
V_{\func{num}}^{L} 
&=
\left(
\begin{array}{ccccc}
 1. & 0. & 0. & 0. & 0. \\
 0. & 0.985598 & 0.161888 & -0.0488206 & 0.00211728 \\
 0. & 0.169076 & -0.94362 & 0.284543 & 0.00548077 \\
 0. & -0.0000241021 & -0.288666 & -0.957407 & 0.00668612 \\
 0. & -0.00301342 & 0.00675933 & 0.00494538 & 0.99996 \\
\end{array}
\right)
\\
V_{\func{num}}^{e}
&=
\left(
\begin{array}{ccccc}
 1. & 0. & 0. & 0. & 0. \\
 0. & 0.986254 & -0.157386 & 0.00134068 & -0.0503047 \\
 0. & 0.148461 & 0.977693 & -0.00738841 & -0.148398 \\
 0. & 0.000610624 & -0.00669703 & -0.999958 & 0.00627442 \\
 0. & 0.0725382 & 0.138932 & 0.00531088 & 0.987627 \\
\end{array}
\right)
\end{split}
\end{equation}
The difference between the mixing matrices can be easily seen by subtracting one from another after taking absolute value.
\begin{equation}
\begin{split}
\lvert V_{\func{ana}}^{L} \rvert - \lvert V_{\func{num}}^{L} \rvert
&=
\left(
\begin{array}{ccccc}
 0 & 0 & 0 & 0 & 0 \\
 0 & 0 & -7.58945\times 10^{-8} & 2.51664\times 10^{-7} & 5.97371\times 10^{-10} \\
 0 & 3.46904\times 10^{-10} & -7.19169\times 10^{-6} & 2.38502\times 10^{-5} & -1.00783\times 10^{-7} \\
 0 & -3.95207\times 10^{-6} & 2.35474\times 10^{-5} & -7.10253\times 10^{-6} & 3.68376\times 10^{-7} \\
 0 & 9.97344\times 10^{-9} & 1.29353\times 10^{-7} & 2.03736\times 10^{-7} & -1.91205\times 10^{-9} \\
\end{array}
\right)
\\
\lvert V_{\func{ana}}^{e} \rvert - \lvert V_{\func{num}}^{e} \rvert
&=
\left(
\begin{array}{ccccc}
 0 & 0 & 0 & 0 & 0 \\
 0 & -2.39211\times 10^{-8} & 4.58026\times 10^{-8} & 9.72974\times 10^{-10} & 3.25661\times 10^{-7} \\
 0 & -1.06479\times 10^{-6} & -2.03783\times 10^{-6} & -1.29588\times 10^{-7} & 1.44968\times 10^{-5} \\
 0 & 5.47244\times 10^{-8} & -2.43141\times 10^{-7} & 0 & 2.58046\times 10^{-7} \\
 0 & 2.50398\times 10^{-6} & 1.42997\times 10^{-5} & 1.84594\times 10^{-7} & -2.19659\times 10^{-6} \\
\end{array}
\right)
\end{split}
\end{equation}
\Antonio{Therefore we have confirmed that the analytic mixing matrix for the charged lepton sector is quite close to one obtained from} 
the numerical SVD diagonalization. Using the numerical mixing matrices derived by the SVD \Antonio{diagonalization, we confirm the following $D_{L,R}^{e \prime}$ matrices of $Z$ couplings with leptons 
of Equation~\ref{eqn:simple_DLep_DRep} as follows (we included here the pre-factor $g/c_w$):}
\begin{equation}
\begin{split}
D_{L}^{e \prime} 
&=
\left(
\begin{array}{ccccc}
 -2.01645\times 10^{-1} & 0. & 0. & 0. & 0. \\
 0. & -2.01643\times 10^{-1} & 4.22223\times 10^{-6} & 5.1508\times 10^{-6} & 7.70341\times 10^{-4} \\
 0. & 4.22223\times 10^{-6} & -2.01634\times 10^{-1} & 1.33333\times 10^{-5} & 1.9941\times 10^{-3} \\
 0. & 5.1508\times 10^{-6} & 1.33333\times 10^{-5} & -2.01629\times 10^{-1} & 2.43264\times 10^{-3} \\
 0. & 7.70341\times 10^{-4} & 1.9941\times 10^{-3} & 2.43264\times 10^{-3} & 1.62175\times 10^{-1} \\
\end{array}
\right)
\\
D_{R}^{e \prime}
&=
\left(
\begin{array}{ccccc}
 1.62204\times 10^{-1} & 0. & 0. & 0. & 0. \\
 0. & 1.62203\times 10^{-1} & 3.60409\times 10^{-6} & 4.87783\times 10^{-4} & -2.59066\times 10^{-6} \\
 0. & 3.60409\times 10^{-6} & 1.62184\times 10^{-1} & -2.68815\times 10^{-3} & 1.4277\times 10^{-5} \\
 0. & 4.87783\times 10^{-4} & -2.68815\times 10^{-3} & -2.01614\times 10^{-1} & 1.93227\times 10^{-3} \\
 0. & -2.59066\times 10^{-6} & 1.4277\times 10^{-5} & 1.93227\times 10^{-3} & 1.62194\times 10^{-1} \\
\end{array}
\right)
\end{split}
\end{equation}
\section{Numerical comparison for the quark sector} \label{app:E}
As we did in Appendix~\ref{app:D}, we carry out the same approach with \Antonio{the} most converged numerical point ($\chi_{\func{CKM}}^2 = 956.828$) for the up- and down-type quark \Antonio{sectors}. 
\begin{equation}
\begin{split}
M^{u} &= 
\begin{pmatrix}
0 & 0 & 0 & 0 & 0 \\[0.5ex]
0 & 0 & 0 & 0 & 14.474 \\[0.5ex]
0 & 0 & 0 & 1206.340 & 277.563 \\[0.5ex]
0 & 0 & 273.503 & -1775.200 & 0 \\[0.5ex]
0 & 550.990 & 434.462 & 0 & -5624.050 
\end{pmatrix}
\\
M^{d} 
&= 
\begin{pmatrix}
0 & 0 & 0 & 0 & -0.938 \\[0.5ex]
0 & 0 & 0 & 0 & -4.041 \\[0.5ex]
0 & 0 & 0 & 1206.340 & -27.427 \\[0.5ex]
0 & 0 & -5.636 & -1775.200 & 0 \\[0.5ex]
0 & 72.915 & -75.760 & 0 & 2623.620
\end{pmatrix}
\end{split}
\end{equation}
For the comparison, we find the analytic mixing matrices $(V_{L,R}^{u,d})_{\func{ana}}$ and the numerical mixing matrices $(V_{L,R}^{u,d})_{\func{num}}$ and then subtract one from another after taking absolute value. The numerical differences are given by:
\begin{equation}
\begin{split}
\lvert (V_{L}^{u})_{\func{ana}} \rvert - \lvert (V_{L}^{u})_{\func{num}} \rvert 
&= 
\left(
\begin{array}{ccccc}
 0 & 0 & 0 & 0 & 0 \\
 0 & -9.63899\times 10^{-10} & -8.4547\times 10^{-6} & 1.12194\times 10^{-5} & 5.10645\times 10^{-7} \\
 0 & 8.60152\times 10^{-7} & -4.562\times 10^{-3} & 6.12434\times 10^{-3} & -3.20031\times 10^{-4} \\
 0 & -4.32016\times 10^{-5} & 6.11329\times 10^{-3} & -4.57314\times 10^{-3} & 3.70205\times 10^{-4} \\
 0 & -6.95382\times 10^{-8} & 3.44869\times 10^{-5} & 4.09197\times 10^{-5} & -2.39443\times 10^{-6} \\
\end{array}
\right)
\\
\lvert (V_{R}^{u})_{\func{ana}} \rvert - \lvert (V_{R}^{u})_{\func{num}} \rvert 
&=
\left(
\begin{array}{ccccc}
 0 & 0 & 0 & 0 & 0 \\
 0 & -5.36541\times 10^{-8} & 4.20284\times 10^{-8} & 3.73725\times 10^{-8} & 4.77149\times 10^{-7} \\
 0 & -1.56456\times 10^{-4} & -4.71406\times 10^{-5} & -7.74521\times 10^{-4} & 1.34741\times 10^{-3} \\
 0 & 1.27331\times 10^{-4} & -7.96888\times 10^{-4} & 6.8947\times 10^{-5} & 5.76999\times 10^{-5} \\
 0 & 1.86671\times 10^{-4} & 1.32945\times 10^{-3} & 2.03335\times 10^{-5} & -1.39892\times 10^{-4} \\
\end{array}
\right)
\\
\lvert (V_{L}^{d})_{\func{ana}} \rvert - \lvert (V_{L}^{d})_{\func{num}} \rvert 
&=
\left(
\begin{array}{ccccc}
 0 & 0 & 0 & 0 & 0 \\
 7.62437\times 10^{-9} & 3.28418\times 10^{-8} & -7.80564\times 10^{-7} & -5.52416\times 10^{-7} & 9.20155\times 10^{-9} \\
 -2.16144\times 10^{-7} & -9.31034\times 10^{-7} & -2.34437\times 10^{-6} & 3.16788\times 10^{-6} & -5.46568\times 10^{-8} \\
 -2.49349\times 10^{-8} & -1.07406\times 10^{-7} & 3.14516\times 10^{-6} & -2.37145\times 10^{-6} & 9.52623\times 10^{-8} \\
 0 & 3.34474\times 10^{-10} & 5.01567\times 10^{-8} & 3.51026\times 10^{-8} & -1.24719\times 10^{-9} \\
\end{array}
\right)
\\
\lvert (V_{R}^{d})_{\func{ana}} \rvert - \lvert (V_{R}^{d})_{\func{num}} \rvert 
&=
\left(
\begin{array}{ccccc}
 0 & 0 & 0 & 0 & 0 \\
 0 & 6.97453\times 10^{-6} & -4.09035\times 10^{-5} & -7.75726\times 10^{-8} & 1.54578\times 10^{-6} \\
 0 & -4.13664\times 10^{-5} & 6.52624\times 10^{-6} & -1.24584\times 10^{-7} & 1.44401\times 10^{-5} \\
 0 & 6.55855\times 10^{-8} & -3.67395\times 10^{-7} & -3.73704\times 10^{-10} & 6.10094\times 10^{-8} \\
 0 & 2.63961\times 10^{-6} & 1.5191\times 10^{-5} & 4.55174\times 10^{-8} & -5.58627\times 10^{-7} \\
\end{array}
\right)
\end{split}
\end{equation} 
Here we can see that the differences for the charged lepton or down-type quark \Antonio{sectors are at most of the order of maximally order of 
$10^{-5}$, whereas the maximal difference for the up-quark sector goes up to the order of $10^{-3}$ due to the sizeable off-diagonal $\mathcal{O}(1)$ element but it is still a good approximation. Now we confirm that the numerical matrices of $Z$ couplings with quarks 
$D_{L,R}^{u,d \prime}$ in the mass basis are given by: }
\begin{equation}
\begin{split}
D_{L}^{u \prime} 
&= 
\left(
\begin{array}{ccccc}
 2.55713\times 10^{-1} & 0 & 0 & 0 & 0 \\
 0 & 2.55711\times 10^{-1} & -3.14988\times 10^{-5} & 3.18699\times 10^{-5} & -7.55591\times 10^{-4} \\
 0 & -3.14988\times 10^{-5} & 2.55083\times 10^{-1} & 6.37525\times 10^{-4} & -1.51148\times 10^{-2} \\
 0 & 3.18699\times 10^{-5} & 6.37525\times 10^{-4} & 2.55068\times 10^{-1} & 1.52929\times 10^{-2} \\
 0 & -7.55591\times 10^{-4} & -1.51148\times 10^{-2} & 1.52929\times 10^{-2} & -1.06859\times 10^{-1} \\
\end{array}
\right)
\\
D_{R}^{u \prime}
&=
\left(
\begin{array}{ccccc}
 -1.08136\times 10^{-1} & 0 & 0 & 0 & 0 \\
 0 & -1.07941\times 10^{-1} & -7.52723\times 10^{-4} & 8.3786\times 10^{-3} & 1.3608\times 10^{-4} \\
 0 & -7.52723\times 10^{-4} & -1.05225\times 10^{-1} & -3.24002\times 10^{-2} & -5.26224\times 10^{-4} \\
 0 & 8.3786\times 10^{-3} & -3.24002\times 10^{-2} & 2.52512\times 10^{-1} & 5.85742\times 10^{-3} \\
 0 & 1.3608\times 10^{-4} & -5.26224\times 10^{-4} & 5.85742\times 10^{-3} & -1.08041\times 10^{-1} \\
\end{array}
\right)
\\
D_{L}^{d \prime}
&=
\left(
\begin{array}{ccccc}
 -3.09781\times 10^{-1} & 0 & 0 & 0 & 0 \\
 0 & -3.0978\times 10^{-1} & 4.35109\times 10^{-6} & 8.49277\times 10^{-6} & 4.62349\times 10^{-4} \\
 0 & 4.35109\times 10^{-6} & -3.09749\times 10^{-1} & 6.287\times 10^{-5} & 3.42266\times 10^{-3} \\
 0 & 8.49277\times 10^{-6} & 6.287\times 10^{-5} & -3.09658\times 10^{-1} & 6.68059\times 10^{-3} \\
 0 & 4.62349\times 10^{-4} & 3.42266\times 10^{-3} & 6.68059\times 10^{-3} & 5.39124\times 10^{-2} \\
\end{array}
\right)
\\
D_{R}^{d \prime}
&=
\left(
\begin{array}{ccccc}
 5.4068\times 10^{-2} & 0 & 0 & 0 & 0 \\
 0 & 5.40678\times 10^{-2} & 4.0453\times 10^{-7} & 2.15284\times 10^{-4} & -3.09181\times 10^{-6} \\
 0 & 4.0453\times 10^{-7} & 5.40667\times 10^{-2} & -6.8355\times 10^{-4} & 9.81685\times 10^{-6} \\
 0 & 2.15284\times 10^{-4} & -6.8355\times 10^{-4} & -3.09705\times 10^{-1} & 5.22435\times 10^{-3} \\
 0 & -3.09181\times 10^{-6} & 9.81685\times 10^{-6} & 5.22435\times 10^{-3} & 5.39929\times 10^{-2} \\
\end{array}
\right)
\end{split}
\label{eqn:numerical_DLdp}
\end{equation}
\Antonio{where the pre-factor $g/c_w$ was included in those matrices.} The most interesting case of Equation~\ref{eqn:numerical_DLdp} is $D_{L}^{d \prime}$ since we know that the left-handed down-type quark sector can access to all mixings among the three SM generations. The used numerical mixing matrix $V_{L}^{d}$ is given by:
\begin{equation}
V_{L}^{d}
=
\left(
\begin{array}{ccccc}
 9.74095\times 10^{-1} & -2.26141\times 10^{-1} & 0 & 0 & 0 \\
 2.26\times 10^{-1} & 9.73488\times 10^{-1} & -2.81626\times 10^{-2} & -2.12226\times 10^{-2} & 1.27099\times 10^{-3} \\
 7.97153\times 10^{-3} & 3.43371\times 10^{-2} & 7.98097\times 10^{-1} & 6.01423\times 10^{-1} & 9.40883\times 10^{-3} \\
 -6.67815\times 10^{-6} & -2.87659\times 10^{-5} & 6.01585\times 10^{-1} & -7.98598\times 10^{-1} & 1.83648\times 10^{-2} \\
 -3.62201\times 10^{-4} & -1.56017\times 10^{-3} & -1.85253\times 10^{-2} & 9.03631\times 10^{-3} & 9.99786\times 10^{-1} \\
\end{array}
\right),
\label{eqn:a_num_VCKM}
\end{equation}
and we can confirm all elements of the first row and column of $D_{L}^{d \prime}$ cancel each other, so identifying the given result in Equation~\ref{eqn:numerical_DLdp}. What we found in this Appendix verifies the fact that the SM $Z$ physics does not get affected by any \Antonio{specific choice of basis.} 

\end{document}